\documentclass{aa}

\usepackage{natbib}
\bibpunct{(}{)}{;}{a}{}{,} % to follow the A&A style

\usepackage{latexsym,exscale,amssymb,amsmath}
\usepackage{eurosym}
\usepackage{color}
\usepackage{graphicx}
\usepackage[percent]{overpic}
\usepackage{hyperref}

\mathchardef\mhyphen="2D

\title{\bf Expanding bubbles in Orion A: [C\,{\sc ii}] observations of M42, M43, and NGC 1977}
\titlerunning{Bubbles of the Orion Nebula, M43, and NGC 1977}
\author{C.H.M. Pabst\inst{\ref{inst1}} \and J.R. Goicoechea\inst{\ref{inst2}} \and D. Teyssier\inst{\ref{inst3}} \and O. Bern\'{e}\inst{\ref{inst4}} \and R.D. Higgins\inst{\ref{inst6}} \and E. T. Chambers\inst{\ref{inst7}} \and S. Kabanovic\inst{\ref{inst6}} \and R. G\"{u}sten\inst{\ref{inst5}} \and J. Stutzki\inst{\ref{inst6}} \and A.G.G.M. Tielens\inst{\ref{inst1}} }

\institute{Leiden Observatory, Leiden University, P.O. Box 9513, 2300 RA Leiden, Netherlands\label{inst1}; \href{mailto:pabst@strw.leidenuniv.nl}{\texttt{pabst@strw.leidenuniv.nl}}
\and Instituto de Fisica Fundamental, CSIC, Calle Serrano 121-123, 28006 Madrid, Spain \label{inst2} 
\and Telespazio Vega UK Ltd. for ESA/ESAC, Urbanizacion Villafranca del Castillo, 28691 Madrid, Spain \label{inst3}
\and IRAP, Universit\'{e} de Toulouse, CNRS, CNES, UPS, 9 Av. colonel Roche, 31028 Toulouse Cedex 4, France \label{inst4}
\and I. Physikalisches Institut der Universit\"{a}t zu K\"{o}ln, Z\"{u}lpicher Strasse 77, 50937 K\"{o}ln, Germany\label{inst6}
\and Max-Planck-Institut f\"{u}r Radioastronomie, Auf dem H\"{u}gel 69, 53121 Bonn, Germany\label{inst5}
\and USRA/SOFIA, NASA Ames Research Center, Mail Stop 232-12, Building N232, P.O. Box 1, Moffett Field, CA 94035-0001, USA\label{inst7}
}
\date{Received 23 January 2020 / Accepted 22 April 2020}

\abstract{{\bf\sf Context}: The Orion Molecular Cloud is the nearest massive-star forming region. Massive stars have profound effects on their environment due to their strong radiation fields and stellar winds. Stellar feedback is one of the most crucial cosmological parameters that determine the properties and evolution of the interstellar medium in galaxies.\\
{\bf\sf Aims}: We aim to understand the role that feedback by stellar winds and radiation play in the evolution of the interstellar medium. Velocity-resolved observations of the [C\,{\sc ii}] $158\,\mu\mathrm{m}$ fine-structure line allow us to study the kinematics of UV-illuminated gas. Here, we present a square-degree-sized map of [C\,{\sc ii}] emission from the Orion Nebula complex at a spatial resolution of $16\arcsec$ and high spectral resolution of $0.2\,\mathrm{km\,s^{-1}}$, covering the entire Orion Nebula (M42) plus M43 and the nebulae NGC 1973, 1975, and 1977 to the north. We compare the stellar characteristics of these three regions with the kinematics of the expanding bubbles surrounding them.\\
{\bf\sf Methods}: We use [C\,{\sc ii}] $158\,\mu\mathrm{m}$ line observations over an area of $1.2\,\mathrm{deg}^2$ in the Orion Nebula complex obtained by the upGREAT instrument onboard SOFIA.\\
{\bf\sf Results}: The bubble blown by the O7V star $\theta^1$ Ori C in the Orion Nebula expands rapidly, at $13\,\mathrm{km\,s^{-1}}$. Simple analytical models reproduce the characteristics of the hot interior gas and the neutral shell of this wind-blown bubble and give us an estimate of the expansion time of $0.2\,\mathrm{Myr}$. M43 with the B0.5V star NU Ori also exhibits an expanding bubble structure, with an expansion velocity of $6\,\mathrm{km\,s^{-1}}$. Comparison with analytical models for the pressure-driven expansion of H\,{\sc ii} regions gives an age estimate of $0.02\,\mathrm{Myr}$. The bubble surrounding NGC 1973, 1975, and 1977 with the central B1V star 42 Orionis expands at $1.5\,\mathrm{km\,s^{-1}}$, likely due to the over-pressurized ionized gas as in the case of M43. We derive an age of $0.4\,\mathrm{Myr}$ for this structure.\\
{\bf\sf Conclusions}: We conclude that the bubble of the Orion Nebula is driven by the mechanical energy input by the strong stellar wind from $\theta^1$ Ori C, while the bubbles associated with M43 and NGC 1977 are caused by the thermal expansion of the gas ionized by their central later-type massive stars.
\vspace{0.5cm}
}


 \hypersetup{draft}

\begin{document}

\maketitle

\section{Introduction}

Stellar feedback, that is injection of energy and momentum from stars, is one of the most important input parameters in cosmological models that simulate and explain the evolution of our universe. Even small variations in this crucial parameter can lead to drastic changes in the results. Too much stellar feedback of early stars disrupts the ambient gas in too early a stage to form more stars and, eventually, planetary systems that allow for the formation of life. Too little feedback fails to prevent the interstellar gas from gravitational collapse to dense, cold clumps. Stellar feedback is often quantified as the star-formation rate (SFR), the rate at which (mostly low-mass) stars are formed. Studies on the interaction of massive stars with their environment generally focus on the effects of supernova explosions injecting mechanical energy into their environment and the radiative interaction leading to the ionization and thermal expansion of the gas. The explosion of a massive star ejects some $10\,M_{\sun}$ at velocities of some $10,000\,\mathrm{km\,s^{-1}}$, injecting about $10^{51}\,\mathrm{erg}$ into the surrounding medium. The hot plasma created by the reverse shock drives the expansion of the supernova remnant. The concerted effects of many supernovae in an OB association create superbubbles that expand perpendicular to the plane and may break open, releasing the hot plasma and any entrained colder cloud material into the lower halo \citep{McKeeOstriker1977, McCray1987, MacLowMcCray1988, NormanIkeuchi1989, Hopkins2012}. 

However, there is evidence that also stellar winds have a profound impact on the interstellar medium (ISM). Observations and models have long attested to the importance of stellar winds from massive stars as sources of mechanical energy that can have profound influence on the direct environment \citep{Castor1975, Weaver1977, Wareing2018, Pabst2019}. These stellar winds drive a strong shock into its surroundings, which sweeps up ambient gas into dense shells. At the same time, the reverse shock stops the stellar wind, creating a hot tenuous plasma. Eventually, the swept up shells break open, venting their hot plasma into the surrounding medium.

Feedback by ionization of the gas surrounding a massive star will also lead to the disruption of molecular clouds \citep{WilliamsMcKee1997, Freyer2006, Dale2013}. Initially, this will be through the more or less spherical expansion driven by the high pressure of the H\,{\sc ii} region \citep{Spitzer1978}. Once this expanding bubble breaks open into the surrounding low density material, a champagne flow will be set up \citep{BedijnTenorio-Tagle1981}, rapidly removing material from the molecular cloud.

Bubble structures are ubiquitous in the ISM \citep[e.g.,][]{Churchwell2006}. Most of these are caused by expanding H\,{\sc ii} regions \citep{Walch2013, Ochsendorf2014}. However, due to the clumpy structure of the ISM, and molecular clouds in particular, the expanding shock fronts can be highly irregular. Depending on the morphology of the shock front, that is the formation of a shell and more or less massive clumps, massive star formation can be either triggered or hindered \citep[e.g.,][]{Walch2012}.

The relative importance of these three different feedback processes (SNe, stellar winds, thermal expanding H\,{\sc ii} regions) is controversial. Given the peculiar motion of massive stars, SN explosions may occur at relatively large distances from their natal clouds and therefore have little effect on the cloud as the energy is expended in rejuvenating plasma from previous SN explosions, sweeping up tenuous intercloud material and transporting it to the superbubble walls \citep{McKeeOstriker1977, MacLowMcCray1988, NormanIkeuchi1989, Ochsendorf2015}. While the mechanical energy output in terms of stellar winds is much less than from SN explosions, they will act directly on the natal cloud.

The Orion molecular cloud is the nearest region of massive star formation and has been studied in a wide range of wavelengths. The dense molecular cloud is arranged in an integral-shaped filament (ISF), which has fragmented into four cores, OMC1, 2, 3, and 4, that are all active sites of star formation. At the front side of the most massive core, OMC1, the massive O7V star $\theta^1$ Ori C has ionized the well-known H\,{\sc ii} region M42 (NGC 1976), that is known as the Orion Nebula. The Orion Nebula, appearing as the middle "star" in the sword of Orion at a distance of $d\simeq 414\,\mathrm{pc}$ from us \citep{Menten2007}, is one of the most picturesque structures of our universe. The strong stellar wind from this star has created a bubble filled with hot plasma that is rapidly expanding into the lower-density gas located at the front of the core \citep{Guedel2008, Pabst2019}. In addition, two slightly less massive stars, B0.5V-type NU Ori and B1V-type 42 Orionis, have created their own ionized gas bubbles, M43 and NGC 1977. These stars are not expected to have strong stellar winds and likely the thermal pressure of the ionized gas dominates the expansion of these two regions.

\begin{figure*}[tb]
\includegraphics[width=\textwidth, height=0.367\textwidth]{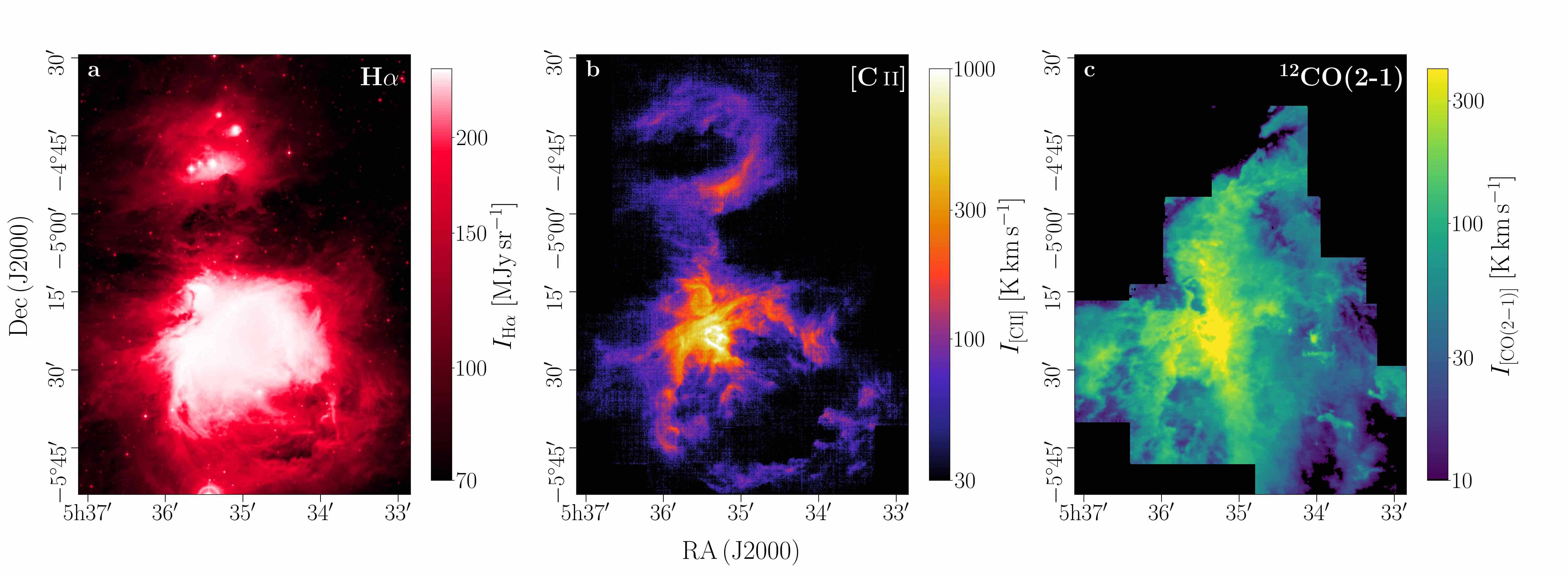}
\caption{Overview of the Orion Nebula (M42), M43, and NGC 1973, 1975, and 1977 in different wavelenghts. Left: DSS2 H$\alpha$ emission (ESO Archive). Center: [C\,{\sc ii}] line-integrated emission \citep{Pabst2019}. Right: Line-integrated $^{12}$CO(2-1) emission \citep{Berne2014, Goicoechea2020}. H$\alpha$ emission stems from the ionized gas ($T\sim 10^4\,\mathrm{K}$), the [C\,{\sc ii}] line is emitted by mostly neutral gas ($T\sim 100\,\mathrm{K}$), whereas CO traces the molecular gas ($T\sim 30\,\mathrm{K}$). M42 and M43 have a substantial amount of molecular gas in the background, while NGC 1977 seems to be devoid of it (the coverage of the CO map is not sufficient, however, see discussion of the expansion characteristics below). All three regions comprise ionized gas within their neutral limb-brightened shells.}
\label{Fig.overview}
\end{figure*}

The current picture is as follows\footnote{For a 3D journey through the Orion Nebula see \texttt{https://www.jpl.nasa.gov/news/news.php?feature=7035\\ \&utm\_source=iContact\&utm\_medium=email\&utm\_campaign=\\NASAJPL\& utm\_content=daily20180111-4}.}: The Trapezium cluster including $\theta^1$ Ori C, the most massive stars in the Orion Nebula complex, is situated at the surface of its natal molecular cloud, the OMC1 region of the Orion A molecular cloud. Presumably, the Trapezium stars live in a valley of the molecular cloud, having swept away the covering cloud layers (see \cite{ODell2009} for a detailed discussion of the structure of the inner Orion Nebula). In the immediate environment of the Trapezium stars, the gas is warm and ionized, constituting a dense H\,{\sc ii} region, the Huygens Region. It is confined towards the back by the dense molecular cloud core and this is exemplified by the prominent ionization front to the south, the Orion Bar, a narrow and very dense structure \cite{Tielens1993, Goicoechea2016}. To the west of the Trapezium stars, embedded in the molecular gas of OMC1, one finds two dense cores with violent emission features, strong outflows and shocked molecular gas, that are the sites of active intermediate- and high-mass star formation, the Becklin-Neugebauer/Kleinmann-Low (BN/KL) region and Orion South (S).

Further outward, the gas is still irradiated by the central Trapezium stars, granting us the beautiful images of the Orion Nebula. The outward regions were dubbed the Extended Orion Nebula \citep[EON,][]{Guedel2008}. It is, in fact, a closed circular structure of about $5\,\mathrm{pc}$ in diameter surrounding the bright Huygens Region, the latter being offset from its center. In the line of sight near the Trapezium stars in the Huygens Region, Orion's Veil \citep{ODell2000} is observed as a layer of (atomic and molecular) foreground gas at $1\mhyphen 3\,\mathrm{pc}$ distance from the blister of ionized gas that forms the bright Orion Nebula. H\,{\sc i} and optical/UV absorption lines reveal multiple velocity components, some with velocities that link them to the expanding bubble part of the Veil\footnote{We refer to Orion's Veil as the Veil in short. We will refer to the large-scale expanding bubble as the Veil shell, being part of the same structure that covers the Huygens Region.} seen across the EON \citep{VanderWerf2013, ODell2018, Abel2019}.

In the vicinity of the Orion Nebula, we find De Mairan's Nebula (NGC 1982 or M43) just to the north of the Huygens Region, and the Running-Man Nebula (NGC 1973, 1975, and 1977) further up north. M43 hosts the central B0.5V star NU Ori and is shielded from ionizing radiation from the Trapezium cluster by the Dark Bay \citep{ODell2010, Simon-Diaz2011}. NGC 1973, 1975, and 1977 possess as brightest star B1V star 42 Orionis in NGC 1977, which is the main ionizing source of that region \citep{Peterson2008}. The structure surrounding NGC 1973, 1975, and 1977 is dominated by the dynamics induced by 42 Orionis, hence, we will denote it by NGC 1977 only in the following.

The hot plasma filling stellar wind bubbles and the ionized gas of H\,{\sc ii} regions are both largely transparent for far-UV radiation. These non-ionizing photons are instead absorbed in the surrounding neutral gas, creating a warm layer of gas, the photodissociation region (PDR), which cools through atomic fine-structure lines \citep{HollenbachTielens1999}. The [C\,{\sc ii}] $158\,\mu\mathrm{m}$ fine-structure line of ionized carbon is the dominant far-infrared (FIR) cooling line of warm, intermediate density gas ($T\sim 50\mhyphen 300 \,\mathrm{K}$, $n \sim 10^3\mhyphen 10^4\,\mathrm{cm^{-3}}$). It can carry up to 2\% of the total FIR emission of the ISM, most of the FIR intensity arising from re-radiation of UV photons by interstellar dust grains. Velocity-resolved line observations provide a unique tool for the study of gas dynamics and kinematics. Velocity-resolved [C\,{\sc ii}] and [$^{13}$C\,{\sc ii}] observations towards the Huygens Region/OMC1, an area of about $7.5\arcmin\times 11.5\arcmin$, were obtained and analyzed by \cite{Goicoechea2015}. Here, we use a large-scale, $60\arcmin\times 80\arcmin$, study of velocity-resolved [C\,{\sc ii}] emission from the Orion Nebula M42, M43 and NGC 1973, 1975 and 1977\footnote{Two movie presentations of the velocity-resolved [C\,{\sc ii}] data are made available at \texttt{http://ism.strw.leidenuniv.nl/research.html\#CII}.}.

This paper is organized as follows. In Section 2, we review the observations that we used in the present study. In Section 3, we discuss the morphology of the shells associated with M42, M43, and NGC 1977 and derive gas masses and expansion velocities. Section 4 contains a discussion of the results presented in Section 3. We compare the observed shell kinematics with analytical models. We conclude with a summary of our results in Section 5.

\section{Observations}

\subsection{[C\,{\sc ii}] observations}

Velocity-resolved [C\,{\sc ii}] line observations towards the Orion Nebula complex, covering M42, M43 and NGC 1973, 1975, and 1977, were obtained during 13 flights in November 2016 and February 2017 using the 14-pixel high-spectral-resolution heterodyne array of the German Receiver for Astronomy at Terahertz Frequencies (upGREAT\footnote{upGREAT is a development by the MPI f\"ur Radioastronomie (Principal Investigator: R. G\"usten) and KOSMA/Universit\"at zu K\"oln, in cooperation with the MPI f\"ur Sonnensystemforschung and the DLR Institut f\"ur Optische Sensorsysteme.}, \cite{Risacher2016}) onboard the Stratospheric Observatory for Infrared Astronomy (SOFIA). We produced a fully-sampled map of a 1.2 square-degree-sized area at a angular resolution of $16\arcsec$. The full map region was observed in the array on-the-fly (OTF) mode in 78 square tiles, each $435.6\arcsec$ wide. Each tile consists of 84 scan lines separated by $5.2\arcsec$, covered once in both $x$ and $y$ direction. Each scan line consists of 84 dumps of $0.3\,\mathrm{s}$, resulting in a root-mean-square noise of $T_{\mathrm{mb}}\simeq 1.14\,\mathrm{K}$ per pixel at a spectral resolution of $0.3\,\mathrm{km\,s^{-1}}$. The original data at a native spectral resolution of $0.04\,\mathrm{km\,s^{-1}}$ were rebinned to $0.3\,\mathrm{km\,s^{-1}}$ channels to increase the signal-to-noise ratio. 90\% of the total 2.2 million spectra required no post-processing, while most problematic spectra could be recovered using a spline baselining approach. A catalogue of splines is generated from data containing no astronomical signal. These splines could then be scaled to the astronomical data and more effectively remove the baselines than a polynomial fit. For a detailed description of the observing strategy and data reductions steps see Higgins et al. (in prep.).

\subsection{CO observations}

We also make use of $^{12}$CO \mbox{$J$\,=\,2-1} (230.5\,GHz) and $^{13}$CO \mbox{$J$\,=\,2-1} (220.4\,GHz) line maps taken with the IRAM\,30\,m radiotelescope (Pico Veleta, Spain) at a native angular resolution of $10.7\arcsec$. The central region ($1^{\circ}\times0.8^{\circ}$) around OMC1 was originally mapped in 2008 with the HERA receiver array. \cite{Berne2014} presented the on-the-fly (OTF) mapping and data reduction strategies. In order to cover the same areas mapped by us in the [C\,{\sc ii}] line, we started to enlarge these CO maps using the new EMIR receiver and FFTS backends. These fully-sampled maps are part of the Large Program ``Dynamic and Radiative Feedback of Massive Stars'' \citep[see the observing strategy and calibration in][]{Goicoechea2020}.

\cite{Goicoechea2020} provide details on how the old HERA and new EMIR CO maps were merged. Line intensities were converted from antenna temperature ($T_{\rm A}^{*}$) scale to main-beam temperature ($T_{\rm{mb}}$) using appropriate beam and forward efficiencies. Finally, and in order to properly compare with the velocity-resolved [C\,{\sc ii}] maps, we smoothed the \mbox{CO(2-1)} data to an angular resolution of $16\arcsec$. The typical root-mean-square noise level in the \mbox{CO(2-1)} map is $0.16\,\mathrm{K}$ in $0.4\,\mathrm{km\,s^{-1}}$ velocity resolution channels.

\subsection{Dust maps}

In order to estimate the mass of the [C\,{\sc ii}]-traced gas in the expanding shells, we make use of FIR photometry obtained by {\it Herschel}. \cite{Lombardi2014} present a dust spectral energy distribution (SED) fit from {\it Herschel}/Photoconductor Array Camera and Spectrometer (PACS, \cite{Poglitsch2010}) and Spectral and Photometric Imaging Receiver (SPIRE, \cite{Griffin2010}) photometric images across a vast region of the Orion molecular cloud. However, they comment that the short-wavelength PACS $70\,\mu\mathrm{m}$ might be optically thick towards the dense molecular cores, hence they exclude it. Since we expect the expanding shell, we are mostly interested in, to mainly consist of optically thin, warm dust, we use the shorter-wavelength bands of PACS at $70\,\mu\mathrm{m}$, $100\,\mu\mathrm{m}$, and $160\,\mu\mathrm{m}$, tracing the warm dust; the longer-wavelength SPIRE bands at $250\,\mu\mathrm{m}$, $350\,\mu\mathrm{m}$, and $500\,\mu\mathrm{m}$, also included in the fit, are dominated by emission from cold (background) dust. We let the dust temperature $T_{\mathrm{d}}$ and the dust optical depth $\tau_{160}$ be free parameters, and fit SEDs using a modified blackbody for fixed grain emissivity index $\beta$:
\begin{align}
I_{\lambda} = B(\lambda,T_{\rm d})\, \tau_{160}\left(\frac{160\,\mu\mathrm{m}}{\lambda}\right)^{\beta}. \label{eq.I}
\end{align}
We convolve and re-grid the PACS and SPIRE maps to the spatial resolution of the SPIRE $500\,\mu\mathrm{m}$ image, that is $36\arcsec$, at a pixel size of $14\arcsec$. SED fits to individual pixels are shown in Fig. \ref{Fig.SED-points}. We note that $\beta=0$ results in the least residual in the PACS bands, but according to \cite{HollenbachTT1991}, at least $\beta=1$ should be used. Often (e.g., in \cite{Goicoechea2015} for OMC1) $\beta=2$ is adopted. The resulting dust temperature and dust optical depth vary considerably with $\beta$. Decreasing $\beta$ from 2 to 1, decreases the dust optical depth by half, as does decreasing $\beta$ from 1 to 0. The longer-wavelength SPIRE bands can only be fitted with $\beta=1\mhyphen2$. Since we use the conversion factors from $\tau$ to gas mass of \cite{Draine2001}, we revert to $\beta=2$ as suggested by their models. \cite{Lombardi2014} employ the $\beta$ map obtained by Planck at $5\arcmin$ resolution, which amounts to $\beta\simeq 1.6$. We note that $\tau_{160}$ is biased towards the warm dust, which is beneficial for our purposes. We could have constrained the SED fits to the three PACS bands, using $\beta=2$. This leaves the dust temperature and the dust optical depth in the shells mostly unchanged. The only exception is the southern part of the limb-brightened Veil shell, where the dust optical depth turns out to be 30\% lower. In addition, the dust optical depth towards the molecular background is reduced by half. From the dust optical depth $\tau_{160}$, we can compute the gas column density:
\begin{align}
N_{\mathrm{H}} \simeq \frac{100\,\tau_{160}}{\kappa_{160}m_{\mathrm{H}}}\simeq 6\cdot 10^{24}\,\mathrm{cm}^{-2}\;\tau_{160},
\end{align}
where we have used a gas-to-dust mass ratio of 100 and assumed a theoretical absorption coefficient $\kappa_{160}\simeq 10.5\,\mathrm{cm^2\,g^{-1}}$ appropriate for $R_{\mathrm{V}}=5.5$ \citep{Weingartner2001}.

To estimate the contribution of emission from very small grains (VSGs) in the $70\,\mu\mathrm{m}$ band, we compare the latter with the {\it Spitzer}/Multiband Imaging Photometer (MIPS) $24\,\mu\mathrm{m}$ image. VSGs in PDRs are stochastically heated and obtain temperatures that are higher than that of larger grains. In the shells of the Orion Veil, M43, and NGC 1977, their contribution is small. In the bubble interiors, however, the emission from hot dust tends to dominate: dust in these H\,{\sc ii} region becomes very warm due to absorption of ionizing photons and resonantly trapped Ly$\alpha$ photons and radiates predominantly at $24\,\mu\mathrm{m}$ \citep{Salgado2016}. For dust in the PDR, the observed flux at wavelength shorter than $24\,\mu\mathrm{m}$ is due to emission by fluctuating grains and PAH molecules and we defer this analysis to a future study.

\subsection{H$\alpha$ observations}
We make use of three different H$\alpha$ observations: the Very Large Telescope (VLT)/Multi Unit Spectroscopic Explorer (MUSE) image taken of the Huygens Region \citep{Weilbacher2015}, the image taken by the Wide Field Imager (WFI) on the European Southern Observatory (ESO) telescope at La Silla of the surrounding EON \citep{DaRio2009}, and a ESO/Digitized Sky Survey 2 (DSS-2) image (red band) covering the entire area observed in [C\,{\sc ii}]. The DSS-2 image is saturated in the inner EON, which is basically the coverage of the WFI image. We use the MUSE image to calibrate the WFI image in units of $\mathrm{MJy\,sr^{-1}}$ with a $\log$ fit of the correlation. The units of the MUSE observations are given as $10^{-12}\,\mathrm{erg\,s^{-1}\,cm^{-3}}$, which we convert to $\mathrm{MJy\,sr^{-1}}$ with the central wavelength $\lambda\simeq 656.3\,\mathrm{nm}$ and a pixel size of $0.2\,\arcsec$. We in turn use the thus referenced WFI image to calibrate the DSS-2 image with a fit of the form $y = a(1-\exp(-bx))$, where $x$, $y$ are the two intensities scaled by the fit parameters. To calculate the surface brightness from the spectral brightness we use a $\Delta \lambda=0.85\,\mathring{\mathrm{A}}$, that is $I = I_{\lambda}\Delta\lambda = I_{\nu}\Delta\nu$.

We only use the thus calibrated DSS-2 image in NGC 1977 for quantitative analysis. The H$\alpha$ surface brightness in the EON is largely due to scattered light from the bright Huygens Region \citep{ODell2010}. Also H$\alpha$ emission in M43 has to be corrected for a contribution of scattered light from the Huygens Region \citep{Simon-Diaz2011}.

We correct for extinction towards NGC 1977, using $R_{\mathrm{V}}=5.5$, suitable for Orion. The reddening is $E(B-V)=0.08$ \citep{Knyazeva1998}. For M43, the reddening is $E(B-V)=0.49$ \citep{Megier2005}. Extinction towards M42 was studied by \cite{ODell2000}, for example, and more recently by \cite{Weilbacher2015}.

\section{Analysis}
\label{analysis}

\subsection{Global morphology}
\label{global-morphology}

\begin{figure}[tb]
\includegraphics[width=0.5\textwidth, height=0.54\textwidth]{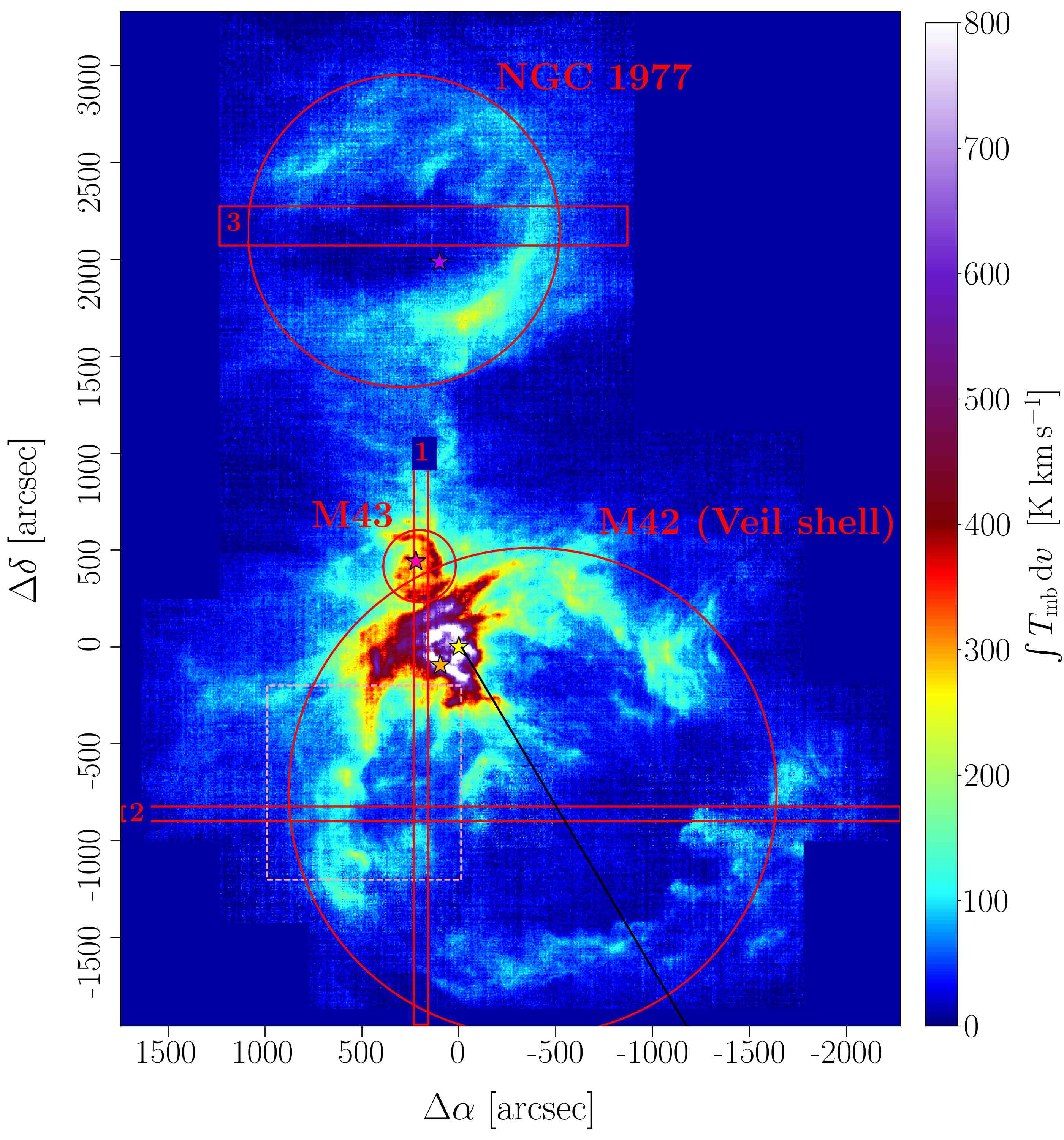}
\caption{[C\,{\sc ii}] line-integrated intensity from the Orion Nebula (M42), M43, and NGC 1977. The red rectangles 1 and 2 indicate the positions of the pv diagrams shown in Figs. \ref{Fig.pv-diagram} (horizontal) and \ref{Fig.3-pv-diagrams} (vertical). Rectangle 3 indicates the position of the pv diagram shown in Fig. \ref{Fig.pv-NGC1977}. The red circles delineate the approximate extent of the expanding bubbles of gas that are the Veil shell, M43, and NGC 1977, and within which the respective masses and luminosities are computed. The light-red dashed rectangle circumscribes the area of the eastern bright arm shown in Figs. \ref{Fig.east_shell_0-5} and \ref{Fig.east_shell_0-15}. The black line indicates the position of the line cut in Fig. \ref{Fig.cross-cut-south}. The stars mark the position of $\theta^1$ Ori C (yellow), $\theta^2$ Ori A (orange), NU Ori (pink), and 42 Orionis (purple). Unless noted otherwise, all coordinate offsets are given with respect to the position of $\theta^1$ Ori C, $(\alpha,\delta) = (5\mathrm{h}35\arcmin 16.46\arcsec, -5^{\circ}23\arcmin22.85\arcsec)$.}
\label{Fig.map}
\end{figure}

From the comparison of the three gas tracers in Fig. \ref{Fig.overview}, we see that [C\,{\sc ii}] emission stems from different structures than either H$\alpha$, tracing the ionized gas, or $^{12}$CO(2-1), tracing the cold molecular gas. In fact, from Fig. 1 in \cite{Pabst2019} we conclude that it stems from the same regions as the warm-dust emission and PAH emission. We can quantify this by correlation plots of the respective tracers (Pabst et al., in prep.). We clearly see bubble-like structures in [C\,{\sc ii}], that are limb-brightened towards the edges. The southern red circle in Fig. \ref{Fig.map} indicates the outlines of the bubble of M42. We note that the Trapezium stars, causing the bubble structure, are offset from the bubble center; in fact, they seem to be located at the northern edge of the bubble. We estimate a bubble radius of $r\simeq 2.7\,\mathrm{pc}$. However, when taking into account the offset of the Trapezium stars, the gas at the southern edge of the bubble is some $4\,\mathrm{pc}$ from the stars. The geometry of the bubble indicates that there is a significant density gradient from the north, where the Trapezium stars are located, to the south, the ambient gas there being much more dilute.

In M43, the star NU Ori is located at the center of the surrounding bubble. 42 Orionis in NGC 1977 is slightly offset from the respective bubble center (cf. Fig. \ref{Fig.map}). Both bubbles are filled with ionized gas as observed in H$\alpha$ emission (cf. Fig. \ref{Fig.overview}) as well as radio emission \citep{Subrahmanyan2001}. In the main part of this section, we derive the mass, physical conditions and the the expansion velocity of the expanding shells associated with M42, M43, and NGC 1977.

\subsection{The expanding Veil shell -- M42}
\subsubsection{Geometry, mass and physical conditions}

The prominent shell structure of the Orion Nebula is surprisingly symmetric, although it is offset from the Trapezium cluster. From the pv diagrams (cf. Sec. \ref{Sec.M42_expv} and App. \ref{App.pv-diagrams}), we estimate its geometric center at $(\Delta\alpha,\Delta\delta)=(-520\arcsec,-550\arcsec)$ and its radius with $r=1360\arcsec=0.38\deg$, which corresponds to $r\simeq 2.7\,\mathrm{pc}$ at the distance of the Orion Nebula; the geometric center of the bubble has a projected distance of $1.5\,\mathrm{pc}$ from the Trapezium cluster. Hence, the gas in the southern edge of the bubble is about $4\,\mathrm{pc}$ away from the Trapezium stars, whereas to the north the bubble outline is more ellipsoid and the gas in the shell is at only $0.5\,\mathrm{pc}$ distance.

The limb-brightened shell of the expanding Veil bubble is mainly seen in the velocity range (with respect to the Local Standard of Rest (LSR)) $v_{\mathrm{LSR}}=5\mhyphen 8\,\mathrm{km\,s^{-1}}$. The emission from the bubble itself can be found down to $v_{\mathrm{LSR}}\simeq -7\,\mathrm{km\,s^{-1}}$ (cf. Sec. \ref{Sec.M42_expv}). The bright main component, that originates from the surface of the background molecular cloud, lies at $v_{\mathrm{LSR}}\simeq 8\,\mathrm{km\,s^{-1}}$; in the bright Huygens Region significant emission extends up to $v_{\mathrm{LSR}}\sim 15\,\mathrm{km\,s^{-1}}$. Here, we also detect the [$^{13}$C\,{\sc ii}] $F=2\mhyphen 1$ line in individual pixels, corresponding to gas moving at $v_{\mathrm{LSR}}\sim 8\,\mathrm{km\,s^{-1}}$ (see previous detections of [$^{13}$C\,{\sc ii}] lines in \cite{BoreikoBetz1996,Ossenkopf2013,Goicoechea2015}).

\begin{figure}[tb]
\includegraphics[width=0.5\textwidth, height=0.33\textwidth]{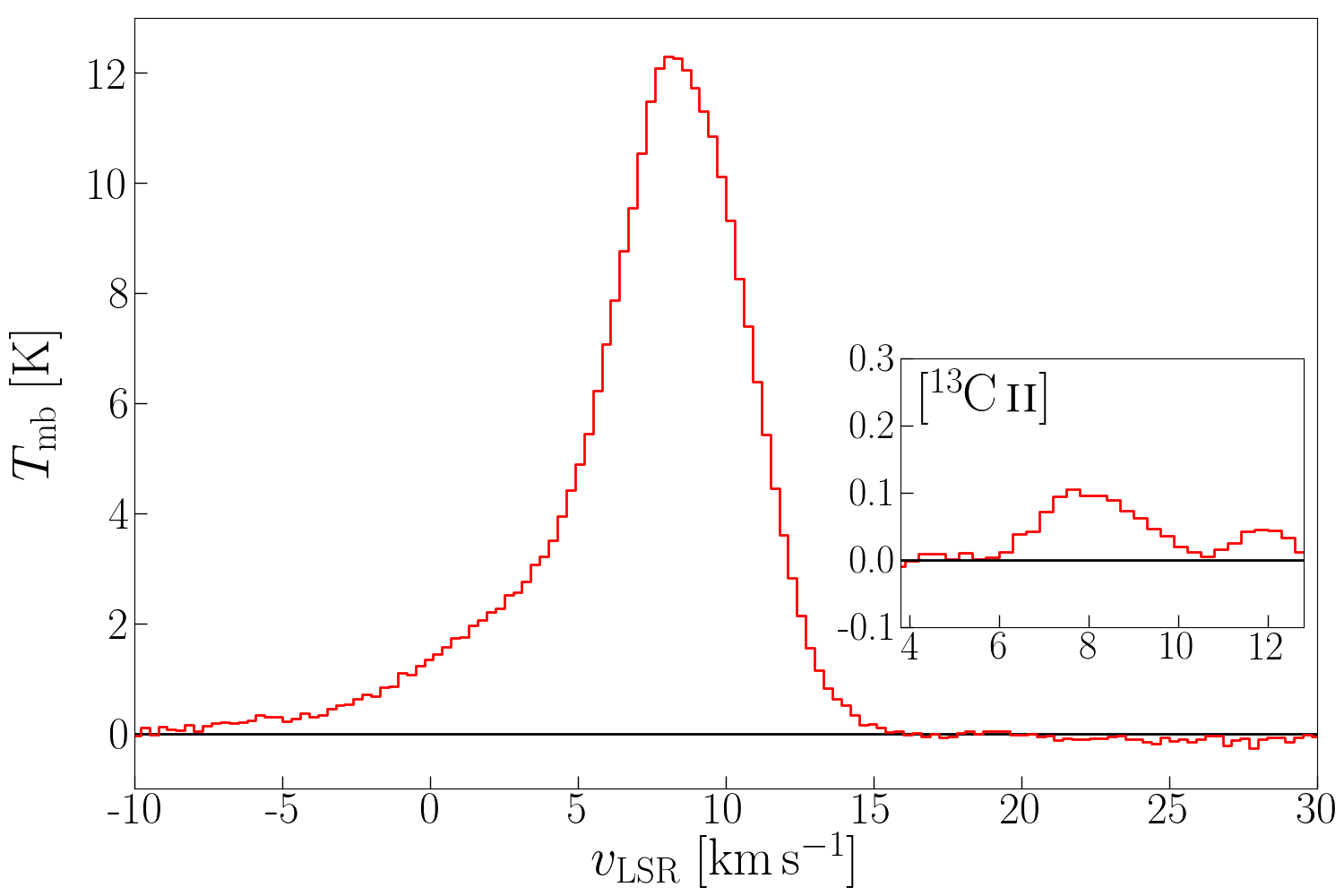}
\caption{Average [C\,{\sc ii}] spectrum towards the Veil shell without OMC1 and ISF. The inlaid panel shows the residual of the spectrum, in the systemic velocity of the [$^{13}$C\,{\sc ii}] $F=2\mhyphen 1$ line, after subtracting the [$^{12}$C\,{\sc ii}] fit. The [$^{12}$C\,{\sc ii}] line can be fitted by two Gaussians with $T_{\mathrm{P}}\simeq 11.2\pm 0.3\,\mathrm{K}$, $v_{\mathrm{P}}\simeq 8.8\pm 0.1\,\mathrm{km\,s^{-1}}$, $\Delta v_{\mathrm{FWHM}}\simeq 5.0\pm 0.1\,\mathrm{km\,s^{-1}}$ and $T_{\mathrm{P}}\simeq 2.3\pm 0.1\,\mathrm{K}$, $v_{\mathrm{P}}\simeq 4.0\pm 0.4\,\mathrm{km\,s^{-1}}$, $\Delta v_{\mathrm{FWHM}}\simeq 9.3\pm 0.5\,\mathrm{km\,s^{-1}}$; the [$^{13}$C\,{\sc ii}] $F=2\mhyphen 1$ component is fitted by a Gaussian with $T_{\mathrm{P}}\simeq 0.11\pm 0.02\,\mathrm{K}$, $v_{\mathrm{P}}\simeq 8.2\pm 0.3\,\mathrm{km\,s^{-1}}$, $\Delta v_{\mathrm{FWHM}}\simeq 2.1\pm 0.6\,\mathrm{km\,s^{-1}}$. The main line could also be fitted with three Gaussian components, but we compare the respective combined peak temperatures of the [$^{12}$C\,{\sc ii}] and [$^{13}$C\,{\sc ii}] line for a first-order estimate of the excitation conditions.}
\label{Fig.spectrum-M42}
\end{figure}

Figure \ref{Fig.spectrum-M42} shows the average [C\,{\sc ii}] spectrum towards the Veil shell without OMC1 and the ISF. The [$^{13}$C\,{\sc ii}] $F=2\mhyphen 1$ line, one of the three [$^{13}$C\,{\sc ii}] fine-structure lines, that is shifted by $11.2\,\mathrm{km\,s^{-1}}$ with respect to the [$^{12}$C\,{\sc ii}] line, is marginally detected. We estimate the detection significance over the integrated line from the fit errors at $5\sigma$. From the [$^{13}$C\,{\sc ii}] $F=2\mhyphen 1$ line, we can compute the [C\,{\sc ii}] optical depth $\tau_{\mathrm{[C\,\textsc{ii}]}}$ and the excitation temperature $T_{\mathrm{ex}}$:
\begin{align}
\frac{1-\exp(-\tau_{\mathrm{[C\,\textsc{ii}]}})}{\tau_{\mathrm{[C\,\textsc{ii}]}}} &\simeq \frac{0.625 T_{\mathrm{P}}([^{12}\mathrm{C\,\textsc{ii}}])}{[^{12}\mathrm{C}/^{13}\mathrm{C}] T_{\mathrm{P}}([^{13}\mathrm{C\,\textsc{ii}}], F=2\mhyphen 1)}, \label{eq.tauCII} \\
T_{\mathrm{ex}} &= \frac{91.2\,\mathrm{K}}{\ln(1+\frac{91.2\,\mathrm{K}(1-\exp(-\tau_{\mathrm{[C\,\textsc{ii}]}}))}{T_{\mathrm{P}}([^{12}\mathrm{C\,\textsc{ii}}])+T_{\mathrm{c}}})} \label{eq.Tex},
\end{align}
where $[^{12}\mathrm{C}/^{13}\mathrm{C}] \sim 67$ is the isotopic ratio for Orion \citep{Langer1990} and $0.625$ is the relative strength of the [$^{13}$C\,{\sc ii}] $F=2\mhyphen 1$ line \citep{Ossenkopf2013}; $T_{\mathrm{c}} = \frac{91.2\,\mathrm{K}}{\exp(91.2\,\mathrm{K}/T_{\mathrm{d}})-1}$ is the continuum brightness temperature of the dust background \citep{Goicoechea2015}. In general, $T_{\mathrm{P}}([^{12}\mathrm{C\,\textsc{ii}}]) \gg T_{\mathrm{c}}$ in the limb-brightened shells. The measured peak temperatures yield $\tau_{\mathrm{[C\,\textsc{ii}]}}\simeq 0.1$ and $T_{\mathrm{ex}}\simeq 144\,\mathrm{K}$ for the averaged spectrum. We obtain an average C$^+$ column density of $N_{\mathrm{C}^+}\simeq 3 \cdot 10^{17}\,\mathrm{cm}^{-2}$. These results have to be interpreted with caution, since we average over a large region with very different conditions ($G_0$, $n$) and the signal is likely dominated by the bright eastern Rim of the Veil shell (see App. \ref{App.east-shell} for a discussion thereof). Also, baseline removal for the [$^{13}$C\,{\sc ii}] line is problematic. From the dust optical depth, we estimate a much larger column density towards the Veil shell, $N_{\mathrm{C}^+}\gtrsim 10^{18}\,\mathrm{cm}^{-2}$.

If we assume the [C\,{\sc ii}] emission from the limb-brightened shell to be (marginally) optical thick, that is $\tau_{\mathrm{[C\,\textsc{ii}]}} \gtrsim 1$, we can also estimate the excitation temperature from the peak temperature of the [$^{12}$C\,{\sc ii}] spectra by eq. \ref{eq.Tex}. We obtain excitation temperatures of $T_{\mathrm{ex}}\simeq 100\,\mathrm{K}$ in the Eastern Rim (and similar gas temperatures), and $T_{\mathrm{ex}}\simeq 50\,\mathrm{K}$ in the far edge of the shell. Here, the density presumably is somewhat lower, resulting in a higher gas temperature. The gas (or kinetic) temperature is given by:
\begin{align}
T_{\mathrm{gas}} = \frac{T_{\mathrm{ex}}}{1-\frac{T_{\mathrm{ex}}}{91.2\,\mathrm{K}}\ln(1+\frac{n_{\mathrm{cr}}}{n})},
\end{align}
where $n_{\mathrm{cr}}\simeq 3\cdot 10^3\,\mathrm{cm^{-3}}$ is the critical density for C${^+}$-H collisions \citep{Goldsmith2012, Pabst2017}. With the density estimates below, the gas temperature in the southern Veil shell is $T_{\mathrm{gas}}\simeq 70\mhyphen 125\,\mathrm{K}$. 

From the dust SEDs, \cite{Pabst2019} estimate $M \simeq 2600\substack{+800 \\ -900}\,M_{\odot}$ for the mass of the gas in the Veil shell after accounting for projection effects. This is in good agreement with their mass estimate from the [$^{13}$C\,{\sc ii}] line in the brightest parts of the shell. The bright eastern part of the shell, however, might not be representative of the entire limb-brightened shell of the Veil shell. From a line cut through this Eastern Rim (Fig. \ref{Fig.M42_CO21_CII}), we estimate a density of $n \simeq 9\cdot 10^3\,\mathrm{cm^{-3}}$ from the spatial separation of the peaks in [C\,{\sc ii}] and CO emission (see App. \ref{App.east-shell}), assuming an $A_{\mathrm{V}}$ of 2 for the C$^+$/C/CO transition and $N_{\mathrm{H}}/A_{\mathrm{V}} \simeq 2\cdot 10^{21}\,\mathrm{cm^{-2}}$ \citep{Bohlin1978}. Although continuously connected to the limb-brightened Veil shell in space and velocity, this Eastern Rim might be a carved-out structure in the background molecular cloud, as has been argued from the presence of foreground scattered light from the Trapezium stars \citep{ODellGoss2009}. We note that the [C\,{\sc ii}] emission from the expanding shell connected to the Eastern Rim is confined to within the Rim, supporting the view that the Eastern Rim is a static confinement of the expanding bubble.

Morphologically, the Eastern Rim fits well in with the rest of the Veil shell, following its curvature well and forming one, seemingly coherent structure. In contrast, there is no connection to the ISF that spans up the dense core of the Orion molecular cloud in the submillimeter maps (cf. Fig. \ref{Fig.overview}). It is then hard to conceive that this part of the nebula is a fortuitous coincidence of the expanding Veil shell encountering a background or foreground molecular cloud structure. On the other hand, the channel maps do not reveal the shift of the shell with increasing velocity in the Eastern Rim that are the signature of an expanding shell in the other parts of the Veil shell (cf. Fig. \ref{Fig.shell-outlines}). Excluding the area associated directly with the dense PDR around the Trapezium (indicated by the circle in Extended Data Fig. 8 in \cite{Pabst2019}), analysis of the dust SED yields $M\simeq 2600\,M_{\odot}$ for the mass of the Veil shell. If we also exclude the Eastern Rim, this drops to $M\simeq 1500\,M_{\odot}$.

From the visual extinction towards the Trapezium cluster, $A_{\mathrm{V}}\simeq 1.8$ \citep{ODellGoss2009}, corresponding to a hydrogen column density of $N_{\mathrm{H}}\simeq 3.6\cdot 10^{21}\,\mathrm{cm^{-2}}$, the mass of the half-shell can be estimated as $M \simeq 2\pi r^2 N_{\mathrm{H}} \mu m_{\mathrm{H}}$, where $\mu=1.4$ is the mean molecular weight. With $r\simeq 2.5\,\mathrm{pc}$, this gives $M \simeq 1600\,M_{\odot}$. From H$_2$ and carbon line observations towards the Trapezium cluster, \cite{Abel2016} estimate a density of $n \simeq 2.5\cdot 10^3\,\mathrm{cm^{-3}}$ and a thickness of the large-scale Veil component B of $d\simeq 0.4\,\mathrm{pc}$, resulting in a visual extinction of $A_{\mathrm{V}}\simeq 1.5$. In a new study of optical emission and absorption lines plus [C\,{\sc ii}] and H\,{\sc i} line observations, \cite{Abel2019} determine $n \simeq 1.6\cdot 10^3\,\mathrm{cm^{-3}}$ for the density in the Veil towards the Trapezium cluster, their Component III(B). However, the central part of the Veil might not be representative of the entire Veil shell, since the [C\,{\sc ii}] emission from the Veil in the vicinity of the Trapezium stars is less distinct than in the farther parts of the Veil shell, suggesting somewhat lower column density in the central part. The visual extinction, as seen by MUSE \citep{Weilbacher2015}, varies across the central Orion Nebula and decreases to only $A_{\mathrm{V}}\simeq 0.7$ to the west of the Trapezium cluster. Using this value reduces the mass estimate of the Veil shell to $M \sim 600\,M_{\odot}$.

X-ray observations of the hot plasma within the EON indicate varying extinction due to the foreground Veil shell. \cite{Guedel2008} derive an absorbing hydrogen column density of $N_{\mathrm{H}}\simeq 4\cdot 10^{20}\,\mathrm{cm}^{-2}$ towards the northern EON and $N_{\mathrm{H}}\leq 10^{20}\,\mathrm{cm}^{-2}$ towards the southern X-ray emitting region. Assuming again $T_{\mathrm{ex}}\simeq 50\,\mathrm{K}$, these column densities correspond to [C\,{\sc ii}] peak temperatures of $T_{\mathrm{P}}\simeq 2\,\mathrm{K}$ and $T_{\mathrm{P}}\simeq 0.5\,\mathrm{K}$, respectively, which would be below the noise level of our observations. Yet, we do detect the [C\,{\sc ii}] Veil shell towards these regions (cf. Figs. \ref{Fig.pv-diagrams-x_wolines} and \ref{Fig.pv-diagrams-y_wolines}). From the lack of observable X-rays towards the eastern EON and the more prominent [C\,{\sc ii}] Veil shell in this region, we suspect that the absorbing column of neutral gas is higher in the eastern Veil shell.

With the MUSE estimate for the column density, we can estimate the density and check for consistency with our [C\,{\sc ii}] observations. With a shell thickness of $d\simeq 0.3\,\mathrm{pc}$ and a visual extinction of $A_{\mathrm{V}}\simeq 0.7$, we estimate a density in the Veil shell of $n\simeq 1.5\cdot 10^3\,\mathrm{cm^{-3}}$. We will use this density estimate in the following, but we note that we consider this a lower limit as the density estimate from a typical dust optical depth, $\tau_{160}\simeq 2\cdot 10^{-3}$, in the southern Veil shell and a typical line of sight, $l\simeq r/2$, gives $n\simeq 4\cdot 10^3\,\mathrm{cm^{-3}}$. However, from the lack of detectable wide-spread CO emission from the Veil shell, also in the southern limb-brightened shell, we can set an upper limit for the visual extinction of $A_\mathrm{V}\sim 2$, equivalent to a gas column of $N_{\mathrm{H}}\sim 4\cdot 10^{21}\,\mathrm{cm^{-2}}$ \citep{Goicoechea2020}.

With a shell thickness of $d\simeq 0.3\,\mathrm{pc}$ and a density of $n\simeq 1.5\cdot 10^3\,\mathrm{cm^{-3}}$, the shell has a radial C$^{+}$ column density of $N_{\mathrm{C}^+}\simeq 2\cdot 10^{17}\,\mathrm{cm}^{-2}$. Adopting this latter value and an excitation temperature of $T_{\mathrm{ex}}= 50\,\mathrm{K}$, the calculated [C\,{\sc ii}] optical depth, given by:
\begin{align}
\tau_{\mathrm{[C\,\textsc{ii}]}} = \frac{A\lambda^3}{4\pi b}N_{\mathrm{C}^+}\frac{\exp\left(\frac{\Delta E}{k_{\mathrm{B}} T_{\mathrm{ex}}}\right) -1}{\exp\left(\frac{\Delta E}{k_{\mathrm{B}} T_{\mathrm{ex}}}\right) +2}\label{eq.tauC+},
\end{align}
in the shell seen face-on is $\tau_{\mathrm{[C\,\textsc{ii}]}}\simeq 0.4$, with a line width of $\Delta v_{\mathrm{FWHM}} = 2\sqrt{\ln 2} b \simeq 4\,\mathrm{km\,s^{-1}}$. The expected peak temperature then,
\begin{align}
T_{\mathrm{P}} = \frac{91.2\,\mathrm{K} (1-\exp(-\tau_{\mathrm{[C\,\textsc{ii}]}}))}{\exp\left(\frac{91.2\,\mathrm{K}}{T_{\mathrm{ex}}}\right) - 1},
\end{align}
is $T_{\mathrm{P}}\simeq 6\,\mathrm{K}$, which is in reasonably good agreement with the peak temperatures of the narrow expanding components in spectra in Fig. \ref{Fig.pv-diagram-gauss}.

We can also calculate the mass of the [C\,{\sc ii}]-emitting gas from the luminosity of the shell, $L_{\mathrm{[C\,\textsc{ii}]}} \simeq 170\,L_{\odot}$ in the limb-brightened Veil shell, assuming that the line is effectively optically thin. By
\begin{align}
M = \frac{\mu m_{\mathrm{H}}}{\mathcal{A}_{\mathrm{C}} A \Delta E} \left(\frac{g_{\mathrm{l}}}{g_{\mathrm{u}}}\exp\left(\frac{\Delta E}{k_{\mathrm{B}} T_{\mathrm{ex}}}\right) + 1 \right)\,L_{\mathrm{[C\,\textsc{ii}]}},
\end{align}
where $\mathcal{A}_{\mathrm{C}}\simeq 1.6\cdot 10^{-4}$ is the carbon gas-phase abundance \citep{Sofia2004}, $A\simeq 2.3\cdot 10^{-6}\,\mathrm{s^{-1}}$ the Einstein coefficient for spontaneous emission, $\Delta E/k_{\mathrm{B}} \simeq 91.2\,\mathrm{K}$ the level separation of the two levels with statistical weights $g_{\mathrm{l}}=2$ and $g_{\mathrm{u}}=4$, we obtain $M\simeq 680\,M_{\odot}$, assuming an excitation temperature of $T_{\mathrm{ex}}= 50\,\mathrm{K}$. As we expect the line to be (marginally) optically thick in the limb-brightened shell, $\tau_{\mathrm{[C\,\textsc{ii}]}}\simeq 1\mhyphen 2$, this value constitutes a lower limit of the shell mass. Correcting with this [C\,{\sc ii}] optical depth would yield a mass estimate of at least $M\simeq 1100\,M_{\odot}$. For further analysis we will use the mass estimate from the dust optical depth, where we have subtracted the mass of the Eastern Rim, $M\simeq 1500\,M_{\odot}$. This mass estimate is robust when considering only the PACS bands in the SED, reducing contamination by cold background gas. We recognize that the column density of the shell is highly variable and is lowest in the foreground shell. This is as expected given the strong pressure gradient from the molecular cloud surface towards us (cf. Sec. \ref{sec.pressure}).

\begin{figure}[tb]
\includegraphics[width=0.5\textwidth, height=0.33\textwidth]{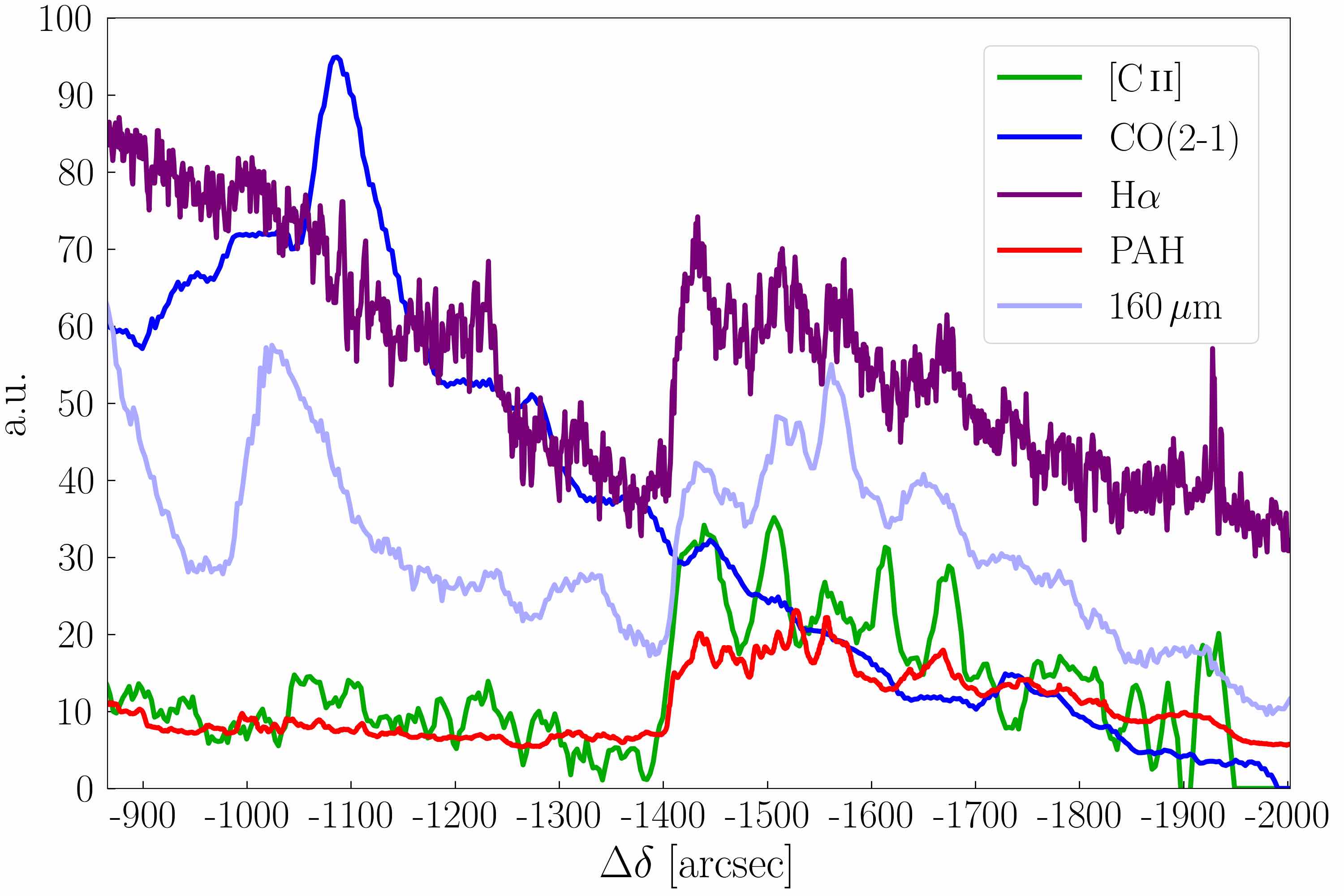}
\caption{Line cut through the southwest of the Veil shell, indicated in Fig. \ref{Fig.map}. The onset of the shell is marked by a steep increase in all tracers, except CO, at $\Delta\delta\simeq -1400\arcsec$. With a line angle of $30^{\circ}$ with respect to the vertical axis, this is at a distance of $3.25\,\mathrm{pc}$ from the Trapezium stars.}
\label{Fig.cross-cut-south}
\end{figure}

We judge that the large-scale arc-structured [C\,{\sc ii}] line emission mainly stems from the outer dense neutral shell indeed, rather than from the contained ionized gas. Evidence for this is the rather small line width where there is a significant signal, which is comparable to the line width of the main component from the neutral surface layer of the molecular cloud ($\Delta v_{\mathrm{FWHM}}\simeq 3\mhyphen4\,\mathrm{km\,s^{-1}}$, cf. Fig. \ref{Fig.pv-diagram-gauss}). If the [C\,{\sc ii}] line originated from the hot ($T\sim 10^4\,\mathrm{K}$) ionized gas, we would expect it to be broader, $\Delta v_{\mathrm{FWHM}}\sim 10\,\mathrm{km\,s^{-1}}$, as observed in the H\,{\sc ii} region bordering the Horsehead Nebula \citep{Pabst2017}. Where the signal from the expanding shell is weak, we indeed observe such broad lines, hence part of the shell might still be ionized (see Sec. \ref{Sec.wind-bubbles} for a discussion of the ionization structure of wind-blown bubbles). The limb-brightened shell is also visible in H$\alpha$, indicating that the surface is ionized. This is also illustrated by the line cut in Fig. \ref{Fig.cross-cut-south}, where the onset of the shell is marked by a steep increase in all tracers except CO. The relative lack of CO indicates low extinction of FUV photons within the southern shell and gives an upper limit on the density in this part, $n \lesssim 3\cdot 10^{3}\mathrm{cm^{-3}}$. From the correlation with other (surface) tracers ([C\,{\sc ii}], $8\,\mu\mathrm{m}$, $160\,\mu\mathrm{m}$) along the line cut we conclude that the shell has a corrugated surface structure, as there are multiple emission peaks in these tracers behind the onset of the shell.

The broader lines ($\Delta v_{\mathrm{FWHM}}>5\,\mathrm{km\,s^{-1}}$) we observe where the shell arc is barely visible, agree with an estimation of the expected [C\,{\sc ii}] emission from ionized gas in the shell. Here, we have \citep{Pabst2017}:
\begin{align}
I_{\mathrm{[C\,\textsc{ii}]}}\simeq 10\,\mathrm{K\;km\,s^{-1}} \frac{n_e}{10^2\,\mathrm{cm}^{-3}} \, \frac{{l}}{1\,\mathrm{pc}},
\end{align}
which is, with an electron density of $n_e\sim 50\mhyphen 100\,\mathrm{cm^{-3}}$ \citep{ODell2010} and a line of sight of $l\sim 0.3\,\mathrm{pc}$, what we find in the spectra from the pv diagram in Fig. \ref{Fig.pv-diagram-gauss}. We do not expect much [C\,{\sc ii}] emission from the inner region of the EON, since the dominant ionization stage of carbon in the vicinity of a O7 star is C$^{2+}$.

\subsubsection{Expansion velocity}
\label{Sec.M42_expv}

\begin{figure}[tb]
\includegraphics[width=0.5\textwidth, height=0.3\textwidth]{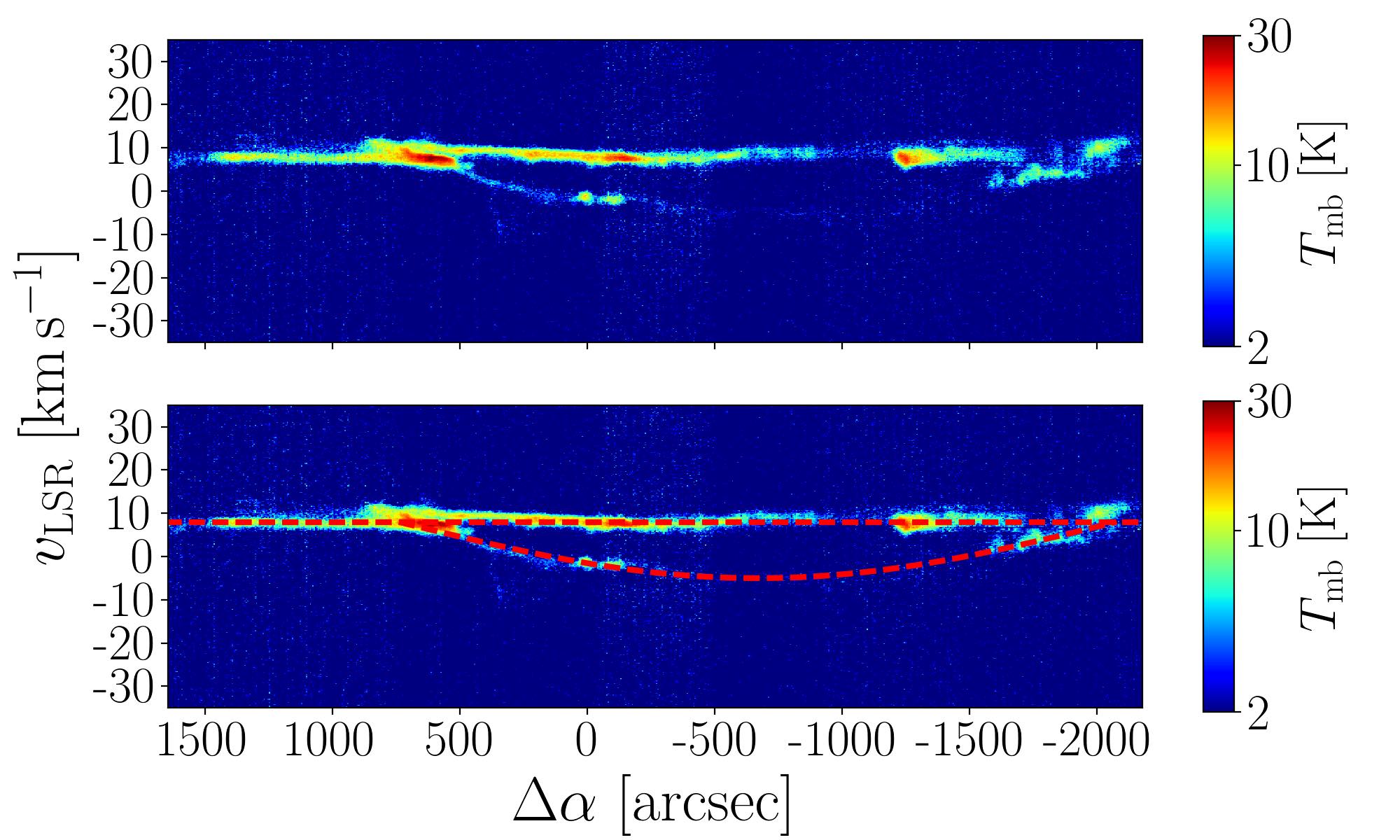}
\caption{[C\,{\sc ii}] pv diagram along horizontal position cut 2 indicated in Fig. \ref{Fig.map} ($\Delta\delta = -907\arcsec\mhyphen -831\arcsec$). The lower panel shows the same cut with the arc structure for an expansion velocity of $13\,\mathrm{km\,s^{-1}}$ on a background velocity of $8\,\mathrm{km\,s^{-1}}$ (red dashed lines).}
\label{Fig.pv-diagram}
\end{figure}

\begin{figure}[tb]
\includegraphics[width=0.5\textwidth, height=0.33\textwidth]{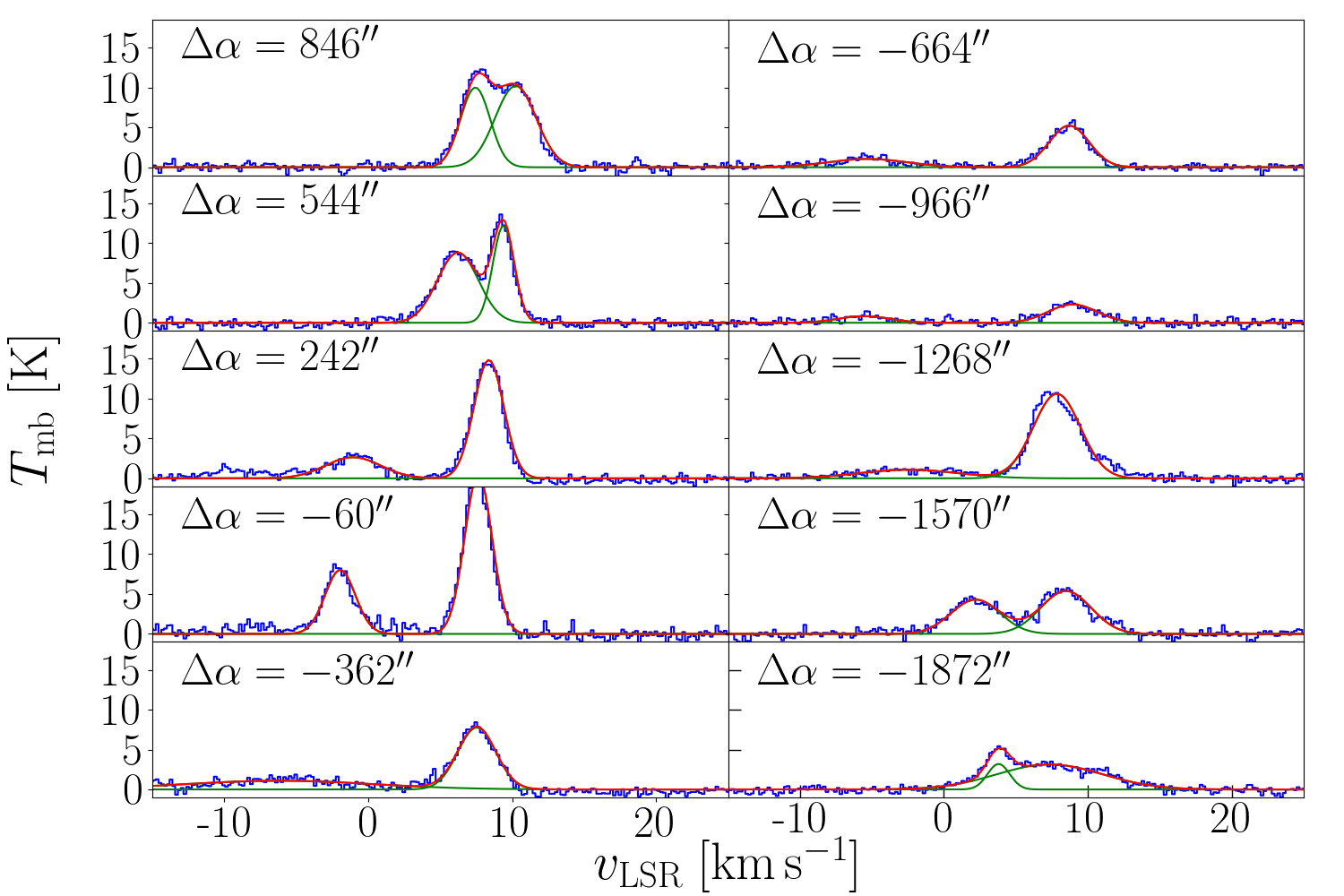}
\caption{[C\,{\sc ii}] spectra in the Veil shell taken from pv diagram in Fig. \ref{Fig.pv-diagram} with Gaussian fits. Spectra at $\Delta\alpha = -362\arcsec$ to $\Delta\alpha = -1268\arcsec$ are averaged over $151\arcsec\times 75.5\arcsec$ in order to improve the signal-to-noise ratio of the faint shell component to be fitted, others are averaged over $75.5\arcsec\times 75.5\arcsec$. Coordinate offsets indicate the lower left-hand corner of the rectangle over which the spectra are averaged; $\Delta\delta = -907\arcsec$. The fit parameters are given in Table \ref{Tab.fit_M42}.}
\label{Fig.pv-diagram-gauss}
\end{figure}

Persistent arc structures observed in position-velocity (pv) diagrams are the signature of bubbles that form in the ISM. By the aid of such diagrams one can estimate the expansion velocity of the associated dense shell. The pv diagrams for the Veil shell reveal the presence of a half shell. Figure \ref{Fig.pv-diagram} shows one such pv diagram as an example. Other cuts are shown in App. \ref{App.pv-diagrams}. \cite{Pabst2019} estimate an expansion velocity of $v_{\mathrm{exp}}\simeq 13\,\mathrm{km\,s^{-1}}$ for this half shell. This value is consistent with Gaussian fits to single (averaged) spectra taken from the pv diagram where this is possible (Fig. \ref{Fig.pv-diagram-gauss}).

\begin{table}[tb]
\begin{tabular}{cc|ccc}
$\Delta\alpha$ & comp. &  $T_{\mathrm{P}}$ & $v_{\mathrm{P}}$ & $\Delta v_{\mathrm{FWHM}}$ \\ 
 & & $[\mathrm{K}]$ & $[\mathrm{km\,s^{-1}}]$ & $[\mathrm{km\,s^{-1}}]$ \\ \hline
$846^{\prime\prime}$ & 1 & $10.2\pm 0.2$ & $10.2\pm 0.1$ & $3.4\pm 0.2$ \\
$846^{\prime\prime}$ & 2 & $10.0\pm 0.5$ & $7.4\pm 0.1$ & $2.4\pm 0.1$ \\ \hline
$544^{\prime\prime}$ & 1 & $12.2\pm 0.2$ & $9.4\pm 0.1$ & $1.7\pm 0.1$ \\
$544^{\prime\prime}$ & 2 & $8.8\pm 0.2$ & $6.2\pm 0.1$ & $3.4\pm 0.1$ \\ \hline
$242^{\prime\prime}$ & 1 & $14.8\pm 0.2$ & $8.4\pm 0.1$ & $2.4\pm 0.1$ \\
$242^{\prime\prime}$ & 2 & $2.6\pm 0.2$ & $-1.1\pm 0.1$ & $4.5\pm 0.3$ \\ \hline
$-60^{\prime\prime}$ & 1 & $20.0\pm 0.3$ & $7.6\pm 0.1$ & $2.2\pm 0.1$ \\
$-60^{\prime\prime}$ & 2 & $8.0\pm 0.2$ & $-2.0\pm 0.1$ & $2.5\pm 0.1$ \\ \hline
$-362^{\prime\prime}$ & 1 & $7.8\pm 0.2$ & $7.5\pm 0.1$ & $3.1\pm 0.1$ \\
$-362^{\prime\prime}$ & 2 & $1.1\pm 0.1$ & $-6.2\pm 0.4$ & $15.4\pm 1.2$ \\ \hline
$-664^{\prime\prime}$ & 1 & $5.2\pm 0.1$ & $8.7\pm 0.1$ & $3.3\pm 0.1$ \\
$-664^{\prime\prime}$ & 2 & $1.0\pm 0.1$ & $-5.3\pm 0.3$ & $6.4\pm 0.6$ \\ \hline
$-966^{\prime\prime}$ & 1 & $2.3\pm 0.2$ & $8.9\pm 0.1$ & $4.0\pm 0.3$ \\
$-966^{\prime\prime}$ & 2 & $0.8\pm 0.2$ & $-5.5\pm 0.3$ & $4.2\pm 0.7$ \\ \hline
$-1268^{\prime\prime}$ & 1 & $10.6\pm 0.2$ & $7.9\pm 0.1$ & $3.9\pm 0.1$ \\
$-1268^{\prime\prime}$ & 2 & $1.1\pm 0.1$ & $-2.5\pm 0.3$ & $8.0\pm 0.8$ \\ \hline
$-1570^{\prime\prime}$ & 1 & $5.4\pm 0.2$ & $8.5\pm 0.1$ & $4.1\pm 0.2$ \\
$-1570^{\prime\prime}$ & 2 & $4.3\pm 0.2$ & $2.2\pm 0.1$ & $4.0\pm 0.2$ \\ \hline
$-1872^{\prime\prime}$ & 1 & $3.1\pm 0.1$ & $7.4\pm 0.2$ & $8.7\pm 0.3$ \\
$-1872^{\prime\prime}$ & 2 & $3.2\pm 0.2$ & $3.8\pm 0.1$ & $1.8\pm 0.2$ \\
\end{tabular}
\caption{Gaussian fit parameters of spectra in Fig. \ref{Fig.pv-diagram-gauss}. The expanding Veil shell is captured in component 2.}
\label{Tab.fit_M42}
\end{table}

It is important to note that the bubble shell traced by [C\,{\sc ii}] emission expands in one direction only, which is towards us. Towards the back, the gas is confined by the background molecular cloud hindering rapid expansion. The prominent $8\,\mathrm{km\,s^{-1}}$ emission originates from this dense background gas.

\begin{figure}[tb]
\includegraphics[width=0.5\textwidth, height=0.4\textwidth]{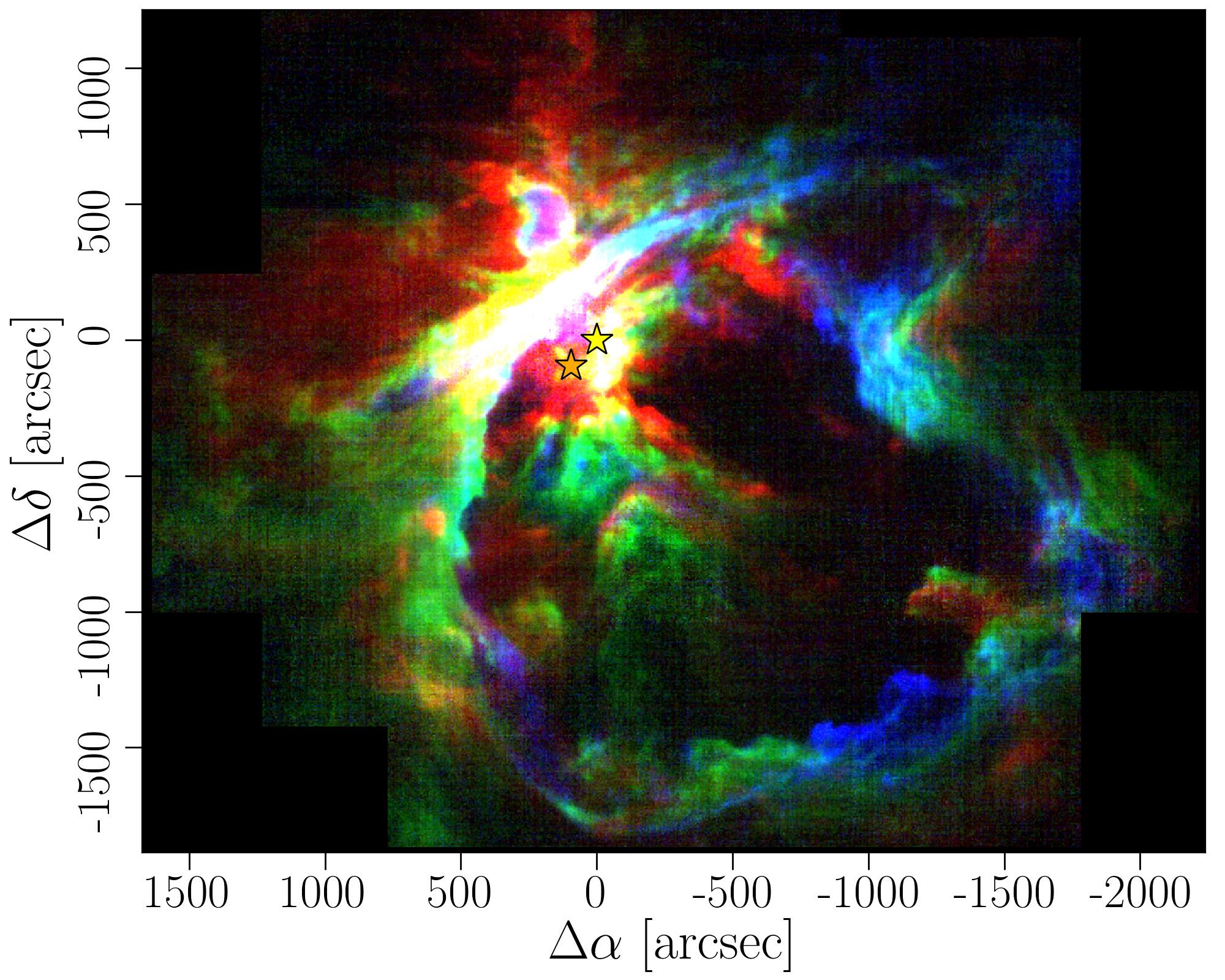}
\caption{Three-color image of velocity channels towards M42 and M43. Blue is the velocity channel $v_{\mathrm{LSR}}=4\mhyphen5\,\mathrm{km\,s^{-1}}$, green $v_{\mathrm{LSR}}=6\mhyphen7\,\mathrm{km\,s^{-1}}$, and red $v_{\mathrm{LSR}}=9\mhyphen10\,\mathrm{km\,s^{-1}}$. With increasing $v_{\mathrm{LSR}}$, the limb-brightened Veil shell filament is displaced outward, away from the bubble center. The gas of M43 that is expanding towards us can be observed in the blue channel; the limb-brightened shell of M43 has higher $v_{\mathrm{LSR}}$.}
\label{Fig.shell-outlines}
\end{figure}

An alternative way to estimate the expansion velocity of the bubble is from channel maps of this region. With increasing $v_{\mathrm{LSR}}$, the limb-brightened shell filament of M42 is observed to move outward, away from the Trapezium stars, as can be seen from Fig. \ref{Fig.shell-outlines}. The projected geometry of the filaments in different channels provides an estimate of $v_{\mathrm{exp}}$. However, this only works well in the bright eastern arm of the shell, the so-called Rim, which is comparatively narrow (cf. App. \ref{App.east-shell}, Figs. \ref{Fig.east_shell_0-5} and \ref{Fig.east_shell_0-15}). In other parts of the limb-brightened shell, the south and west, the filaments in different velocities do not line up consistently. To the north of the Trapezium stars, we do not see outward expansion at all. From these alternative methods (fitting Gaussians to single spectra, outlines of the limb-brightened shell) we find values agreeing with those computed above from the pv diagram, $v_{\mathrm{exp}}\simeq 10\mhyphen 15\,\mathrm{km\,s^{-1}}$. Also the majority of pv diagrams shown in Figs. \ref{Fig.pv-diagrams-x_wolines} and \ref{Fig.pv-diagrams-y_wolines} are consistent with an overall expansion velocity of $v_{\mathrm{exp}}\simeq 13\,\mathrm{km\,s^{-1}}$.

\begin{figure}[tb]
\includegraphics[width=0.5\textwidth, height=0.375\textwidth]{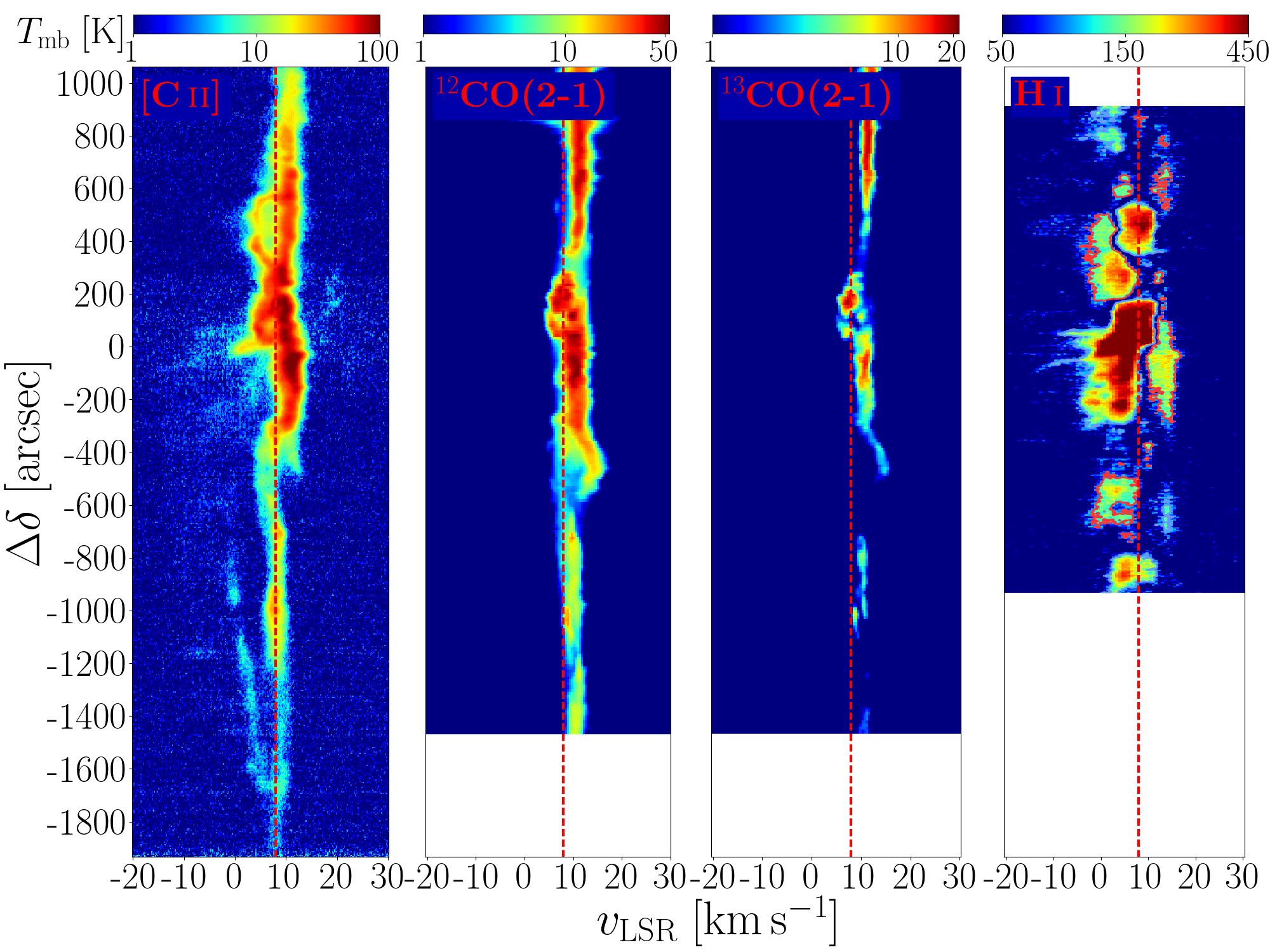}
\caption{PV diagrams along vertical position cut 1 indicated in Fig. \ref{Fig.map} ($\Delta\alpha = 167\arcsec\mhyphen 242\arcsec$) in [C\,{\sc ii}], \mbox{$^{12}$CO(2-1)}, \mbox{$^{13}$CO(2-1)} \citep{Berne2014, Goicoechea2020}, and H\,{\sc i} \citep{VanderWerf2013}.}
\label{Fig.3-pv-diagrams}
\end{figure}

Figure \ref{Fig.3-pv-diagrams} compares the velocity structures of [C\,{\sc ii}], \mbox{$^{12}$CO(2-1)}, \mbox{$^{13}$CO(2-1)}, and H\,{\sc i} along a vertical cut at $\Delta\alpha = 167\arcsec\mhyphen 242\arcsec$. This figure illustrates the complex dynamics in the region of the Orion Nebula and contrasts the various kinematic structures in the different tracers. It gives an impression of the variety of emission structures and the potential synergies between the four data sets. The cut shows arc structures close to the central region (Huygens Region) that coincide with structures detected in H\,{\sc i} and discussed in \cite{VanderWerf2013}. Also, the shell surrounding M43 is distinctly visible (cf. Fig \ref{Fig.spectra-M43}). We note that to the north the bubble-like arc of the Veil shell is disrupted by apparently violent gas dynamics in that region, even far to the west of the Huygens Region. Closer examination of pv diagrams in this region reveal multiple arcs indicative of bubble structures (see App. \ref{App.pv-diagrams} for all pv diagrams through the Veil shell). South of OMC1, the observed difference in velocity ($\sim 1\,\mathrm{km\,s^{-1}}$) between the [C\,{\sc ii}] background and the molecular cloud as traced by CO is typical for the advection flow through the PDR \citep[ch. 12]{Tielens}. From our analysis, we conclude that the systematic uncertainty of the mass estimate are of the order of 50\%, systematic uncertainties of the estimates of the extent of the shell and of the expansion velocity are about 30\%.

\subsection{M43}
\subsubsection{Geometry, mass and physical conditions}

\begin{figure}[tb]
\includegraphics[width=0.5\textwidth, height=0.5\textwidth]{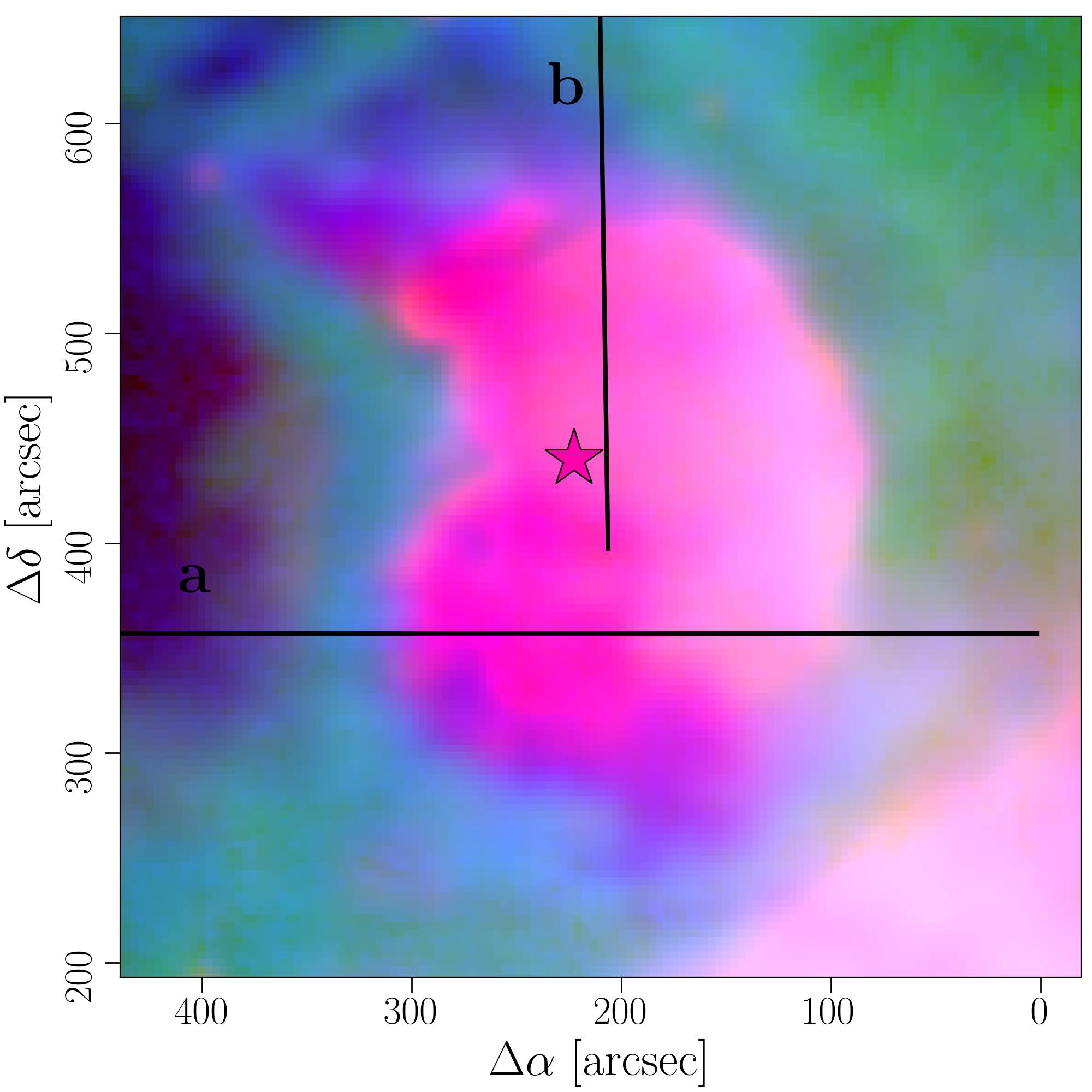}
\caption{Three-color image of M43. Red is H$\alpha$ emission from the ionized gas, green is line-integrated CO(2-1) emission from molecular gas, and blue is the [C\,{\sc ii}] line-integrated intensity, tracing the neutral gas. The line cuts in Fig. \ref{Fig.cross-cuts_M43} are taken along lines a and b.}
\label{Fig.M43_rgb}
\end{figure}

\begin{figure}[tb]
\begin{minipage}{0.5\textwidth}
\includegraphics[width=\textwidth, height=0.4\textwidth]{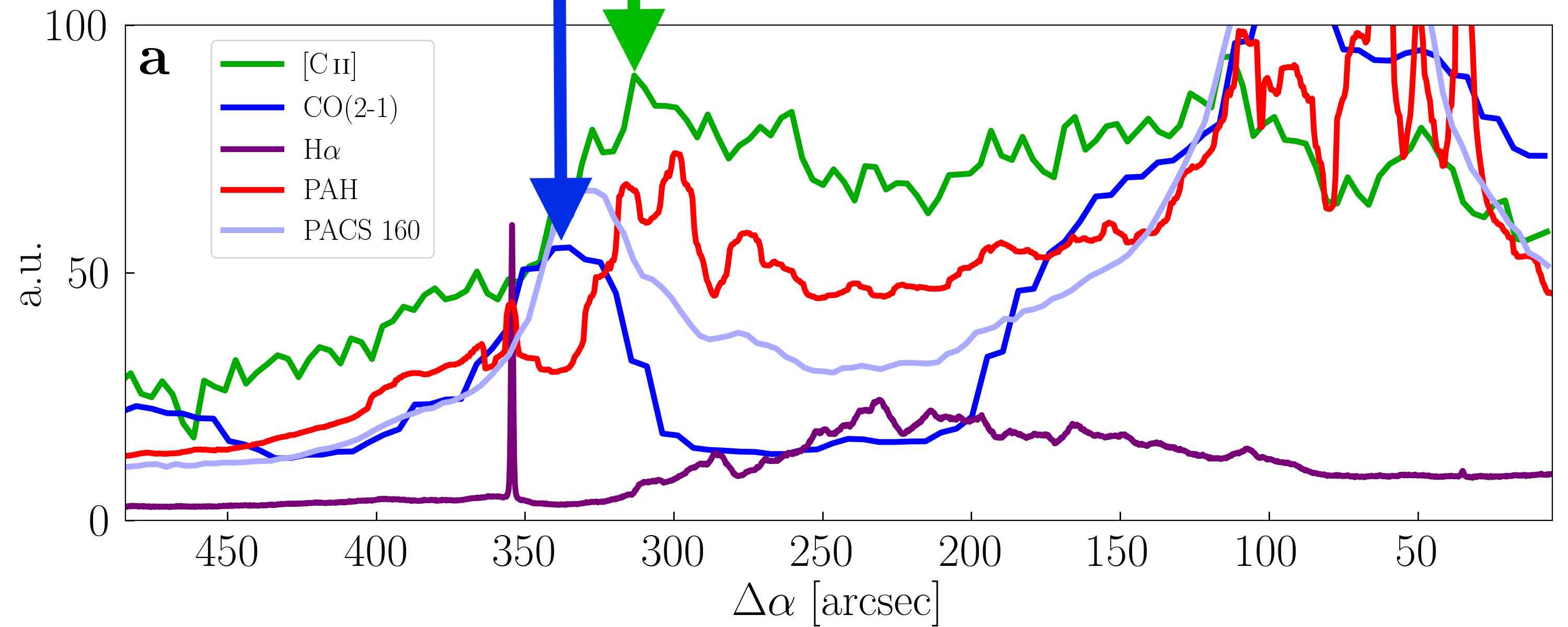}
\end{minipage}

\begin{minipage}{0.5\textwidth}
\includegraphics[width=\textwidth, height=0.4\textwidth]{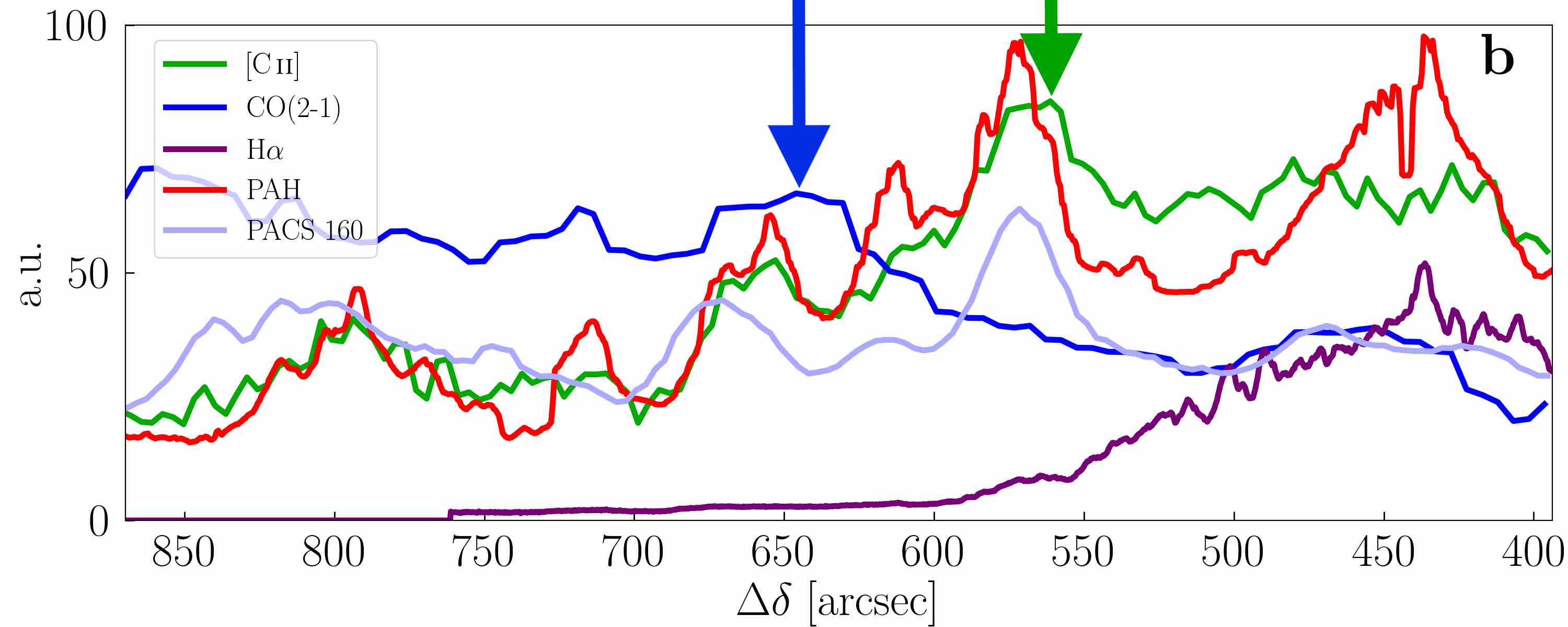}
\end{minipage}
\caption{Line cuts along lines a and b, indicated in Fig. \ref{Fig.M43_rgb}. The arrows indicate the peaks in [C\,{\sc ii}] emission (green) and CO emission (blue) that correspond to the shell.}
\label{Fig.cross-cuts_M43}
\end{figure}

The nebula M43 hosts the central star NU Ori, which is located approximately at the geometrical center of the limb-brightened shell. The shell radius is $r\simeq 150\arcsec \simeq 0.3\,\mathrm{pc}$ with a thickness of $\Delta r \simeq 0.05\,\mathrm{pc}$. Fig. \ref{Fig.M43_rgb} shows a three-color image of M43. H$\alpha$ stems from the center of M43, constrained by a thin shell of [C\,{\sc ii}] and CO in the east and north. To the south lies the massive bulk of OMC1. The H\,{\sc ii} region of M43 appears as a region of higher dust temperature, the shell is distinctly visible in [C\,{\sc ii}], IRAC $8\,\mu\mathrm{m}$ emission, FIR emission from warm dust (cf. Fig. 1 in \cite{Pabst2019}), and CO. The limb-brightened shell reveals the structure of a PDR with $8\,\mu\mathrm{m}$ PAH emission, [C\,{\sc ii}] emission, and $160\,\mu\mathrm{m}$ warm dust emission originating at the inner, illuminated side while the CO emission originate from deeper within the shell (Fig. \ref{Fig.cross-cuts_M43}). M43 is located just eastwards of the ridge of the molecular ISF, whose illuminated, [C\,{\sc ii}] emitting outliers form the background against which the half-shell expands.

From our dust SEDs, we derive a mass of $M \simeq 210\,M_{\odot}$. However, this mass is likely contaminated by the mass of the molecular background, whose surface is visible in [C\,{\sc ii}] (cf. Fig. \ref{Fig.3-pv-diagrams}). Using only the PACS bands in the SED, reduces the mass estimate to $M \simeq 110\,M_{\odot}$. From the [C\,{\sc ii}] luminosity, $L_{\mathrm{[C\,\textsc{ii}]}} \simeq 24\,L_{\odot}$ and assuming $T_{\mathrm{ex}}\simeq 90\,\mathrm{K}$ (see below), the mass is estimated to be $M \simeq 55\,M_{\odot}$.

We note that the limb-brightened shell of M43 consists of two segments. The eastern shell lies closer to the central star and therefore is probably denser. We can derive the density of the shell by the distance between the [C\,{\sc ii}] and the CO peak in a line cut through that region (Fig. \ref{Fig.cross-cuts_M43}a), $\Delta d \simeq 20\arcsec \simeq 0.04\,\mathrm{pc}$ between the [C\,{\sc ii}] peak at $\Delta\alpha\simeq 310\arcsec$ and the CO peak at $\Delta\alpha\simeq 330\arcsec$. Assuming a typical PDR structure with $\Delta A_{\mathrm{V}}\simeq 2$ between the PDR front, traced by [C\,{\sc ii}], and the CO peak, we arrive at a density of $n\simeq 3\cdot 10^4\,\mathrm{cm^{-3}}$. In the northern shell (Fig. \ref{Fig.cross-cuts_M43}b) we derive a density of $n\simeq 8\cdot 10^3\,\mathrm{cm^{-3}}$ with $\Delta d \simeq 80\arcsec \simeq 0.16\,\mathrm{pc}$ between the [C\,{\sc ii}] peak at $\Delta\delta\simeq 560\arcsec$ and the CO peak at $\Delta\delta\simeq 640\arcsec$. With the shell extent above, that is $V=\frac{2\pi}{3}\left((0.3\,\mathrm{pc})^3-(0.25\,\mathrm{pc})^3\right)$, and the latter density estimate, we compute a mass of the expanding hemisphere of $M \simeq 7\,M_{\odot}$, which is significantly lower than the mass derived from the dust optical depth and the [C\,{\sc ii}] luminosity. Given the potential contamination of these mass estimates by background molecular cloud material, we elected to go with the latter estimate for the gas mass.

\begin{figure}[tb]
\includegraphics[width=0.5\textwidth, height=0.33\textwidth]{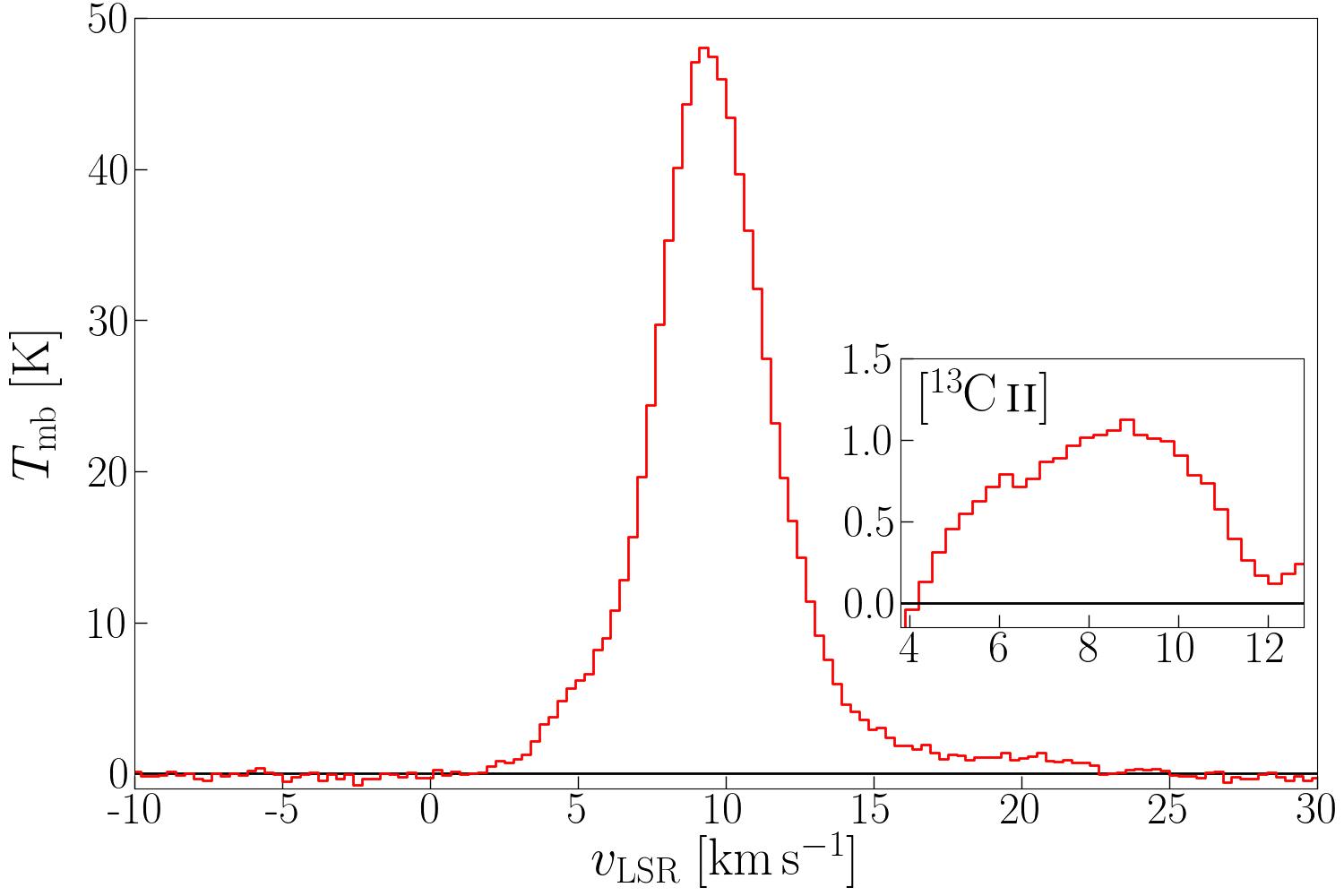}
\caption{Average [C\,{\sc ii}] spectrum towards the eastern shell of M43. The inlaid panel shows the residual of the spectrum, in the systemic velocity of the [$^{13}$C\,{\sc ii}] $F=2\mhyphen 1$ line, after subtracting the [$^{12}$C\,{\sc ii}] fit. The [$^{12}$C\,{\sc ii}] line can be fitted by a Gaussian with $T_{\mathrm{P}}\simeq 46.2\pm 0.3\,\mathrm{K}$, $v_{\mathrm{P}}\simeq 9.6\pm 0.1\,\mathrm{km\,s^{-1}}$, $\Delta v_{\mathrm{FWHM}}\simeq 4.6\pm 0.1\,\mathrm{km\,s^{-1}}$; the [$^{13}$C\,{\sc ii}] $F=2\mhyphen 1$ component is fitted by a Gaussian with $T_{\mathrm{P}}\simeq 1.1\pm 0.2\,\mathrm{K}$, $v_{\mathrm{P}}\simeq 8.6\pm 0.3\,\mathrm{km\,s^{-1}}$, $\Delta v_{\mathrm{FWHM}}\simeq 4.8\pm 0.8\,\mathrm{km\,s^{-1}}$. We note that a better fit of the  [$^{12}$C\,{\sc ii}] line can be obtained by fitting two components with $T_{\mathrm{P}}\simeq 30.9\pm 1.3\,\mathrm{K}$, $v_{\mathrm{P}}\simeq 9.6\pm 0.1\,\mathrm{km\,s^{-1}}$, $\Delta v_{\mathrm{FWHM}}\simeq 3.3\pm 0.1\,\mathrm{km\,s^{-1}}$ and $T_{\mathrm{P}}\simeq 17.5\pm 1.3\,\mathrm{K}$, $v_{\mathrm{P}}\simeq 9.8\pm 0.1\,\mathrm{km\,s^{-1}}$, $\Delta v_{\mathrm{FWHM}}\simeq 6.8\pm 0.2\,\mathrm{km\,s^{-1}}$, but both fits leave significant residuals.}
\label{Fig.spectrum-M43}
\end{figure}

When averaging over the eastern limb-brightened shell of M43, we detect the [$^{13}$C\,{\sc ii}] $F=2\mhyphen 1$ line at a $4\sigma$ level, as shown in Fig. \ref{Fig.spectrum-M43}. From this\footnote{We use the peak temperature of the single-component fit, assuming that the [$^{13}$C\,{\sc ii}] component corresponds to the combination of the two main components, as suggested by the similar line widths of the single-component fits.}  we obtain an average $\tau_{\mathrm{[C\,\textsc{ii}]}} \simeq 2.4\substack{+1.0 \\ -1.4}$ and an excitation temperature of $T_{\mathrm{ex}}\simeq 89\substack{+20 \\ -5}\,\mathrm{K}$. This translates into a C$^{+}$ column density of $N_{\mathrm{C}^+}\simeq 2.7\substack{+0.9 \\ -1.2}\cdot 10^{18}\,\mathrm{cm}^{-2}$ along the line of sight. Assuming that most of the material is located in the limb-brightened shell, we can estimate a density in the shell with the assumed length of the line of sight of $l \sim r/2$. With $r\simeq 0.3\,\mathrm{pc}$, this gives $n\simeq 3 \cdot 10^4\,\mathrm{cm}^{-3}$, which is about the same as the previous estimate from the [C\,{\sc ii}]-CO peak separation. With this density we compute a gas temperature of $T_{\mathrm{gas}}\simeq 90\,\mathrm{K}$.

If we assume the [C\,{\sc ii}] emission from the limb-brightened shell to be (marginally) optical thick, we can estimate the excitation temperature from the [C\,{\sc ii}] peak temperature. This gives an average of $T_{\mathrm{ex}}\simeq 110\,\mathrm{K}$, and hence a slightly higher gas temperature than from the [$^{13}$C\,{\sc ii}] estimate.

\subsubsection{Expansion velocity}

\begin{figure}[tb]
\includegraphics[width=0.5\textwidth, height=0.3\textwidth]{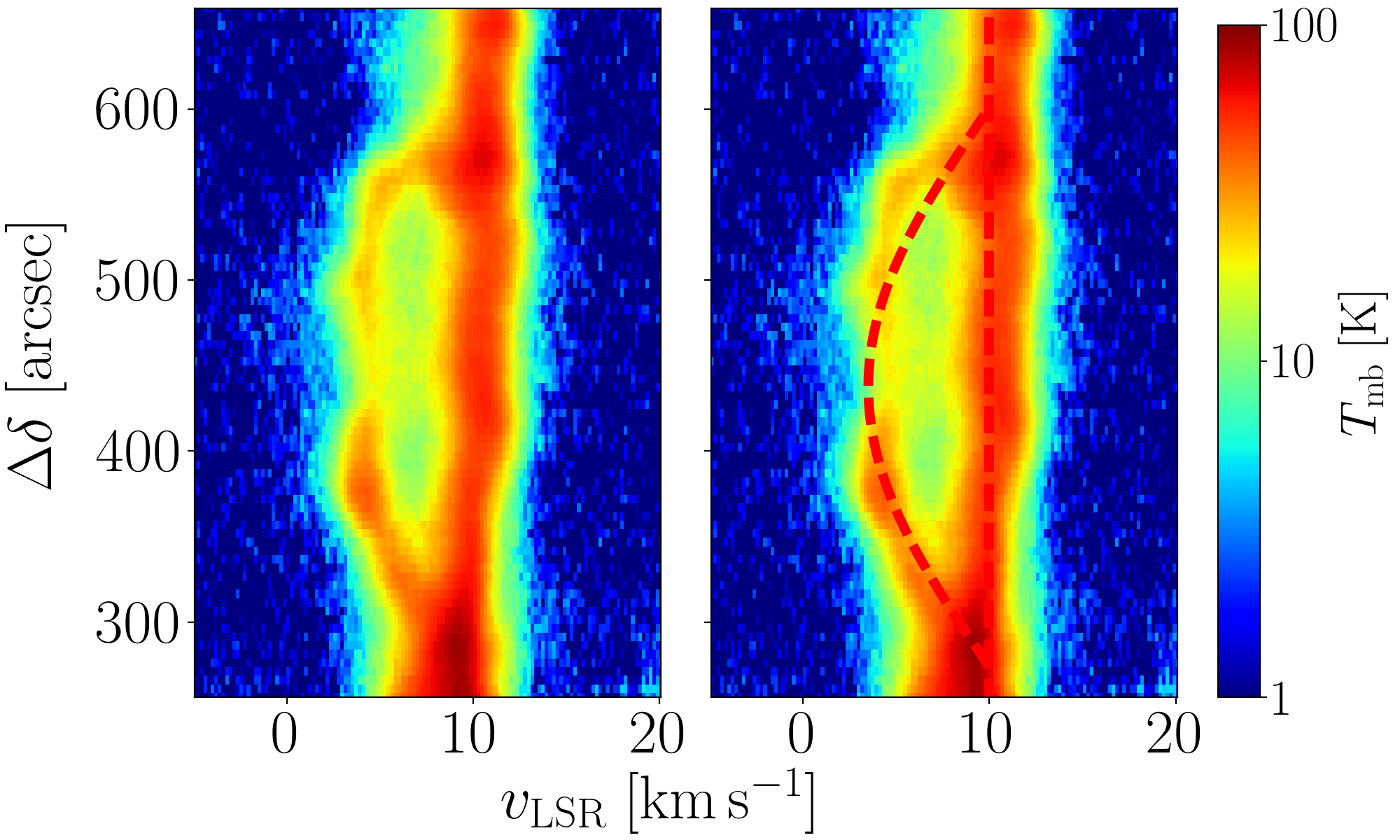}
\caption{[C\,{\sc ii}] pv diagram along vertical position cut 1 indicated in Fig. \ref{Fig.map} ($\Delta\alpha = 167\arcsec\mhyphen 242\arcsec$), zoomed in to M43 (cf. Fig. \ref{Fig.3-pv-diagrams}). The right panel shows the same cut with the arc structure for an expansion velocity of $6\,\mathrm{km\,s^{-1}}$ on a background velocity of $8\,\mathrm{km\,s^{-1}}$ (red dashed lines).}
\label{Fig.spectra-M43}
\end{figure}

The [C\,{\sc ii}] pv diagram running through M43 clearly exhibits a half shell structure (Fig. \ref{Fig.spectra-M43}). From the spectra taken towards this region (Fig. \ref{Fig.spectra-M43}), we measure an expansion velocity of $v_{\mathrm{exp}}\simeq 6\,\mathrm{km\,s^{-1}}$. The bubble only expands towards us, away from the background molecular cloud. We do not see an expanding CO counterpart (cf. Fig. \ref{Fig.3-pv-diagrams}). The pv diagram is consistent with the radius of the shell of $r\simeq 150\arcsec \simeq 0.3\,\mathrm{pc}$ and the expansion velocity of $v_{\mathrm{exp}}\simeq 6\,\mathrm{km\,s^{-1}}$, both values are much less than those derived for the Trapezium wind-blown bubble. Also, in Fig. \ref{Fig.spectra-M43}, we observe [C\,{\sc ii}] emission from within the shell arc, presumably stemming from the expanding ionized gas.

\begin{figure}[tb]
\includegraphics[width=0.5\textwidth, height=0.33\textwidth]{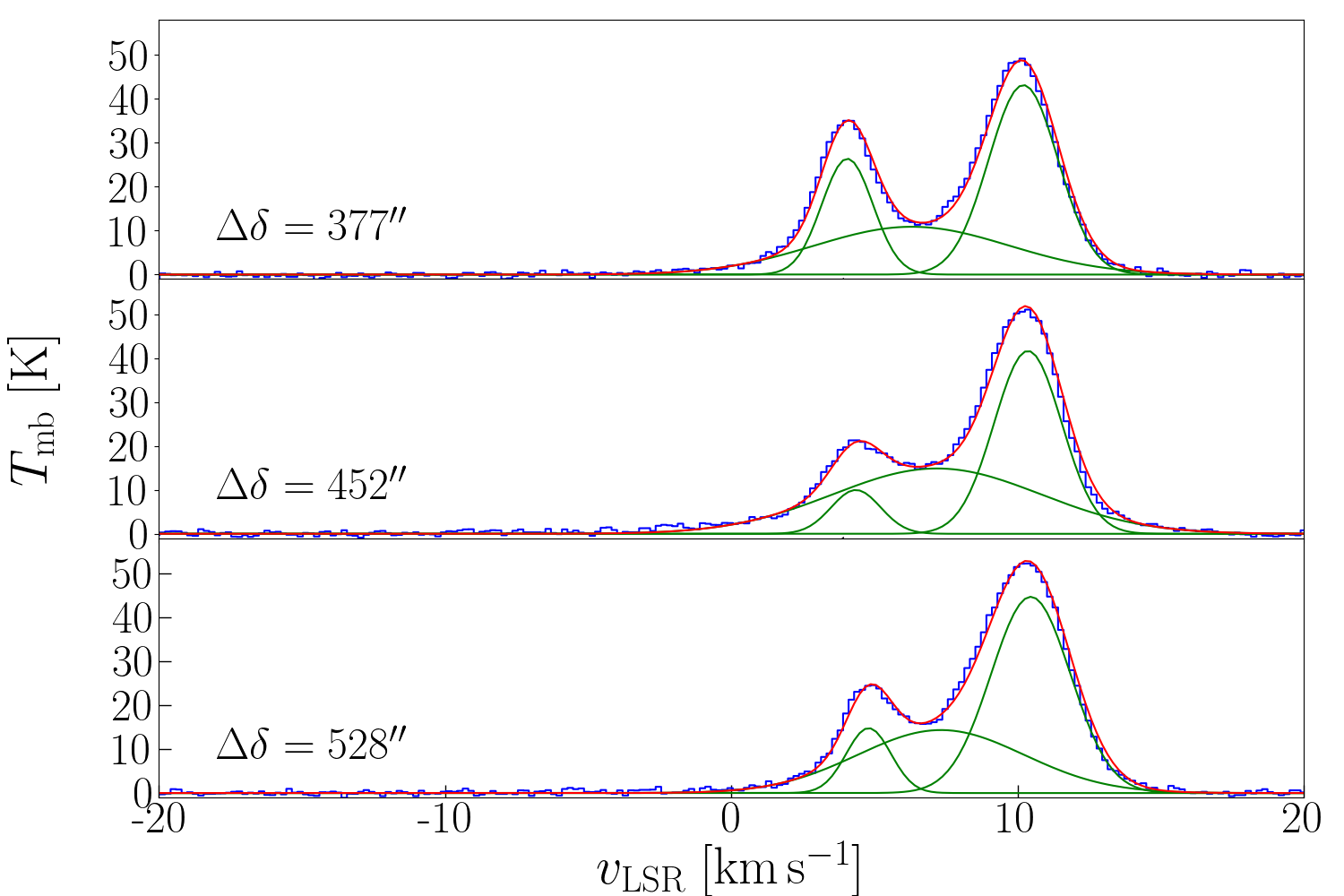}
\caption{[C\,{\sc ii}] spectra along position cut 1 (cf. Fig. \ref{Fig.spectra-M43}) in M43 with Gaussian fits. Each spectrum is averaged over $75.5\arcsec\times 25.2\arcsec$. The expanding shell has a distinguished emission peak. Additionally to the main component and the shell component, we fit a third, broad, component, that possibly stems from the ionized gas within the shell. Coordinate offsets indicate the lower left-hand corner of the rectangle over which the spectra are averaged; $\Delta\alpha = 242\arcsec$. The fit parameters are given in Table \ref{Tab.fit_M43}.}
\label{Fig.spectra-M43_gauss}
\end{figure}

\begin{table}[tb]
\begin{tabular}{cc|ccc}
$\Delta\delta$ & comp. &  $T_{\mathrm{P}}$ & $v_{\mathrm{P}}$ & $\Delta v_{\mathrm{FWHM}}$ \\
 & & $[\mathrm{K}]$ & $[\mathrm{km\,s^{-1}}]$ & $[\mathrm{km\,s^{-1}}]$ \\ \hline
$377^{\prime\prime}$ & 1 & $43.1\pm 0.4$ & $10.2\pm 0.1$ & $2.9\pm 0.1$ \\
$377^{\prime\prime}$ & 2 & $26.4\pm 0.3$ & $4.1\pm 0.1$ & $2.1\pm 0.1$ \\ 
$377^{\prime\prime}$ & 3 & $10.9\pm 0.3$ & $6.3\pm 0.2$ & $8.0\pm 0.1$ \\ \hline
$452^{\prime\prime}$ & 1 & $41.6\pm 0.4$ & $10.4\pm 0.1$ & $2.8\pm 0.1$ \\
$452^{\prime\prime}$ & 2 & $10.0\pm 0.3$ & $4.4\pm 0.1$ & $2.0\pm 0.1$ \\ 
$452^{\prime\prime}$ & 3 & $14.9\pm 0.3$ & $7.2\pm 0.1$ & $8.5\pm 0.1$ \\ \hline
$528^{\prime\prime}$ & 1 & $44.7\pm 0.9$ & $10.5\pm 0.1$ & $3.3\pm 0.1$ \\
$528^{\prime\prime}$ & 2 & $14.9\pm 0.4$ & $4.8\pm 0.1$ & $1.8\pm 0.1$ \\ 
$528^{\prime\prime}$ & 3 & $14.3\pm 0.4$ & $7.3\pm 0.2$ & $6.9\pm 0.2$ \\
\end{tabular}
\caption{Gaussian fit parameters of spectra in Fig. \ref{Fig.spectra-M43_gauss}.}
\label{Tab.fit_M43}
\end{table}

The eastern shell of M43, that lies closer to the shell center, moves outward at a velocity of about $v_{\mathrm{exp}}\simeq 2.5\,\mathrm{km\,s^{-1}}$ (estimated from channel maps, see above and App. \ref{App.east-shell}). This corroborates the thesis that this is a denser region compared to other structures in M43. Again, the systematic uncertainty of the mass estimate is of the order of 50\%, systematic uncertainties of the estimates of the extent of the shell and of the expansion velocity are about 30\%.

\subsection{NGC 1977}
\subsubsection{Geometry, mass and physical conditions}

\begin{figure}[tb]
\includegraphics[width=0.5\textwidth, height=0.417\textwidth]{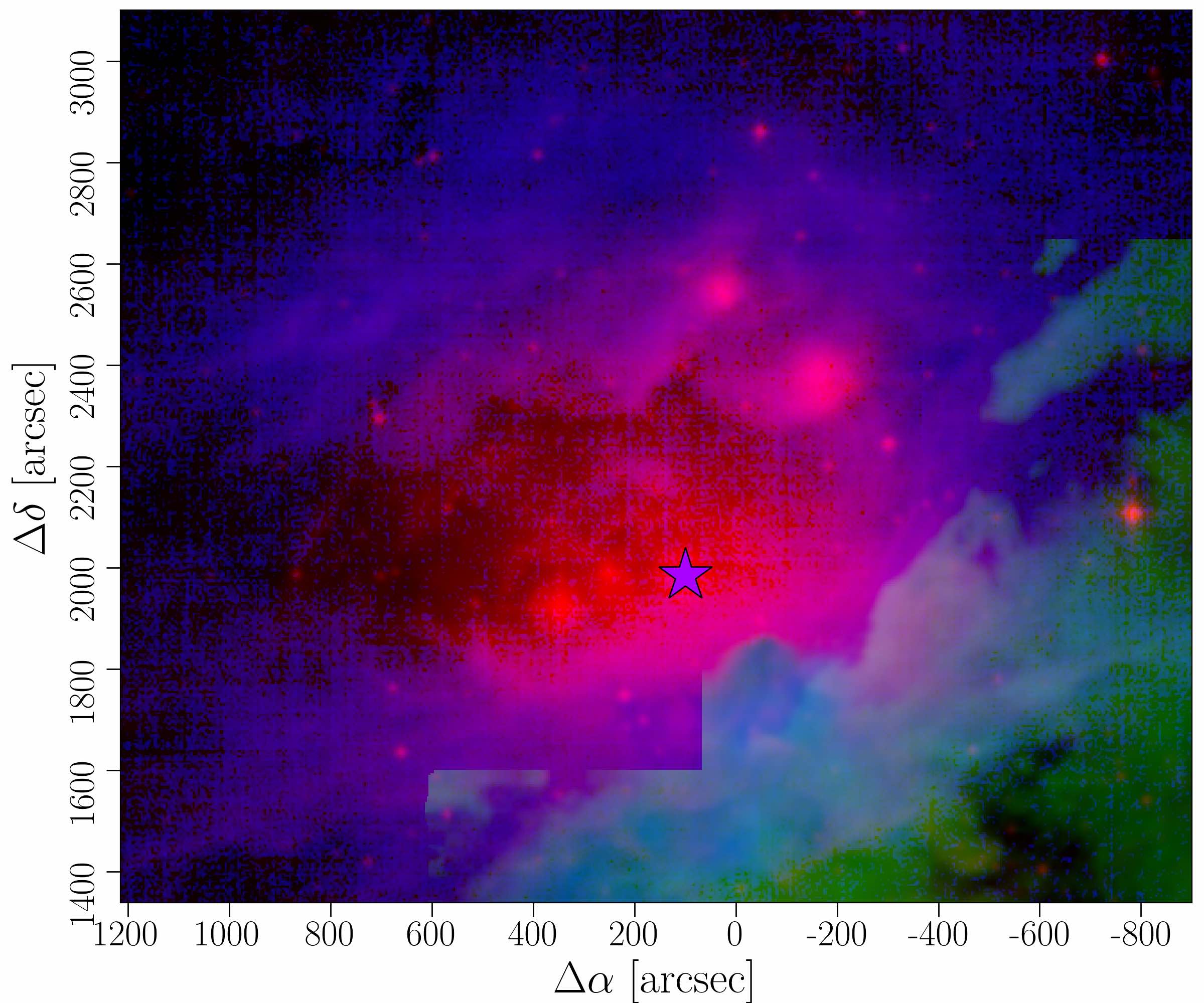}
\caption{Three-color image of NGC 1977. Red is H$\alpha$ emission from the ionized gas, green is line-integrated CO(2-1) emission from molecular gas (coverage only in OMC3), and blue is the [C\,{\sc ii}] line-integrated intensity, tracing the neutral gas.}
\label{Fig.NGC1977_rgb}
\end{figure}

As for M42 and M43, the shell structure of NGC 1977 is almost spherical, but again it is offset from the central star(s). We estimate its geometric center at $(\Delta\alpha,\Delta\delta)=(195\arcsec,155\arcsec)$ (from 42 Orionis, $(\Delta\alpha,\Delta\delta)=(-100.4\arcsec,1984.7\arcsec)$ from $\theta^1$ Ori C) and its outer radius with $r=800\arcsec=0.22\deg$, which corresponds to $r\simeq 1.6\,\mathrm{pc}$, assuming a distance of $414\,\mathrm{pc}$ \citep{Menten2007}. Hence, the projected distance of the bubble center from 42 Orionis is $0.5\,\mathrm{pc}$. The thickness of the limb-brightened shell is $\Delta r = 7'\simeq 0.8\,\mathrm{pc}$, somewhat thicker than the shell of M42, but more dilute as well. Figure \ref{Fig.NGC1977_rgb} shows a three-color image of NGC 1977. Visible H$\alpha$ stems from the center of NGC 1977, constituting an H\,{\sc ii} region, surrounded by a shell of [C\,{\sc ii}] and constrained by a bulk CO cloud (OMC3) in the southwest.

The limb-brightened shell is seen in [C\,{\sc ii}] emission in the velocity range $v_{\mathrm{LSR}}=9\mhyphen 15\,\mathrm{km\,s^{-1}}$. The brightest [C\,{\sc ii}] component stems from the molecular core OMC3, which hosts a PDR at its surface towards 42 Orionis. The [C\,{\sc ii}] emission associated with this PDR is analyzed in detail by Kabanovic et al. (in prep.). With the FUV luminosity given by \cite{Kim2016}, we compute an incident FUV intensity of $G_0\simeq 100$ at the shell surface.

From our $\tau_{160}$ map, we estimate the mass of the shell, where we exclude the region of OMC3. We obtain $M \simeq 700\,M_{\odot}$. With the [C\,{\sc ii}] luminosity $L_{\mathrm{[C\,\textsc{ii}]}} \simeq 140\,L_{\odot}$, we derive a mass of the [C\,{\sc ii}]-emitting gas of $M\simeq 540\,M_{\odot}$ for $T_{\mathrm{ex}}=50\,\mathrm{K}$, somewhat lower than the mass derived from the dust opacity, but still in good agreement\footnote{The excitation temperature is not well-constrained, see analysis below.}.

\begin{figure}[tb]
\includegraphics[width=0.5\textwidth, height=0.33\textwidth]{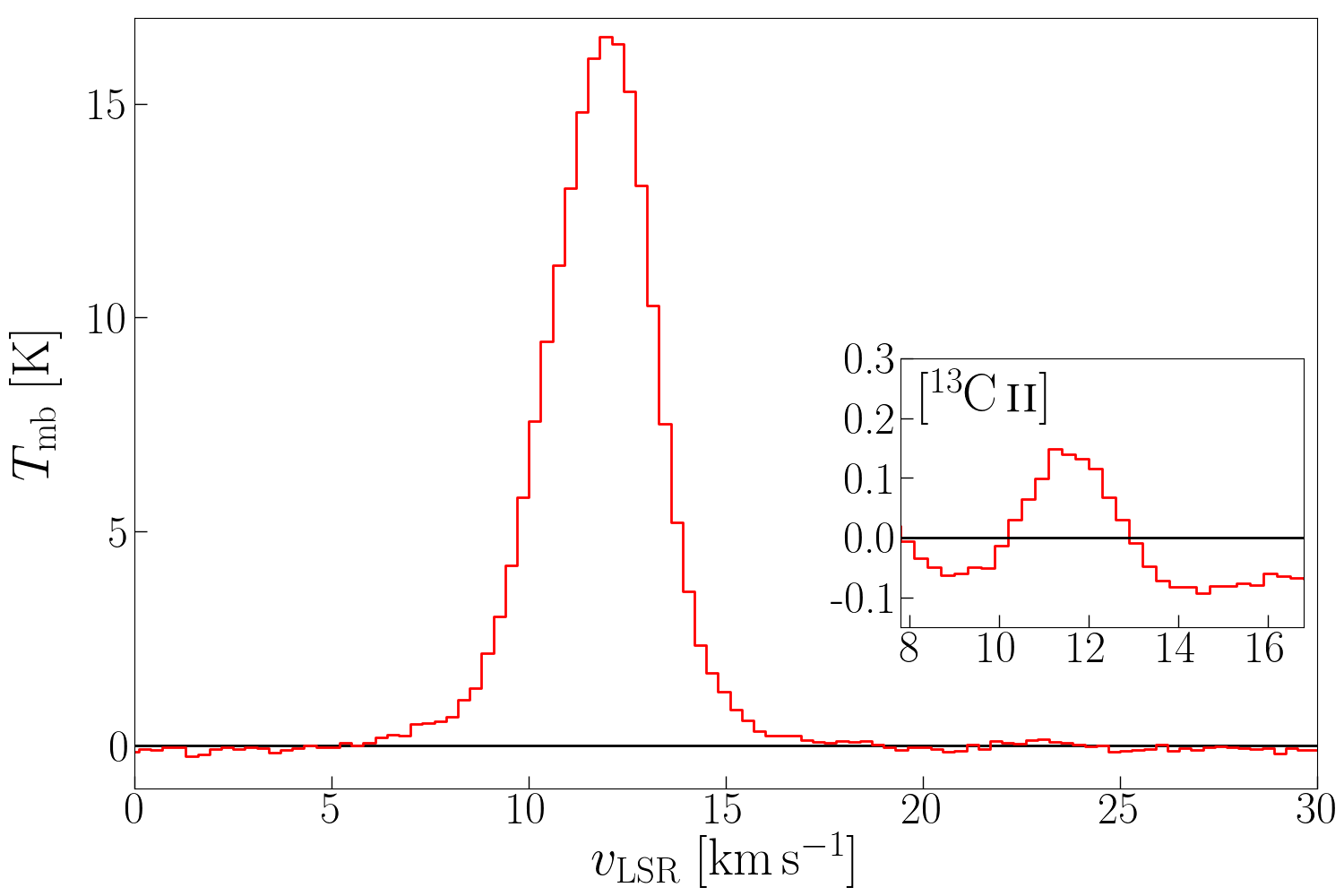}
\caption{Average [C\,{\sc ii}] spectrum towards [C\,{\sc ii}]-bright shell of NGC 1977 (without OMC3). The inlaid panel shows the (smoothed) residual of the spectrum, in the systemic velocity of the [$^{13}$C\,{\sc ii}] $F=2\mhyphen 1$ line, after subtracting the [$^{12}$C\,{\sc ii}] fit. The [$^{12}$C\,{\sc ii}] component can be approximated by a Gaussian with $T_{\mathrm{P}}\simeq 16.5\pm 0.1\,\mathrm{K}$, $v_{\mathrm{P}}\simeq 12.0\pm 0.1\,\mathrm{km\,s^{-1}}$, $\Delta v_{\mathrm{FWHM}}\simeq 3.2\pm 0.1\,\mathrm{km\,s^{-1}}$; the [$^{13}$C\,{\sc ii}] $F=2\mhyphen 1$ component is fitted by a Gaussian with $T_{\mathrm{P}}\simeq 0.21\pm 0.03\,\mathrm{K}$, $v_{\mathrm{P}}\simeq 11.8\pm 0.2\,\mathrm{km\,s^{-1}}$, $\Delta v_{\mathrm{FWHM}}\simeq 2.1\pm 0.5\,\mathrm{km\,s^{-1}}$.}
\label{Fig.spectrum-NGC1977}
\end{figure}

Figure \ref{Fig.spectrum-NGC1977} shows the average spectrum towards the [C\,{\sc ii}]-bright shell of NGC 1977, excluding OMC3. Here, the detection of the [$^{13}$C\,{\sc ii}] $F=2\mhyphen 1$ line is marginal with $\sigma \simeq 3$. As the [C\,{\sc ii}] optical depth and excitation temperature are very sensitive to the exact [$^{13}$C\,{\sc ii}] peak temperature, it is hard to derive reliable values with the uncertainties at hand (fit uncertainties and baseline ripples). With the results of a Gaussian fit as given in the caption of Fig. \ref{Fig.spectrum-NGC1977}, we obtain $\tau_{\mathrm{[C\,\textsc{ii}]}}\simeq 0.7\pm 0.3$ and $T_{\mathrm{ex}}\simeq 70\substack{+30 \\ -10}\,\mathrm{K}$. The resulting C$^+$column density then is $N_{\mathrm{C}^+}\simeq 6\pm 2\cdot 10^{17}\,\mathrm{cm}^{-2}$. If we assume a column length of $r/2\simeq 0.5\,\mathrm{pc}$ in the limb-brightened shell, we estimate a density of $n\simeq 2.5 \cdot 10^3\,\mathrm{cm}^{-3}$. This results in an estimate of the gas temperature of $T_{\mathrm{gas}}\simeq 180\,\mathrm{K}$, which is somewhat higher than expected for the rather moderate radiation field in NGC 1977.

The mean dust optical depth from NGC 1977 (without OMC3) is $\tau_{160}\simeq 2.4\cdot 10^{-3}$. From this, with $r/2\simeq 0.5\,\mathrm{pc}$, we estimate a gas density of $n\simeq 1 \cdot 10^4\,\mathrm{cm}^{-3}$ in the shell. With the average $T_{\mathrm{ex}}\simeq 70\,\mathrm{K}$ (from the dust optical depth and the peak temperature of the [C\,{\sc ii}]), this gives $T_{\mathrm{gas}}\simeq 90\,\mathrm{K}$. From the dust optical depth and the [C\,{\sc ii}] peak temperature, we calculate $\tau_{\mathrm{[C\,\textsc{ii}]}}\sim 3$, so [C\,{\sc ii}] emission is optically thick. This is in disagreement with the result from the [$^{13}$C\,{\sc ii}] $F=2\mhyphen 1$ line. However, the result obtained here seems to be more credible, as it yields a gas temperature that is more in line with a previous [C\,{\sc ii}] study of the Horsehead Nebula at similar impacting radiation field \citep{Pabst2017}. PDR models predict a surface temperature of $\sim 150\,\mathrm{K}$ \citep{Kaufman2006, PoundWolfire2008}, but the [C\,{\sc ii}]-emitting layer is expected to be somewhat cooler.

In the region where the spectrum in Fig. \ref{Fig.spectra-NGC1977} is taken, we have an average $\tau_{160}\simeq 8.6\cdot 10^{-4}$, corresponding to a hydrogen column density of $N_{\mathrm{H}}\simeq 5\cdot 10^{21}\,\mathrm{cm}^{-2}$. From H$\alpha$ emission, we can estimate the electron density in the ionized gas that is contained in the shell. We obtain $n_e\simeq 40 \,\mathrm{cm^{-3}}$. Hence, with a column length of $l\sim 2\,\mathrm{pc}$ the ionized gas is only a fraction of the total column density. Most of the dust emission within the column stems from the denser expanding shell. With a shell thickness of $d\simeq 0.8\,\mathrm{pc}$, appropriate for the limb-brightened shell, we estimate a gas density of $n\simeq 2\cdot 10^3 \,\mathrm{cm^{-3}}$. From the column density\footnote{We use half the column density for each of the two [C\,{\sc ii}] spectral components.} (assuming that all carbon is ionized) and the spectrum we calculate an excitation temperature of $T_{\mathrm{ex}}\simeq 37\,\mathrm{K}$ and a [C\,{\sc ii}] optical depth of $\tau_{\mathrm{[C\,\textsc{ii}]}}\simeq 1$. This corresponds to a gas temperature of $T_{\mathrm{gas}}\simeq 60\,\mathrm{K}$, which is somewhat lower than in the limb-brightened shell as calculated above. However, a slight overestimation of the dust optical depth of the expanding material would lead to a significantly increased estimate of the gas temperature.

\begin{figure}[tb]
\includegraphics[width=0.5\textwidth, height=0.33\textwidth]{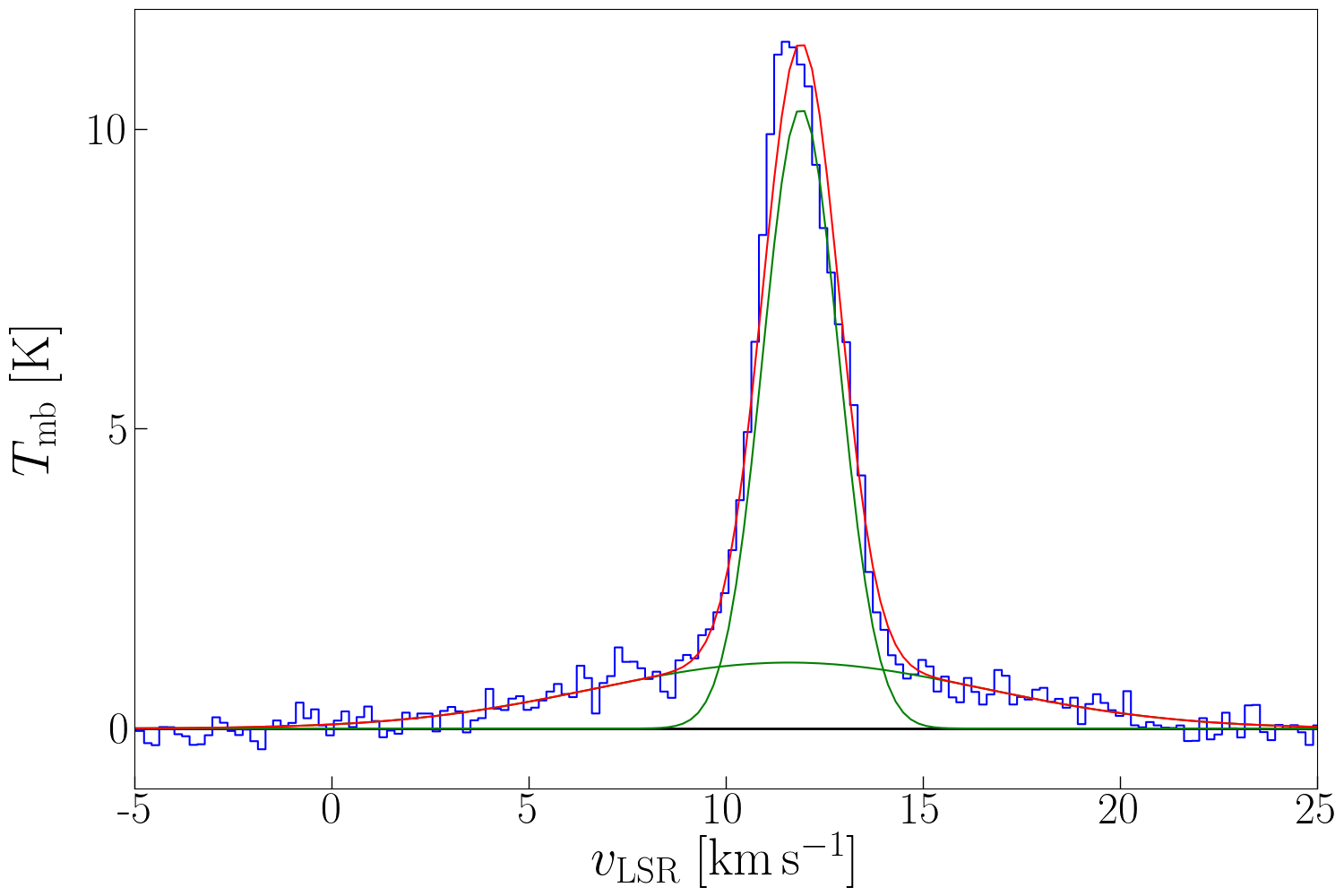}
\caption{[C\,{\sc ii}] spectrum towards NGC 1977, averaged over circular area with radius $125\arcsec$ centered at 42 Orionis ($(\Delta\alpha,\Delta\delta)=(-100.4\arcsec,1984.7\arcsec)$), with Gaussian fits. The fit parameters are given in Table \ref{Tab.fit_NGC1977_center}.}
\label{Fig.spectrum-HII-NGC1977}
\end{figure}

Towards the center of the H\,{\sc ii} region, we observe a very broad component in the [C\,{\sc ii}] spectrum (Fig. \ref{Fig.spectrum-HII-NGC1977}). This component likely stems from the ionized gas. The expected [C\,{\sc ii}] intensity from ionized gas at a temperature of $T\sim 10^4\,\mathrm{K}$ with an electron density of $n_e\simeq 40 \,\mathrm{cm^{-3}}$ and a line of sight of $l\simeq 2\,\mathrm{pc}$ matches the observed intensity very well. We note that the line is broader than what one would expect from thermal broadening alone. The additional broadening might be due to enhanced turbulence and the expansion movement of the gas.

\begin{table}[tb]
\begin{tabular}{c|ccc}
comp. &  $T_{\mathrm{P}}$ & $v_{\mathrm{P}}$ & $\Delta v_{\mathrm{FWHM}}$ \\ 
& $[\mathrm{K}]$ & $[\mathrm{km\,s^{-1}}]$ & $[\mathrm{km\,s^{-1}}]$ \\ \hline
1 & $10.3\pm 0.1$ & $11.9\pm 0.1$ & $2.3\pm 0.1$ \\
2 & $1.1\pm 0.1$ & $11.6\pm 0.2$ & $11.6\pm 0.6$
\end{tabular}
\caption{Gaussian fit parameters of spectrum in Fig. \ref{Fig.spectrum-HII-NGC1977}.}
\label{Tab.fit_NGC1977_center}
\end{table}

\subsubsection{Expansion velocity}

\begin{figure}[tb]
\includegraphics[width=0.5\textwidth, height=0.33\textwidth]{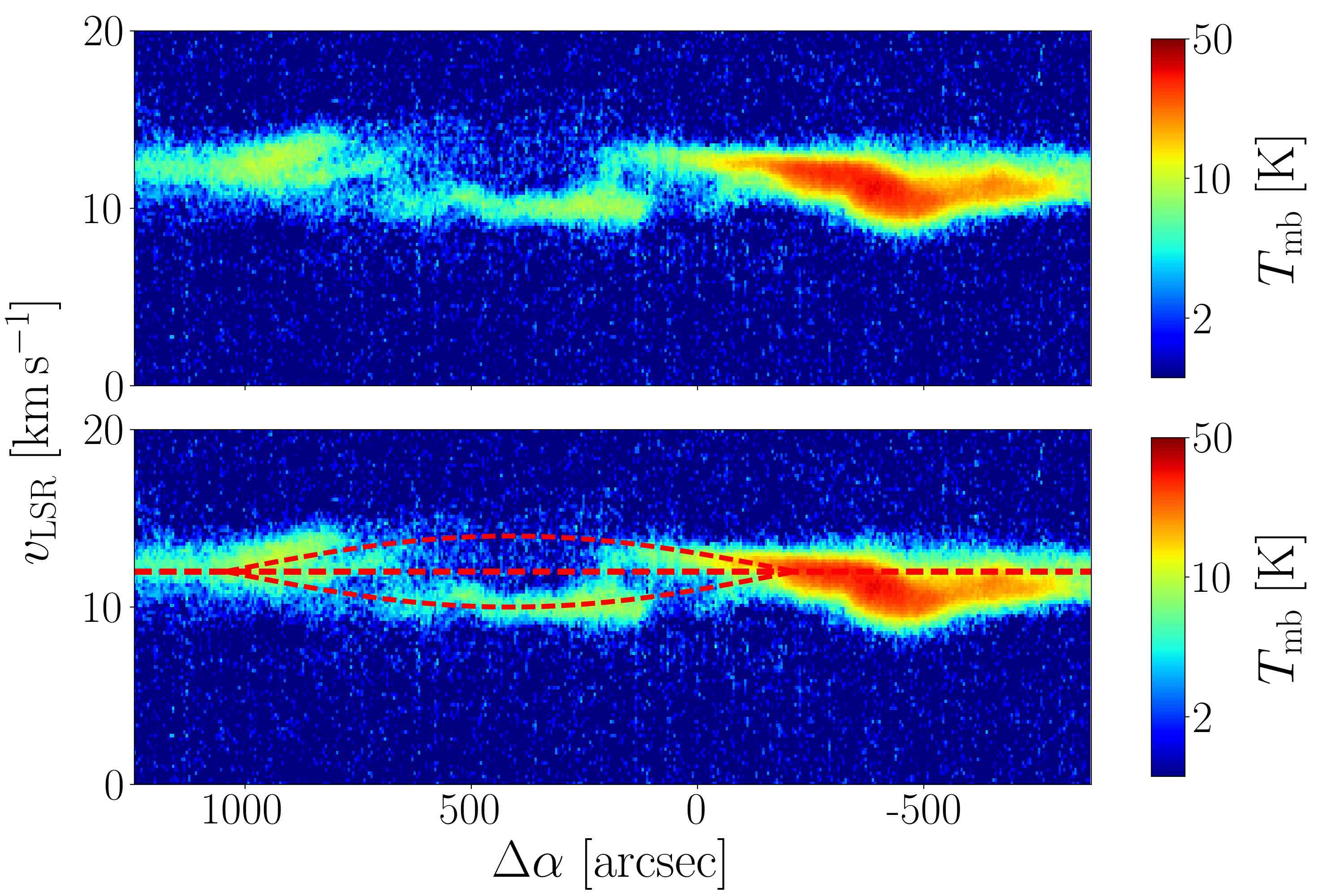}
\caption{[C\,{\sc ii}] pv diagram of NGC 1977 along position cut 3, indicated in Fig. \ref{Fig.map} ($\Delta\delta=2085\arcsec\mhyphen 2259\arcsec$). The lower panel shows the same cut with the arc structure for an expansion velocity of $\pm 1.5\,\mathrm{km\,s^{-1}}$ (red dashed lines).}
\label{Fig.pv-NGC1977}
\end{figure}

Figure \ref{Fig.pv-NGC1977} shows a [C\,{\sc ii}] pv diagram through the center of the bubble associated with NGC 1977. We recognize evidence of expansion but the arc structure is much fainter than in M42 and M43 and disrupted. As opposed to M42 and M43, there is no background molecular cloud constraining the expanding gas: we see the bubble expanding in two directions.

\begin{figure}[tb]
\includegraphics[width=0.5\textwidth, height=0.33\textwidth]{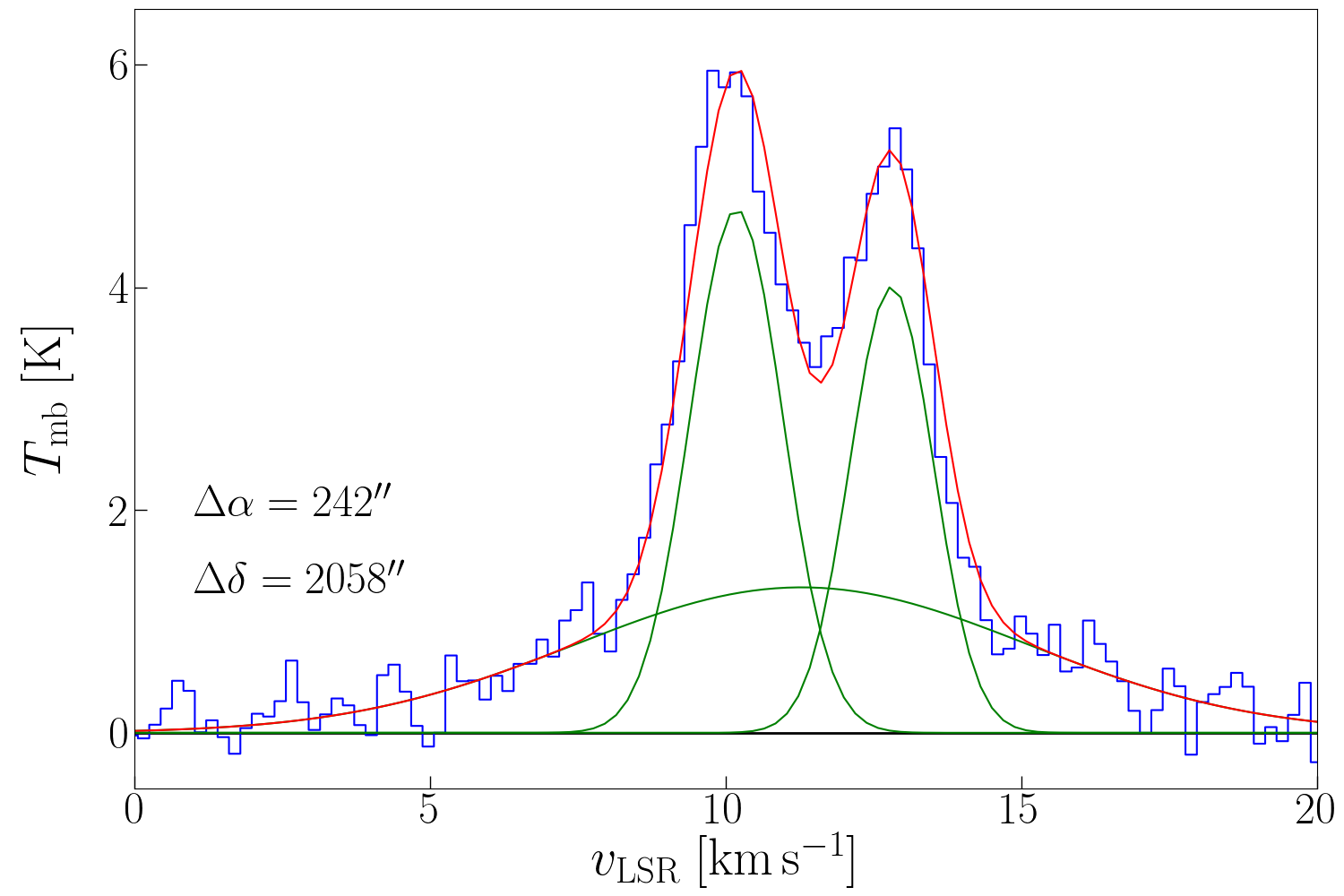}
\caption{[C\,{\sc ii}] spectrum along pv diagram in NGC 1977 (Fig. \ref{Fig.pv-NGC1977}) with Gaussian fits, averaged over $201\arcsec\times 201\arcsec$. Coordinate offsets indicate the lower left-hand corner of the square over which the spectrum is averaged. The two narrow components stem from the expanding shell, while the broad component originates in the ionized gas that is contained within the shell. Since the bubble expands in both directions, the peak-velocity difference is twice the expansion velocity, $v_{\mathrm{exp}}\simeq 1.3\,\mathrm{km\,s^{-1}}$. The fit parameters are given in Table \ref{Tab.fit_NGC1977}.}
\label{Fig.spectra-NGC1977}
\end{figure}

Since the [C\,{\sc ii}] emission from NGC 1977 is fainter than that of M42, we have opted to determine the expansion velocity from spectra towards this region (Fig. \ref{Fig.spectra-NGC1977}), resulting in $v_{\mathrm{exp}}\simeq 1.3\,\mathrm{km\,s^{-1}}$. This is consistent with the pv diagram shown in Fig. \ref{Fig.pv-NGC1977}, from which we obtain the bubble radius $r\simeq 500\arcsec \simeq 1.0\,\mathrm{pc}$ with the expansion velocity of $v_{\mathrm{exp}}\simeq 1.5\,\mathrm{km\,s^{-1}}$. As for the Veil shell and M43, the systematic uncertainty of the mass estimate are of the order of 50\%, systematic uncertainties of the estimates of the extent of the shell and of the expansion velocity are about 30\%.

\begin{table}[tb]
\begin{tabular}{c|ccc}
comp. &  $T_{\mathrm{P}}$ & $v_{\mathrm{P}}$ & $\Delta v_{\mathrm{FWHM}}$ \\ 
 & $[\mathrm{K}]$ & $[\mathrm{km\,s^{-1}}]$ & $[\mathrm{km\,s^{-1}}]$ \\ \hline
1 & $4.0\pm 0.2$ & $10.2\pm 0.1$ & $1.7\pm 0.1$ \\
2 & $4.7\pm 0.2$ & $12.8\pm 0.1$ & $1.8\pm 0.1$ \\
3 & $1.3\pm 0.2$ & $11.3\pm 0.2$ & $9.0\pm 0.7$
\end{tabular}
\caption{Gaussian fit parameters of spectrum in Fig. \ref{Fig.spectra-NGC1977}.}
\label{Tab.fit_NGC1977}
\end{table}

\section{Discussion}

OMC1, including the Trapezium cluster, is the nearest site of active intermediate- and high-mass star formation (e.g., \cite{Hillenbrand1997, Megeath2012, DaRio2014, Megeath2016, Grossschedl2019} for discussions of the stellar content). Most of the stars formed and forming are low-mass stars, but also some high-mass stars, that have profound impact on the characteristics and evolution of their environment. Outside the Huygens Region, a significant massive (triple) star is the O9.5IV star $\theta^2$ Ori A, located just to the southeast of the Orion Bar. The radiation of this star dominates the ionization structure of the gas towards the south-east of the Huygens Region \citep{ODell2017}. It also possesses strong winds. The most dominant star in the Orion Nebula is the O7V star $\theta^1$ Ori C, the most massive Trapezium star. It is itself a binary, with possibly a third companion \citep{Lehmann2010}. While its exact peculiar velocity is controversial, the $\theta^1$ Ori C is plowing away from the molecular cloud, the site of its birth, and will have travelled some $25\,\mathrm{pc}$ before it explodes as a supernova \citep{ODell2009, Kraus2009, Pabst2019}.

While the small, $0.5\,\mathrm{pc}$ sized Huygens Region has been extensively studied, studies on the much fainter EON are less numerous. \cite{ODell2010} determine the temperature of the ionized gas within the large EON H\,{\sc ii} region to be $T\sim 8.3\cdot 10^3\,\mathrm{K}$, while electron densities decrease from $n_e\sim 3000\,\mathrm{cm^{-3}}$ at the Trapezium stars to about $n_e\sim 30\,\mathrm{cm^{-3}}$ $20\arcmin$ away, roughly as $d^{-2}$, with $d$ the projected distance. \cite{Guedel2008} report that the EON exhibits significant X-ray emission, emanating from hot ($T\sim 2\cdot 10^6\,\mathrm{K}$), dilute ($n_e\sim 0.1\mhyphen 0.5\,\mathrm{cm^{-3}}$) gas. While, in principle, ionization followed by thermal expansion of the H\,{\sc ii} region can create a bubble of $\sim 2\,\mathrm{pc}$ size, the hot gas is the tell-tale signature that the stellar wind of massive stars, rather than stellar radiation, is driving the expansion and forming the EON cavity. As \cite{Guedel2008} note, the emission characteristics in conjunction with its structure and young age render it unlikely that the bubble is a supernova remnant. While the observed morphology is in qualitative agreement with simple models for stellar winds from massive stars \citep{Weaver1977}, the observed plasma temperature is lower than expected from refined models and suggests that mass loading of the hot plasma has been important \citep{Arthur2012}.

Recently, the inner shocked wind bubble surrounding $\theta^1$ Ori C has been identified in optical line observations \citep{Abel2019}. This inner shock will heat the gas in the EON that drives the expansion of a larger, outer, shell. These same observations, however, suggest that the thus heated gas is only free to escape through the south-west of the Huygens Region.

The limb-brightened edge of the Veil shell, the dense shell associated with M42, is readily observed edge-on. It is a closed, surprisingly spherically symmetric structure enveloping the inner Huygens Region and the EON \citep{Pabst2019}, and confining the hot X-ray emitting and ionized gas observed by \cite{Guedel2008} and \cite{ODell2010} in the foreground. Likewise, the H\,{\sc ii} regions of M43 and NGC 1977 are surrounded by rather dense shells. The limb-brightened shell of M43 exhibits a PDR-like layered structure (cf. Fig. \ref{Fig.cross-cuts_M43}). At the southern edge of the shell surrounding NGC 1977 one encounters OMC3, which also possesses a PDR-like structure, irradiated by 42 Orionis (Kabanovic et al., in prep.).

For our analysis we adopt a distance of $414\pm 7\,\mathrm{pc}$ \citep{Menten2007} towards the Orion Nebula complex, although more recent results suggest somewhat lower values, $388\pm 5\,\mathrm{pc}$ \citep{Kounkel2017}, which is in agreement with Gaia DR2 results \citep{Grossschedl2018}. In view of other uncertainties in the analysis, we consider this a minor source of uncertainty.

\subsection{The pressure balance}
\label{sec.pressure}

\begin{table*}[tb]
\centering
\begin{tabular}{ll|cccccc}
region & & $n\;\mathrm{[cm^{-3}]}$ & $T_{\mathrm{gas}}\;\mathrm{[K]}$ & $p_{\mathrm{thermal}}/k_{\mathrm{B}}\;\mathrm{[cm^{-3}\,K]}$ & $G_0$ \\ \hline % & $v_{\mathrm{P}}\;\mathrm{[km\,s^{-1}]}$ & $\Delta v_{\mathrm{FWHM}}\;\mathrm{[km\,s^{-1}]}$ \\ \hline
Orion Bar & H\,{\sc ii} & $5\cdot 10^3$ & $8\cdot 10^3$ & $8\cdot 10^7$ & \\ %& & \\
 & PDR & $6\cdot 10^5$ & $500$ & $3\cdot 10^8$ & $2\cdot 10^4$ \\ %& & \\
Veil shell & plasma & 0.3 & $2\cdot 10^6$ & $1\cdot 10^6$ & \\ %& & \\
 & H\,{\sc ii} & $50$ & $8\cdot 10^3$ & $8\cdot 10^5$ & \\%& -5 & $4 \mhyphen 12$ \\
 & PDR & $10^3\mhyphen 10^4$ & $\sim 100$ & $1\mhyphen 10\cdot 10^5$ & $\sim 100$ \\%& 8 & $2\mhyphen 4$\\
M43 & H\,{\sc ii} & 500 & $7.5\cdot 10^3$ & $8\cdot 10^6$ & \\%& 7 & 8 \\
 & PDR & $10^4$ & 100 & $1\cdot 10^6$ & $\sim 2\cdot 10^3$ \\%& 10 & 2\\
 NGC 1977 & H\,{\sc ii} & 40 & $\sim 10^4$ & $\sim 8\cdot 10^5$ \\%& & 12 & 12 \\
 & PDR & $10^3$ & 90 & $9\cdot 10^4$ & $\sim 100$ %& 13 & 2
\end{tabular}
\caption{Physical conditions of respective H\,{\sc ii} region and adjacent limb-brightened PDR shell in M42, M43, and NGC 1977. The plasma in M42 was analyzed by \cite{Guedel2008}. Data on the M42 H\,{\sc ii} region are from \cite{ODell2010}, data on the M43 H\,{\sc ii} region are from \cite{Simon-Diaz2011}. We note that the density of the M42 H\,{\sc ii} region given here is appropriate for the southern EON, as is $G_0$ in the Veil shell PDR.}
\label{Tab.physical_conditions}
\end{table*}

Table \ref{Tab.physical_conditions} summarizes the physical conditions in the H\,{\sc ii} regions and the limb-brightened PDR shells in M42, M43, and NGC 1977. Table \ref{Tab.pressure} summarizes the various pressure terms in the total pressure $p_{\mathrm{tot}} = p_{\mathrm{thermal}} + p_{\mathrm{magnetic}} + p_{\mathrm{turb}} + p_{\mathrm{lines}} + p_{\mathrm{rad}}$ in the PDRs of M42 (Orion Bar and Veil shell), M43, and NGC 1977.

The thermal pressure, $p_{\mathrm{thermal}}=nkT$, in the Orion Bar is controversial. Constant-pressure and H\,{\sc ii} region models of atomic and molecular lines of \cite{Pellegrini2009} and observations of molecular hydrogen by \cite{Allers2005} ($n\simeq 10^5\,\mathrm{cm^{-3}}$, $T_{\mathrm{gas}}\simeq 500\,\mathrm{K}$) indicate a pressure of $p/k_{\mathrm{B}}\simeq 5\cdot 10^7\,\mathrm{K\,cm^{-3}}$. Observations of [C\,{\sc ii}] and [O\,{\sc i}] emission in the Orion Bar suggest similar values ($n\gtrsim 10^5\,\mathrm{cm^{-3}}$, $T_{\mathrm{gas}}\gtrsim 300\,\mathrm{K}$) \citep{Bernard-Salas2012, Goicoechea2015}. High-J CO observations indicate a higher gas pressure within the Bar, $3\cdot 10^8\,\mathrm{K\,cm^{-3}}$ \citep{Joblin2018}. This latter value is in agreement with observations of carbon radio recombination lines (CRRLs) that measure the electron density in the PDR directly ($n\gtrsim 4\mhyphen 7\cdot 10^5\,\mathrm{cm^{-3}}$) \citep{Cuadrado2019}. The thermal pressures in the EON portion of M42, M43, and NGC 1977 have been calculated from the parameters given in Table \ref{Tab.physical_conditions}.

The magnetic field strength in the Orion Bar has been estimated from the far-IR dust polarization using the Davis-Chandrasekhar-Fermi method to be $300\,\mu\mathrm{G}$ \citep{Chuss2019}. This corresponds to a magnetic field pressure, $p_{\mathrm{magnetic}}=B^2/8\pi$, of $3\cdot 10^7\,\mathrm{K\,cm^{-3}}$. The magnetic field in the Veil in front of the Huygens Region is measured to be $B_{\mathrm{los}} \simeq -50\mhyphen -75\,\mu\mathrm{G}$ from the H\,{\sc i} and OH Zeeman effect \citep{Troland2016}; we compute the magnetic field pressure from the lower value\footnote{The magnetic pressure is computed from $B_{\mathrm{tot}}= 3 B_{\mathrm{los}}^2$.}. The turbulent pressure, $p_{\mathrm{turb}}=\rho \sigma_{\mathrm{turb}}^2$, is calculated from the [C\,{\sc ii}] line width in the average spectra towards the Veil shell, M43, and NGC 1977 (Figs. \ref{Fig.spectrum-M42}, \ref{Fig.spectrum-M43}, and \ref{Fig.spectra-NGC1977}) after correction for thermal broadening at a kinetic temperature of $T_{\mathrm{gas}} \sim 100\,\mathrm{K}$ (cf. Table \ref{Tab.physical_conditions}); with a typical line width of $\Delta v_{\mathrm{FWHM}} \simeq 4\,\mathrm{km\,s^{-1}}$ this gives $\sigma_{\mathrm{turb}} \simeq 1.7\,\mathrm{km\,s^{-1}}$. In the Orion Bar, we use $\sigma_{\mathrm{turb}} \simeq 1.5\,\mathrm{km\,s^{-1}}$ \citep{Goicoechea2015}.

The radiation pressure is in principle given by $p_{\mathrm{rad}}=L_{\star}/(4\pi R_S^2c)$; the luminosities of the central stars are given in Table \ref{Tab.1}. We assume the (projected) distances $R_S\simeq 4\,\mathrm{pc}$ for the (southern) Veil shell, $R_S\simeq 0.25\,\mathrm{pc}$ for M43, and $R_S\simeq 1.0\,\mathrm{pc}$ for NGC 1977. However, for the Orion Bar, a direct measurement of the infrared flux suggests a value that is an order of magnitude lower than the pressure calculated from the stellar luminosity and the projected distance of $R_S\simeq 0.114\,\mathrm{pc}$ \citep{Pellegrini2009}, an incident radiation field of $G_0 = 2.6\cdot 10^4$ \citep{Marconi1998, Salgado2016}. Hence, we choose to calculate the radiation pressure in the Orion Bar from this latter value. We note that in the other cases the radiation pressure given in Table \ref{Tab.pressure} is, thus, an upper limit.

Resonant scattering of Ly$\alpha$ photons constitutes the major contribution to the line pressure term $p_{\mathrm{lines}}$. We expect this term to be at most of the order of the radiation pressure and hence negligible in the cases presented here \citep{Krumholz2009}.

We consider now the PDR associated with the Orion Bar, which is likely characteristic for the dense PDR associated with the OMC1 core. Perusing Tables \ref{Tab.physical_conditions} and \ref{Tab.pressure}, we conclude that the thermal pressure derived from high-J CO lines and CRRLs exceeds the turbulent and magnetic pressure by an order of magnitude. While the [C\,{\sc ii}] line stems from the surface of the PDR, high-J CO lines and CRRLs stem from deeper within the PDR. Models of photoevaporating PDRs suggest that the pressure increases with depth into the PDR \citep{Bron2018}. The [C\,{\sc ii}]-emitting PDR surface may be well described by a lower pressure ($5\cdot 10^7\,\mathrm{K\,cm^{-3}}$), in which case approximate equipartition holds between the three pressure terms. Their combined pressures of $1\cdot 10^8\,\mathrm{K\,cm^{-3}}$ then is balanced by the thermal pressure of the ionized gas, $8\cdot 10^7\,\mathrm{K\,cm^{-3}}$, with a contribution from the radiation pressure.

For the PDR in the Veil shell, the thermal, turbulent and magnetic pressure are again in approximate equipartition. In this case, there is approximate pressure equilibrium between the combined pressures in the PDR gas and the combined pressures of the ionized gas and the hot plasma in the EON. Overall, there is a clear pressure gradient from the dense molecular cloud core behind the Trapezium stars to the Veil shell in front. This strong pressure gradient is responsible for the rapid expansion of the stellar wind bubble towards us and sets up the ionized gas flow, which (almost freely) expands away from the ionization front at about $10\,\mathrm{km\,s^{-1}}$ \citep{Garcia-Diaz2008, ODell2009}. During the initial phase, the thermal pressure of the hot gas drives the expansion of the shell and radiation pressure provides only a minor contribution \citep{Silich2013}. Radiation pressure takes over once the plasma has cooled through energy conduction and the bubble enters the momentum-driven phase.

Turning now towards the H\,{\sc ii} regions, M43 and NGC 1977, we note that the sum of the thermal and turbulent pressures in the PDRs of these two regions is well below the thermal pressure of the ionized gas. While, in principle, the pressure could be balanced by the magnetic field, we expect equipartition to hold as well. Rather, we take this pressure difference to imply that the overpressure of the ionized gas drives the expansion of these H\,{\sc ii} regions. We note that radiation pressure is not important for these two nebulae.

\begin{table}[tb]
\centering
\addtolength{\tabcolsep}{-3pt}
\begin{tabular}{l|cccc}
$p/k_{\mathrm{B}}$ & Orion Bar & Veil shell & M43 & NGC 1977 \\ \hline
thermal & $3\cdot 10^8$ & $1\mhyphen 10\cdot 10^5$ & $1\cdot 10^6$ & $9\cdot 10^5$ \\
magnetic & $3\cdot 10^7$ & $2\cdot 10^6$ & -- & -- \\
turbulence & $3\cdot 10^7$ & $0.5\mhyphen 3\cdot 10^6$ & $8\cdot 10^5$ & $8\cdot 10^4$ \\
radiation & $1\cdot 10^7$ & $1\cdot 10^5$ & $3\cdot 10^6$ & $8\cdot 10^4$
\end{tabular}
\caption{Comparison of pressure terms in PDRs of the Orion Bar, Veil shell, M43, and NGC 1977. In the Veil shell, higher pressures correspond to the limb-brightened edges, while lower pressures apply to the foreground expanding shell. The radiation pressure is computed from the total luminosity of the central star.}
\label{Tab.pressure}
\end{table}

\subsection{Rayleigh-Taylor instability of the Veil shell} 
\begin{figure}[tb]
\includegraphics[width=0.5\textwidth, height=0.33\textwidth]{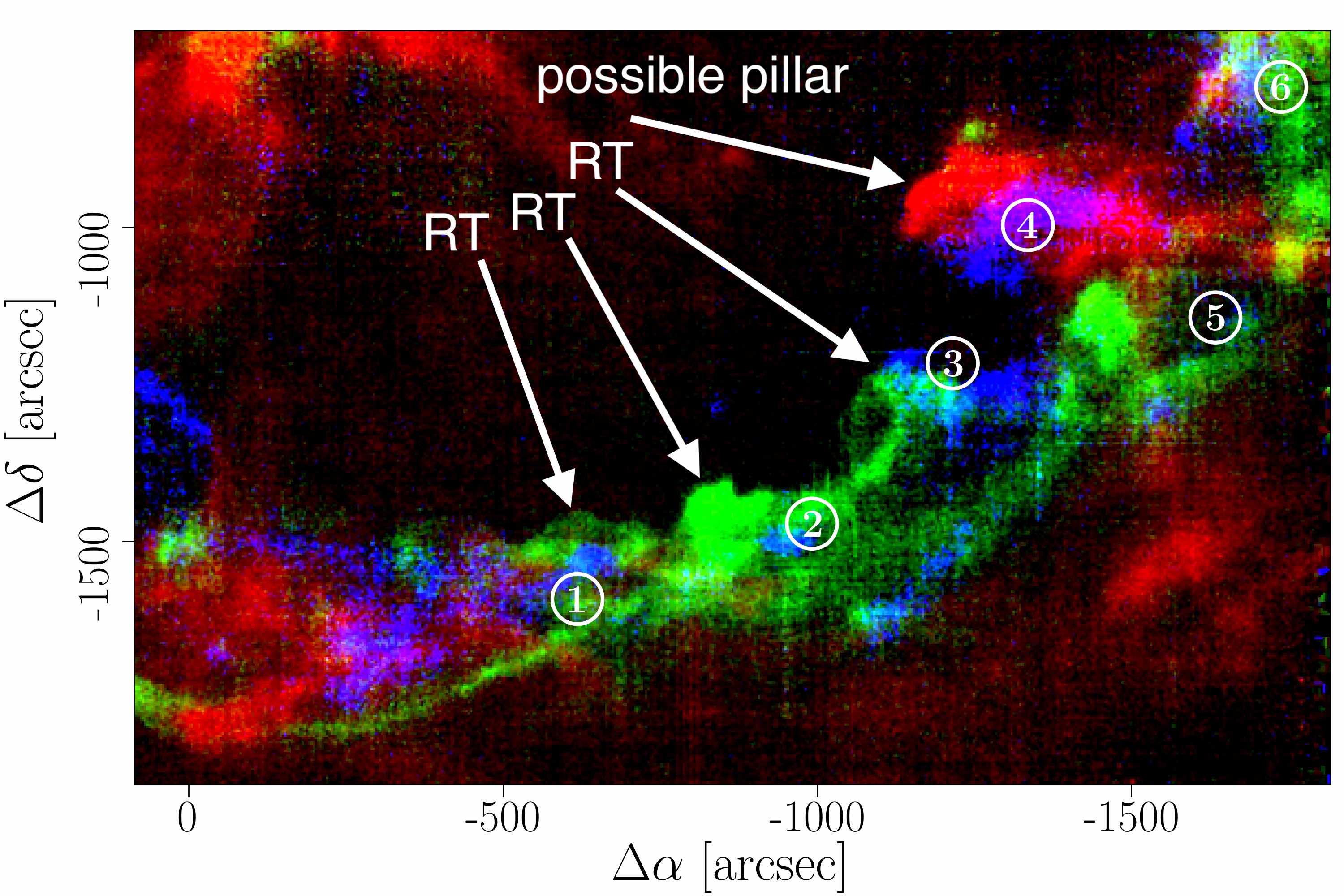}
\caption{Three-color image of [C\,{\sc ii}] velocity channels of the southern Veil shell. Blue is the velocity channel $v_{\mathrm{LSR}}=0\mhyphen2\,\mathrm{km\,s^{-1}}$, green $v_{\mathrm{LSR}}=4\mhyphen6\,\mathrm{km\,s^{-1}}$, and red $v_{\mathrm{LSR}}=8\mhyphen10\,\mathrm{km\,s^{-1}}$. Arrows indicate structures that resemble fingers formed due to the Rayleigh-Taylor (RT) instability. One of these, labeled ``possible pillar'', might also be formed by an overdensity encountered by the expanding shell and subsequently sculpted by the strong stellar radiation field, as it also is  observed in CO (cf. Fig. \ref{Fig.overview}c). The spectra shown in Fig. \ref{Fig.spectra-RT} were extracted towards the areas indicated by the numbered circles.}
\label{Fig.image-RT}
\end{figure}

Molecular clouds, under the influence of massive stars, can develop hydrodynamical instabilities that compress or fragment the cloud. In our [C\,{\sc ii}] data as well as in the H$\alpha$ image, we see evidence that the expanding gas might suffer from Rayleigh-Taylor instability, leading to the formation of elongated structures perpendicular to the expanding surface (cf. Fig. \ref{Fig.overview}a and b). These fingers are particularly clear in the EON towards the south of the Veil shell. Figure \ref{Fig.image-RT} shows this part of the Veil shell with its complex velocity structure. Rayleigh-Taylor instability occurs at the interface of a ``heavy'' fluid (here the expanding neutral [C\,{\sc ii}] gas) floating on top of a ``light'' fluid (here the hot plasma). This instability leads to increased turbulence and mixing of the gas. The presence of magnetic fields may alter the growth behavior of the characteristic structures, but it is yet under debate whether Rayleigh-Taylor instability will be enhanced or suppressed \citep{Stone2007,Carlyle2017}, the details likely depending crucially on the magnetic field configuration, which is hard to determine observationally.

The scale size of the unstable region after a time $t$ is given by:
\begin{align}
h = \alpha \mathcal{A} gt^2,
\end{align}
where $g$ is the (constant) acceleration and the Atwood number $\mathcal{A}=\frac{n_1-n_2}{n_1+n_2}$ is a measure of the density contrast across the contact discontinuity \citep{Chevalier1992, Duffell2016}. In this case, $\mathcal{A}\sim 1$, as the density contrast is large. The parameter $\alpha$ is determined by experiments and numerical simulations to be $\sim 0.05$. Realizing that the radius of the contact discontinuity is $R_{\mathrm{cd}}=\frac{1}{2}gt^2$, experiments and simulations predict that $h/R_{\mathrm{cd}}\sim 0.1$, which is in good agreement with the observed structures, that have a typical height of $\sim 100\arcsec \simeq 0.2\,\mathrm{pc}$ in a shell with a radius of $\sim 2.5\,\mathrm{pc}$. This lends further support for the interpretation of the observed morphology as the result of the Rayleigh-Taylor instability and for the general interpretation of the dominance of the stellar wind in driving the expansion of the shell.

\begin{figure}[tb]
\includegraphics[width=0.5\textwidth, height=0.33\textwidth]{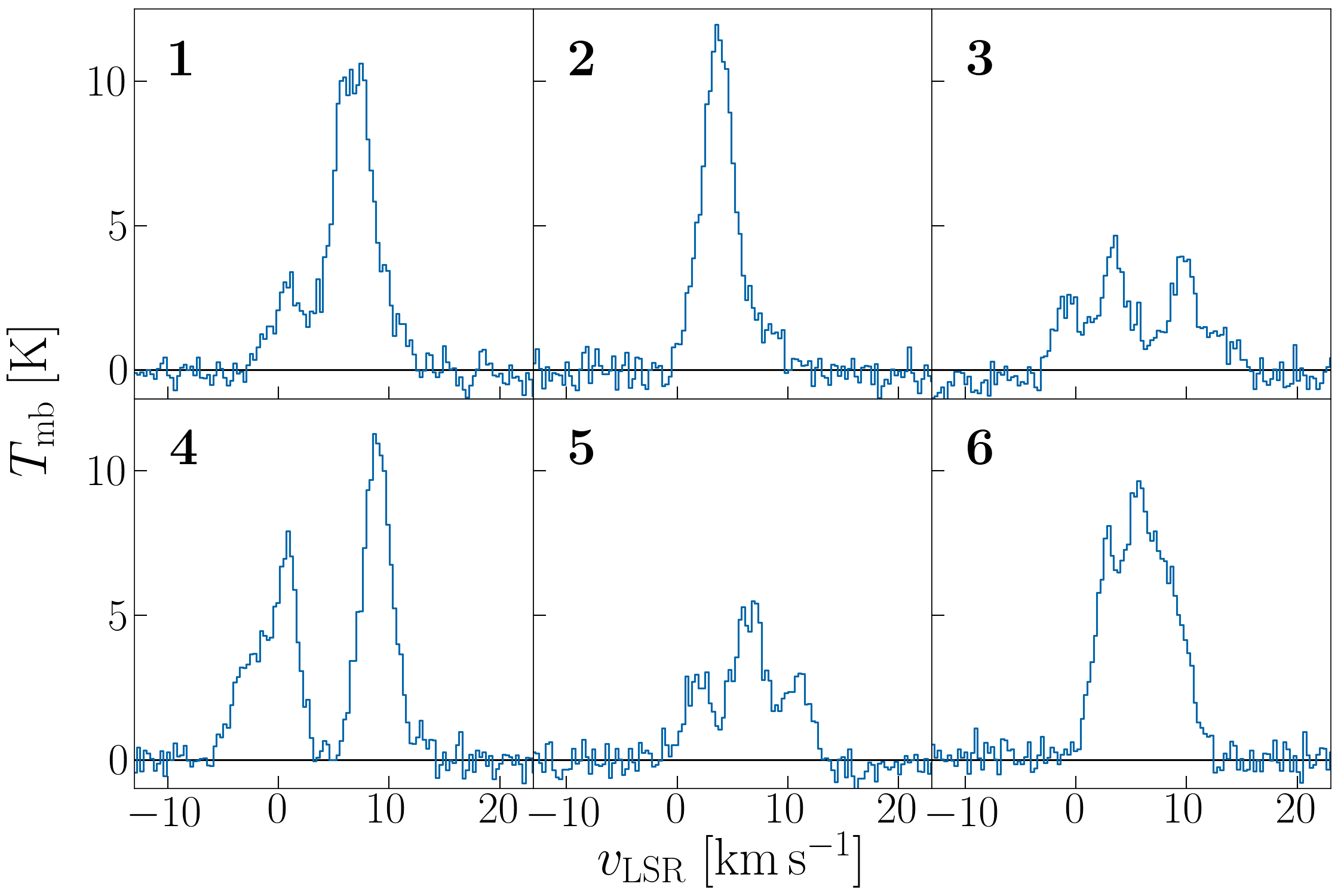}
\caption{[C\,{\sc ii}] spectra towards southern Veil shell. Each spectrum is averaged over a circle with a radius of $40\arcsec$ as indicated in Fig. \ref{Fig.image-RT}. We note that each spectrum consists of multiple line components, which is characteristic of thermodynamic instabilities.}
\label{Fig.spectra-RT}
\end{figure}
%1 -- $(\Delta\alpha,\Delta\delta)=(-622\arcsec,-1592\arcsec)$, 2 -- $(-997\arcsec,-1472\arcsec)$, 3 -- $(-1222\arcsec,-1217\arcsec)$, 4 -- $(-1342\arcsec,-997\arcsec)$, 5 -- $(-1642\arcsec,-1144\arcsec)$, 6 -- $(-1747\arcsec,-777\arcsec)$.

[C\,{\sc ii}] observations provide a unique opportunity to probe the injection of turbulence into a molecular cloud through expanding shock waves, as [C\,{\sc ii}] emission traces the surface layer that transmits the shock wave into the surrounding medium. The morphology and velocity structure of our [C\,{\sc ii}] observations seem to indicate the importance of Rayleigh-Taylor and Kelvin-Helmholtz instabilities\footnote{Kelvin-Helmholtz instability in the Orion Nebula has been studied by \cite{Berne2010, Berne2012}. The characteristic ``ripples'' they detect in the western shell in PAH and CO emission are also present in the [C\,{\sc ii}] emission from that region.} in the transmittance of turbulence into the expanding Veil shell. The simulations of \cite{Nakamura2006} show a similar velocity structure in the synthesized spectra as our observed spectra (Fig. \ref{Fig.spectra-RT}). With a typical velocity dispersion of $3\,\mathrm{km\,s^{-1}}$, the energy injected into the Veil shell by the Rayleigh-Taylor instability can be estimated at $3^2/13^2$ of the kinetic energy of the Veil shell, that is $1.4\cdot 10^{47}\,\mathrm{erg}$. Such macro-turbulence, associated with cloud destruction, will eventually be converted into micro-turbulence of the cloud medium \citep{Nakamura2006}. Post-processing of such numerical simulations in the [C\,{\sc ii}] emission and comparison to the data might be very illuminating.

\subsection{The structure of M43}

\begin{figure}[tb]
\includegraphics[width=0.5\textwidth, height=0.33\textwidth]{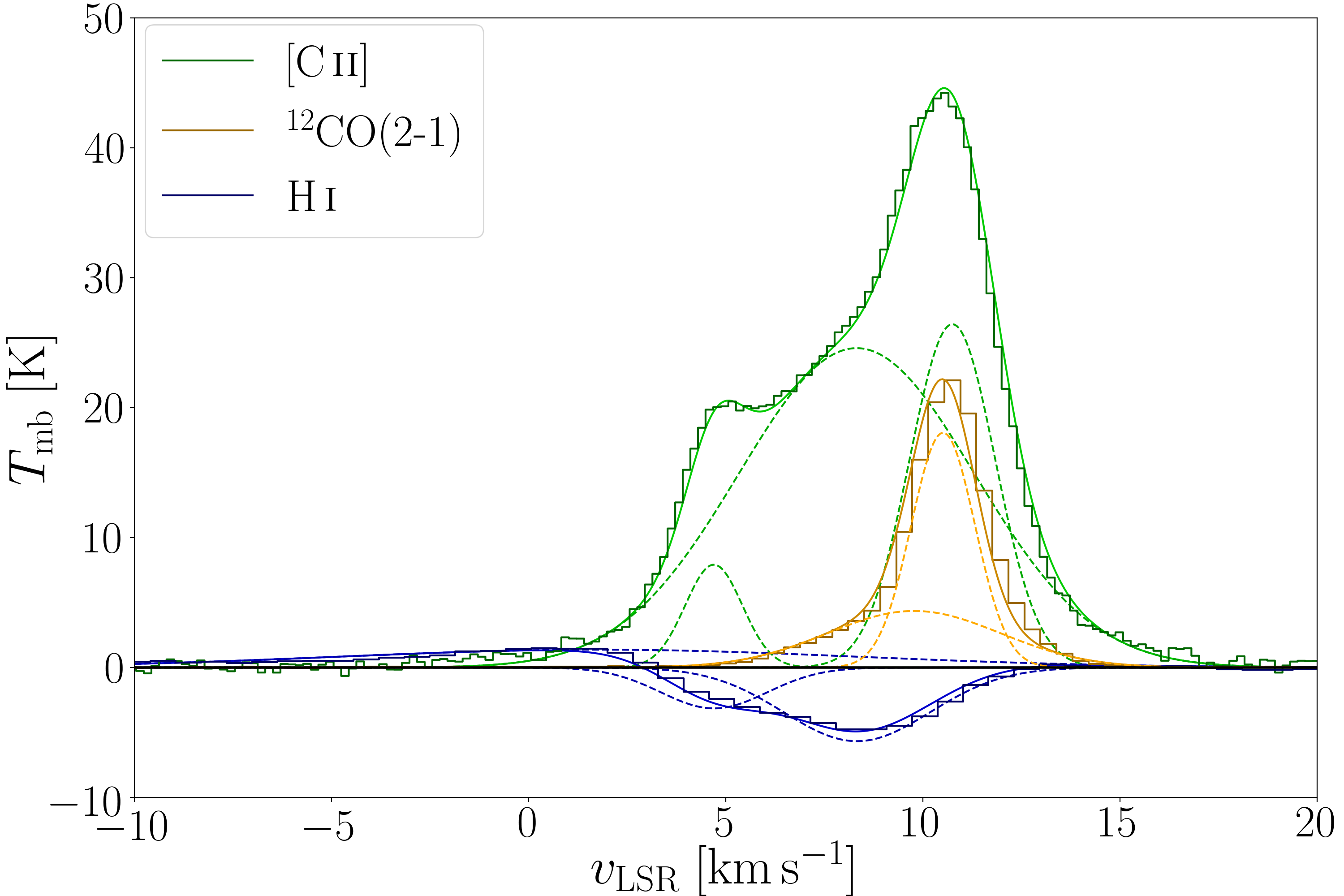}
\caption{[C\,{\sc ii}] (green), ${}^{12}$CO(2-1) (yellow), and H\,{\sc i} (blue) spectra towards the center of M43, averaged over a circle with radius $r\simeq 40\arcsec$ and center $(\Delta\alpha,\Delta\delta)=(-226\arcsec,440\arcsec)$. Plotted are also the results of Gaussian fits (dashed lines). The fit parameters are given in Table \ref{Tab.fit_M43_center}. The H\,{\sc i} spectrum is multiplied by $10^{-2}$ to fit the scale.}
\label{Fig.spectra-M43_3}
\end{figure}

Both [C\,{\sc ii}] and H\,{\sc i} show three velocity components while CO shows only two (Fig. \ref{Fig.spectra-M43_3} and Table \ref{Tab.fit_M43_center}). We attribute the [C\,{\sc ii}] high-velocity component ($10.7\,\mathrm{km\,s^{-1}}$) to emission from the background molecular gas associated with OMC1, that we also recognize in CO emission (cf. Fig. \ref{Fig.3-pv-diagrams}). The $4.7\,\mathrm{km\,s^{-1}}$ [C\,{\sc ii}] component can be attributed to a half-shell expanding towards us. As this component is also visible in H\,{\sc i} absorption against the radio continuum emission from the H\,{\sc ii} region, this component can be placed firmly in front. We note the difference in line width between the H\,{\sc i} and [C\,{\sc ii}] ($3.2\,\mathrm{km\,s^{-1}}$ versus $1.7\,\mathrm{km\,s^{-1}}$), which we attribute to the importance of thermal broadening for the former. This would imply a temperature of the neutral gas of $\sim 150\,\mathrm{K}$.

The $8.3\,\mathrm{km\,s^{-1}}$ [C\,{\sc ii}] component is more enigmatic. Given its large line width, we are tempted to associated this component with emission from the ionized gas. However, a $8.3\,\mathrm{km\,s^{-1}}$ component is also present in absorption in H\,{\sc i} and this would place it in front of the ionized gas. Perhaps, this H\,{\sc i} component is really associated with the Veil partially overhanging M43, whereas the [C\,{\sc ii}] component is clearly confined to the interior of M43 (cf. Fig. \ref{Fig.spectra-M43}).

Furthermore, we do not see a counterpart to the CO $9.8\,\mathrm{km\,s^{-1}}$ component in [C\,{\sc ii}] and this gas must be associated with a structure in OMC1. Finally, the very broad, low velocity ($1.7\,\mathrm{km\,s^{-1}}$) H\,{\sc i} emission component seems to form a coherent structure with H\,{\sc i} emission associated with the Northeast Dark Lane (as can be seen from the comparison of the pv diagrams in Fig. \ref{Fig.3-pv-diagrams}). This H\,{\sc i} component is also seen in dust extinction but there is no counterpart in the [C\,{\sc ii}] line, in the dust continuum at $70\,\mu\mathrm{m}$, or in the PAH emission at $8\,\mu\mathrm{m}$. Hence, this component is not illuminated by either $\theta^1$ Ori C or NU Ori, placing this material firmly in front of the Veil. As has been suggested by \cite{Troland2016}, this molecular gas very likely has been undisturbed by the star-formation process in OMC1.

In the [C\,{\sc ii}] line-integrated map (cf. Fig. \ref{Fig.M43_rgb}), the limb-brightened shell structure of M43 separates into two parts, with a break towards the east. This rupture is reflected in the pv diagram (Fig. \ref{Fig.spectra-M43}) through the center of M43, exhibiting a gap in the middle of the arc. In pv diagrams along the R.A. axis, this disruption, however, is located in the western part of M43 (cf. Fig. \ref{Fig.pv-diagrams-x_wolines})

\begin{table}[tb]
\begin{tabular}{cc|ccc}
line & comp. &  $T_{\mathrm{P}}$ & $v_{\mathrm{P}}$ & $\Delta v_{\mathrm{FWHM}}$ \\
 & & $[\mathrm{K}]$ & $[\mathrm{km\,s^{-1}}]$ & $[\mathrm{km\,s^{-1}}]$ \\ \hline
\text{[C\,{\sc ii}]} & 1 & $26.4\pm 0.3$ & $10.7\pm 0.1$ & $2.6\pm 0.1$ \\
\text{[C\,{\sc ii}]}  & 2 & $24.6\pm 0.2$ & $8.3\pm 0.1$ & $7.0\pm 0.1$ \\ 
\text{[C\,{\sc ii}]}  & 3 & $7.9\pm 0.3$ & $4.7\pm 0.1$ & $1.7\pm 0.1$ \\ \hline
$^{12}$CO(2-1) & 1 & $18.1\pm 0.1$ & $10.5\pm 0.1$ & $1.9\pm 0.1$ \\
$^{12}$CO(2-1) & 2 & $4.4\pm 0.1$ & $9.8\pm 0.1$ & $5.2\pm 0.1$ \\ \hline
H\,{\sc i} & 1 & $-5.7\pm 0.2$ & $8.3\pm 0.2$ & $4.4\pm 0.3$ \\
H\,{\sc i} & 2 & $-3.1\pm 0.3$ & $4.7\pm 0.2$ & $3.2\pm 0.3$ \\ 
H\,{\sc i} & 3 & $1.4\pm 0.2$ & $1.7\pm 0.6$ & $15.5\pm 0.9$ \\
\end{tabular}
\caption{Gaussian fit parameters of spectra in Fig. \ref{Fig.spectra-M43_3}. The H\,{\sc i} peak temperatures have to be multiplied by $10^2$ to obtain the main-beam temperature.}
\label{Tab.fit_M43_center}
\end{table}

\subsection{The Veil stellar-wind bubble}
\label{Sec.wind-bubbles}

\cite{Guedel2008} in their influential paper on the X-ray emission from Orion revealed that the interior of the Veil shell as outlined by the [C\,{\sc ii}] line, the $8\,\mu\mathrm{m}$ PAH emission, and the $70\,\mu\mathrm{m}$ dust emission is filled with a hot ($T\sim 2\cdot 10^6\,\mathrm{K}$), tenuous ($n\sim 0.5\,\mathrm{cm}^{-3}$) plasma with a total mass of $\sim 0.07\,M_{\sun}$. This hot plasma is generated by the reverse shock acting on the fast stellar wind originating from $\theta^1$ Ori C. The observations reported in this study provide the velocity, radius and mass of the shell swept up by the bubble. Following \cite{Guedel2008}, we will use the stellar wind theory of \cite{Castor1975} and \cite{Weaver1977} to analyze these data.

Based on models of \cite{Castor1975} we have obtained the following two time-dependent equations for a thin shell expanding under the influence of stellar winds, under the assumption that the stellar wind is constant; this assumption is valid during most of the lifetime of a star \citep{Haid2018} and also in the so-called snowplow phase that is discussed by \cite{Castor1975}:
\begin{align}
R_S(t) &\simeq 28\, \left(\frac{\dot{M}_6 v_{2000}^2}{n_0}\right)^{1/5}\,\left(\frac{t}{\mathrm{1\,Myr}}\right)^{3/5} \,\mathrm{pc} \label{Eq.R}, \\
\dot{R}_S(t) &\simeq 16.4\, \left(\frac{\dot{M}_6 v_{2000}^2}{n_0}\right)^{1/5}\,\left(\frac{t}{\mathrm{1\,Myr}}\right)^{-2/5} \,\mathrm{km\,s^{-1}} \label{Eq.Rdot},
\end{align}
where $\dot{M}_6$ is the mass-loss rate in $10^{-6}\,M_{\sun}\,\mathrm{yr^{-1}}$,  $v_{2000}$ is the wind velocity in $2000\,\mathrm{km\,s^{-1}}$, and $n_0$ is the density of the ambient, uncompressed gas. For $\theta^1$ Ori C, the wind parameters we adopt are $\dot{M}_6 \simeq 0.4$ and $v_{2000} \simeq 1.25$ \citep{Howarth1989, Stahl1996}.

We thus have two equations ($R_S$, $\dot{R}_S$) and two unknowns ($n_0$, $t$). We denote the solution we find for the time between the onset of expansion and the present time by $t_0$. Re-arranging Eqs. \ref{Eq.R} and \ref{Eq.Rdot} gives
\begin{align}
t_0 &\simeq 0.06\, \left(\frac{R_S}{1\,\mathrm{pc}}\right)\left(\frac{10\,\mathrm{km\,s^{-1}}}{\dot{R}_S}\right)\,\mathrm{Myr} \label{Eq.t}, \\
n_0 &\simeq 3.5\cdot 10^3\, (\dot{M}_6 v_{2000}^2)\left(\frac{1\,\mathrm{pc}}{R_S}\right)^2\left(\frac{10\,\mathrm{km\,s^{-1}}}{\dot{R}_S}\right)^3\,\mathrm{cm^{-3}} \label{Eq.n0}.
\end{align}
From our observations we estimate the present-day values $R_S(t_0)\simeq 4\pm 0.5\,\mathrm{pc}$ and $v_{\mathrm{exp}}=\dot{R}_S(t_0)\simeq 13\pm 2\,\mathrm{km\,s^{-1}}$ at the far (south) side of the expanding shell. With these values, we obtain for the unknowns $n_0 \simeq 5\pm 2\cdot 10^1\,\mathrm{cm^{-3}}$ and  $t_0 \simeq 0.2\pm 0.05\,\mathrm{Myr}$.

With these values, the temperature of the hot plasma inside the bubble and its density are predicted by the model to be $T_{\mathrm{plasma}}\simeq 2.3\pm 0.2\cdot 10^6\,\mathrm{K}$ and $n_{\mathrm{plasma}}\simeq 0.2\pm 0.05\,\mathrm{cm^{-3}}$ \citep{Castor1975}. This is in excellent agreement with the X-ray observations by \cite{Guedel2008}. However, numerical models of \cite{Arthur2012}, that take into account saturation of thermal conduction at the shell interface and mass loading by proplyds in the Orion nebula, produce hotter gas with $T_{\mathrm{plasma}}\sim 10^7\,\mathrm{K}$. Yet, the mixing efficiency of the photo-evaporation flow from the proplyds with the stellar wind material is not well constrained and a major source of uncertainties.

In an alternative approach, we fit the five observables ($R_S$, $v_S$, $M_S$, $T_{\mathrm{plasma}}$, $n_{\mathrm{plasma}}$) = ($2.7\,\mathrm{pc}$, $13\,\mathrm{km\,s^{-1}}$, $1500\,M_{\sun}$, $2\cdot 10^6\,\mathrm{K}$, $0.3\,\mathrm{cm^{-3}}$) with the two parameters $t_0$ and $n_0$, given the expressions by \cite{Castor1975}. From a weighted least-square fit, with relative errors (0.2, 0.2, 0.5, 0.3, 0.5), we find the best fit for $t_0\simeq 0.24\pm 0.05\,\mathrm{Myr}$ and $n_0\simeq 1.7\pm 0.7 \cdot 10^2\,\mathrm{cm^{-3}}$. In this fit, only the last three observables are fitted in good agreement; in contrast, the shell radius is predicted to be $3.8\,\mathrm{pc}$ and the expansion velocity as $9\,\mathrm{km\,s^{-1}}$. The plasma temperature and density do not depend strongly on $t_0$ and $n_0$ \citep{Castor1975}. Adopting a lower gas mass of $600\,M_{\sun}$ results in a better fit with $t_0\simeq 0.17\pm 0.05\,\mathrm{Myr}$ and $n_0\simeq 1.0\pm 0.6 \cdot 10^2\,\mathrm{cm^{-3}}$. In this scheme, however, we rely on the spherical symmetry of the problem, which is not applicable for the Orion Nebula expanding predominantly along its steepest density gradient. Considering expansion in an ambient medium with a density gradient scaling with $R_S^{-2}$ alters the time-dependent behavior of the bubble kinematics considerably \citep[sec. VII]{OstrikerMcKee1988}.

From the morphology of the expansion in the eastern arm of the shell (cf. App. \ref{App.east-shell}, Figs. \ref{Fig.east_shell_0-5} and \ref{Fig.east_shell_0-15}), we obtain with $R_S(t) \simeq 1.75\,\mathrm{pc}$ and $\dot{R}_S(t) \simeq 4.25\,\mathrm{km\,s^{-1}}$ $n_0 \simeq 8\cdot 10^3\,\mathrm{cm^{-3}}$ and $t_0 \simeq 0.25\,\mathrm{Myr}$. Here, the ambient gas density is thus much higher.

The derived lifetime of the bubble, $t_0 \simeq 0.2\,\mathrm{Myr}$, indicates that the onset of the gas coupling to the stellar wind is rather recent, but agrees with the presumed ages of the Trapezium cluster and $\theta^1$ Ori C, which is $<1\,\mathrm{Myr}$ with a median age of the stellar population of $0.3\,\mathrm{Myr}$ \citep{Prosser1994, Hillenbrand1997}. In contrast, \cite{Simon-Diaz2006} give an age of $\theta^1$ Ori C of $2.5\pm 0.5\,\mathrm{Myr}$, derived from optical spectroscopy of the stars in the Trapezium cluster, which is much longer than the expansion time scale of the Veil bubble. We note, though, that the star has to leave the dense core where it was born before large-scale expansion can set on in a relatively dilute ambient medium. We expect this time to be of the order of $0.1\,\mathrm{Myr}$, considering the peculiar velocity of $\theta^1$ Ori C \citep{Vitrichenko2002, Stahl2008, ODell2009}. From the absence of evidence for the destruction of nearby proplyds and interaction between the moving foreground layers, \cite{ODell2009} estimate the time scale since $\theta^1$ Ori C has moved into lower-density material, starting to ionize the EON, is of a few $10^4$ years. This estimate is an order of magnitude younger than the expansion time we find. However, from modelling the disk masses of proplyds close to $\theta^1$ Ori C, \cite{Stoerzer1999} find that $\theta^1$ Ori C can be significantly older than $10^5$ years if the orbits of the proplyds are radial. \cite{Salgado2016} derive an age of $10^5$ years for $\theta^1$ Ori C from the mass of the Orion Bar.

For the density within the shell, the compressed gas, the simple model of \cite{Castor1975} gives
\begin{align}
n_s = \frac{\mu \dot{R}_S^2}{kT_s}\, n_0.
\end{align}
If the temperature in the shell is $T_s=80\,\mathrm{K}$, as assumed by \cite{Castor1975}, and $\mu/m_{\mathrm{H}}=1.3$, assuming atomic H\,{\sc i} gas, the density within the shell becomes $n_s \simeq 200 (\dot{R}_S/10\,\mathrm{km\,s^{-1}})^2\,n_0$. In the southern shell, this then yields $n_s\simeq 400\,n_0 \simeq 2\cdot 10^4\,\mathrm{cm^{-3}}$. In the eastern bright arm of the shell, we expect a shell density of $n_s\simeq 50\,n_0 \simeq 5\cdot 10^5\,\mathrm{cm^{-3}}$: here, the density is much higher but also the shell is much thinner than in the southern part. Both theoretical densities are an order of magnitude higher than what we estimated from observations.

With a mass of the expanding gas shell of $M\simeq 1500\,M_{\sun}$ and an expansion velocity of $v_{\mathrm{exp}}\simeq 13\,\mathrm{km\,s^{-1}}$, the kinetic energy in the shell is computed to be $E_{\mathrm{kin}} \simeq 2.5\cdot 10^{48}\,\mathrm{erg}$. Dividing this energy by the expansion time scale provides us with a measure that we can compare to the wind luminosity of the presumed driving star $\theta^1$ Ori C. This yields $\frac{E_{\mathrm{kin}}}{t_0} \simeq 4\cdot 10^{35}\,\mathrm{erg\,s^{-1}}$. Comparing this number with the wind luminosity, $L_w \simeq 8\cdot 10^{35} \,\mathrm{erg\,s^{-1}}$, we estimate a wind efficiency of about 50\%. Adding the wind of $\theta^2$ Ori A, which is comparable to the stellar wind of $\theta^1$ Ori C, with $L_w \simeq 2\cdot 10^{35} \,\mathrm{erg\,s^{-1}}$, reduces the wind efficiency to 40\%; the stellar wind parameters of $\theta^2$ Ori A are $\dot{M}_6 \sim 0.08$ and $v_{2000} \sim 1.45$ \citep{Petit2012}. Adding both winds in the above analysis, would increase densities by a factor $1.3$, but has no effect on the expansion time.

During the initial stages, the swept up shell will be very thin and the stellar ionizing flux will be able to fully ionize the shell. Once the shell has swept up enough material, the ionization front will become trapped in the shell. The Veil shell has already reached this later stage as evidenced by the morphology of the shell in H$\alpha$ and [C\,{\sc ii}] emission (cf. Fig. \ref{Fig.overview}) and the narrow line width of the [C\,{\sc ii}] line. In addition, we note that bright PAH emission (cf. Fig. 1 in \cite{Pabst2019}) mostly originates from the neutral gas in PDRs and is very weak if associated at all with ionized gas. For the ambient densities we estimated above, that is $n_0\simeq 2\mhyphen 5\cdot 10^1\,\mathrm{cm^{-3}}$, the time when the shell becomes neutral is $t \simeq 0.1\mhyphen 0.2\,\mathrm{Myr}$ according to the \cite{Weaver1977} model\footnote{If the density is actually higher, the ionizing radiation gets trapped earlier.}. This is in good agreement with our estimate of the expansion time.

As the [C\,{\sc ii}] observations reveal, the strong density gradient towards the front of the cloud has produced a rapid expansion of the bubble towards us. We note that the $13\,\mathrm{km\,s^{-1}}$ expansion velocity exceeds the escape velocity from OMC1 ($2\,\mathrm{km\,s^{-1}}$) and eventually the swept-up mass will be lost to the environment. In addition, the bubble will break open, releasing the hot plasma and the expanding ionized gas. Inspection of the pv diagrams (Figs. \ref{Fig.pv-diagrams-x_wolines} and \ref{Fig.pv-diagrams-y_wolines}) indicates that the Veil shell is quite thin in some directions and break out may be ``imminent''. Eventually, the shell material, the hot plasma, and the entrained ionized gas will be mixed into the interior of the Orion-Eridanus superbubble. The next supernova that will go off, one of the Orion Belt stars in $ \sim1\,\mathrm{Myr}$, will sweep up this ``loose'' material and transport it to the walls of the superbubble. The last time this happened, about $1\,\mathrm{Myr}$ ago, this created Barnard's loop.

New numerical simulations show that the effects of stellar wind on their surroundings depends strongly on the characteristics of the medium \citep{Haid2018}. Stellar winds dominate if the bubble expands into an ambient warm ionized medium (WIM, $T\sim 10^4\,\mathrm{K}$, $n\sim 0.1\,\mathrm{cm^{-3}}$). If the ambient medium is colder and neutral (CNM, $T\sim 100\,\mathrm{K}$, $n\sim 100\,\mathrm{cm^{-3}}$), ionizing radiation is the dominant driving force. The warm neutral medium (WNM, $T\sim 2\cdot 10^3\,\mathrm{K}$, $n\sim 1\,\mathrm{cm^{-3}}$) is an intermediate case. However, these models are not descriptive of the situation of $\theta^1$ Ori C. New studies need to investigate the effects of stellar winds of a star in the outskirts of a molecular cloud, an environment that is rather dilute ($n\sim 10\,\mathrm{cm^{-3}}$) and subject to density gradients. We realize that the Veil shell provides an excellent test case for models of stellar wind expansion into a surrounding medium given the large number of observational constraints, including the mass of the shell, the temperature and density of the shell and of the plasma, the velocities of the shell and the dense ionized gas in the Huygens Region.

\subsection{The expanding bubble of $\theta^2$ Ori A}

\begin{figure}[tb]
\centering
\includegraphics[width=0.375\textwidth, height=0.2625\textwidth]{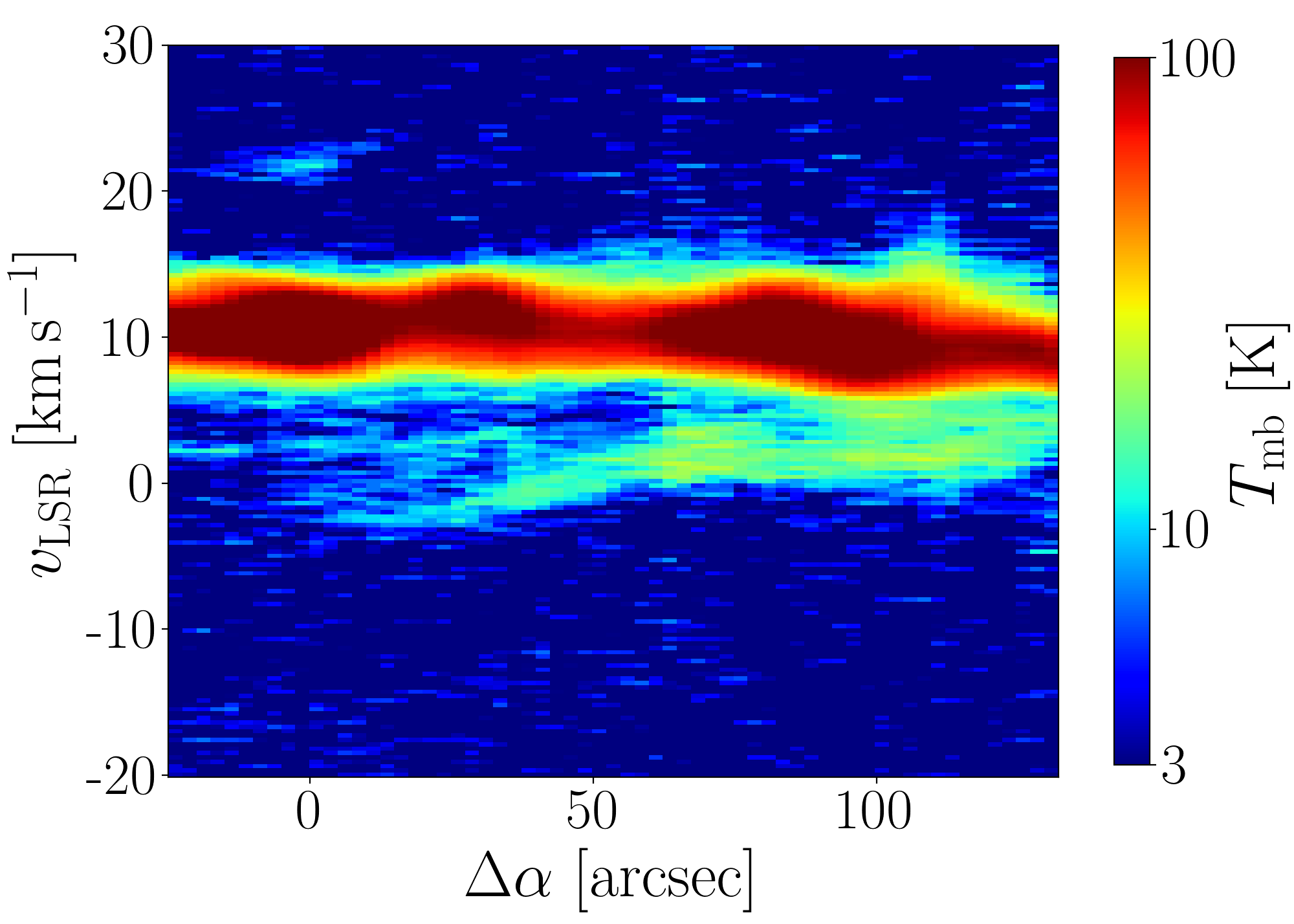}
\caption{PV diagram along position cut shown in Fig. 13 of \cite{VanderWerf2013}, perpendicular to the Orion Bar through the Huygens Region.}
\label{Fig.spectrogram_L}
\end{figure}

\cite{VanderWerf2013} discuss the shell structure that seems to surround $\theta^2$ Ori A, their component L, that can be identified in our [C\,{\sc ii}] data, as well. Figure \ref{Fig.spectrogram_L} shows the same cut for the [C\,{\sc ii}] intensity as Fig. 13 of \cite{VanderWerf2013} for the H\,{\sc i} optical depth. Component L is a rather small arc-like structure, $r\simeq 50\arcsec \simeq 0.1\,\mathrm{pc}$ extending to a LSR velocity of $v_{\mathrm{LSR}}\simeq -3.2\,\mathrm{km\,s^{-1}}$. This would translate into an expansion velocity of $v_{\mathrm{exp}}\simeq 11\,\mathrm{km\,s^{-1}}$. If this was a wind blown bubble, we would obtain from Eqs. \ref{Eq.t} and \ref{Eq.n0}, $t_0\simeq 0.006\,\mathrm{Myr}$ and $n_0\simeq 4\cdot 10^4\mathrm{cm^{-3}}$. This might be reasonable, considering that the star might only very recently have emerged from the dense material were it was born. It might also be that $\theta^2$ Ori A is the ionizing source for this structure, but not its cause. \cite{Schulz2006} estimate the stellar age to be $\sim 0.3\,\mathrm{Myr}$, similar to $\theta^1$ Ori C, but not much younger.

Also the arc-like structure D of \cite{VanderWerf2013} has been discussed as associated with $\theta^2$ Ori A \citep{Garcia-Diaz2007}, but \cite{VanderWerf2013} attribute it to $\theta^2$ Ori B, an early B-type star. However, in the H\,{\sc i} cuts component D only appears with a radius of $r\simeq 100\arcsec$, while we observe a larger structure, $r\simeq 500\arcsec \simeq 1\,\mathrm{pc}$, that has a large negative velocity ($v_{\mathrm{LSR}}\simeq -20\,\mathrm{km\,s^{-1}}$). The latter's geometric center seems to lie more to the southeast, as can be seen in our pv diagrams (cf. \ref{Fig.pv-diagrams-x_wolines}). We do not observe component D in Fig. \ref{Fig.spectrogram_L}, indicating that it is not illuminated by UV radiation and hence a foreground component. Assuming the larger [C\,{\sc ii}] structure to be the wind-blown bubble of $\theta^2$ Ori A, comparison with models would give $t_0\simeq 0.002\,\mathrm{Myr}$ and $n_0\simeq 3\cdot 10^1\mathrm{cm^{-3}}$ at the expansion velocity of $v_{\mathrm{exp}}\simeq 28\,\mathrm{km\,s^{-1}}$. Possibly the wind-blown bubble of $\theta^2$ Ori A has been disrupted by the stellar wind of $\theta^1$ Ori C and merged with the second's bubble.

\subsection{The thermal bubbles of M43 and NGC 1977}
\label{sec.comparison-M43}

The ionizing stars of M43 and NGC 1977 are of spectral types B0.5 and B1, respectively, and such stars have feeble stellar winds that are incapable of driving a bubble into their surroundings. We attribute the bubbles associated with these two nebulae to the thermal expansion of the ionized gas \citep{Spitzer1978}. We will return to the stellar wind aspects later.

\cite{Simon-Diaz2011} conclude that M43 is ionization-bounded. The Str\"{o}mgren radius of the H\,{\sc ii} region environing NU Ori with the rate of ionizing photons $\mathbb{N}\simeq 1.5\cdot 10^{47}\,\mathrm{s^{-1}}$ \citep{Simon-Diaz2011} is $\mathcal{R}_S\simeq 0.2\,\mathrm{pc}\,(10^3\,\mathrm{cm}^{-3}/n)^{2/3}$. Assuming the electron density \cite{Simon-Diaz2011} report, derived from optical line ratios, $n_e\simeq 5\cdot 10^2\,\mathrm{cm^{-3}}$, $\mathcal{R}_S\simeq 0.3\,\mathrm{pc}$, which is in excellent agreement with the radius we measure.

After the initial phase of rapid ionization, the overpressure of the ionized gas compared to its environment drives the thermal expansion of the H\,{\sc ii} region \citep{Krumholz2009, Spitzer1978}. For a homogeneous region, the present-time size and velocity of expansion ($R_S$, $\dot{R}_S$) are related to the expansion time and initial density ($t_0$, $n_0$):
\begin{align}
t_0 &\simeq 0.056 \left(\frac{R_S}{1\,\mathrm{pc}}\right)\left(\frac{\dot{R}_S}{c_s}\right)^{4/3}\left(\left(\frac{c_s}{\dot{R}_S}\right)^{7/3}-1\right)\,\mathrm{Myr},\\
n_0 &\simeq 56 \left(\frac{\mathbb{N}}{10^{47}}\right)^{1/2} \left(\frac{c_s}{\dot{R}_S}\right)^2\left(\frac{1\,\mathrm{pc}}{R_S}\right)^{3/2} \,\mathrm{cm^{-3}},
\end{align}
where $c_s \simeq 10\,\mathrm{km\,s^{-1}}$ is the sound speed in the ionized gas.

For M43, we obtain with $R_S(t_0) \simeq 0.3\pm 0.05\,\mathrm{Myr}$ and $v_{\mathrm{exp}} = \dot{R}_S(t_0) \simeq 6\pm 2\,\mathrm{km\,s^{-1}}$ an expansion time of $t_0 \simeq 0.02\pm 0.01\,\mathrm{Myr}$ and an initial density of $n_0 \simeq 1.4\pm 0.7\cdot 10^3\,\mathrm{cm^{-3}}$. This implies that the star NU Ori is rather young. The expected main-sequence lifetime of a $\sim 20\,M_{\sun}$ star like NU Ori is $\sim 10\,\mathrm{Myr}$.

From the H$\alpha$ emission in NGC 1977, we derive an electron density of $n_e\simeq 35 \,\mathrm{cm^{-3}}$ throughout the region, assuming a radius of $r\simeq 1\,\mathrm{pc}$. With the rate of ionizing photons $\mathbb{N}\simeq 1\cdot 10^{45}\,\mathrm{s^{-1}}$ \citep{Kim2016, Diaz-Miller1998}, the Str\"omgren radius of the H\,{\sc ii} region surrounding 42 Orionis is $\mathcal{R}_S\simeq 0.3\,\mathrm{pc}$. This is smaller than the size of the shell we observe. However, \cite{Hohle2010} give an effective temperature of $T_{\mathrm{eff}}=25400\,\mathrm{K}$ for 42 Orionis. Also from its spectral class, B1V, we expect a much higher flux of ionizing photons, $\mathbb{N}\sim 1\cdot 10^{47}\,\mathrm{s^{-1}}$ \citep{Sternberg2003}. We can estimate the rate of ionizing photons emitted by 42 Orionis (with minor assistance from other stars) from the H$\alpha$ emission in the H\,{\sc ii} region. We have:
\begin{align}
L(\mathrm{H}\alpha) &= \int 4\pi j_{\mathrm{H}\alpha} \,\mathrm{d}V = \int \frac{4\pi j_{\mathrm{H}\alpha}}{n_{\mathrm{p}}n_{\mathrm{e}}}n_{\mathrm{p}}n_{\mathrm{e}} \,\mathrm{d}V, \\
\mathbb{N} &= \int \alpha_{\mathrm{B}}n_{\mathrm{p}}n_{\mathrm{e}}\,\mathrm{d}V.
\end{align}
With $\frac{4\pi j_{\mathrm{H}\alpha}}{n_{\mathrm{p}}n_{\mathrm{e}}}=1.24\cdot 10^{-25}\,\mathrm{erg\,cm^3\,s^{-1}}$, $j_{\mathrm{H}\alpha}/j_{\mathrm{H}\beta}=2.85$ \citep{HummerStorey1987}, and $\alpha_{\mathrm{B}} = 2.6\cdot 10^{-13}\,\mathrm{cm^3\,s^{-1}}$ \citep{StoreyHummer1995} we obtain from the H$\alpha$ luminsity integrated over NGC 1977, $L(\mathrm{H}\alpha) \simeq 45\,L_{\sun}$, a photon rate of $\mathbb{N}\simeq 1\cdot 10^{47}\,\mathrm{s^{-1}}$. With this value, we derive a present-day Str\"omgren radius of $\mathcal{R}_S\simeq 1.5\,\mathrm{pc}$, which is in good agreement with the observed radius of the shell ($r=1.0\mhyphen 1.6\,\mathrm{pc}$). With the extent of the expanding shell derived from the pv diagram in Fig. \ref{Fig.pv-NGC1977}, $R_S(t_0)\simeq 1\,\mathrm{pc}\pm 0.2$, and the expansion velocity $v_{\mathrm{exp}} = \dot{R}_S(t_0) \simeq 1.5\pm 0.5\,\mathrm{km\,s^{-1}}$ the expansion time is $t_0 \simeq 0.4\pm 0.2\,\mathrm{Myr}$ and the initial density $n_0\simeq 2.5\pm 1.2\cdot 10^3\,\mathrm{cm^{-3}}$. This, as well, suggests that 42 Orionis is rather young, compared to its expected lifetime of $\sim 10\,\mathrm{Myr}$.

\subsection{Stellar wind versus thermal expansion of bubbles}
\label{Sec.wind_vs_hii}
In the previous sections, we analyze the kinematics of the bubbles around M42, M43, and NGC 1977 in terms of stellar-wind driven and thermal-pressure driven expansions, respectively. Here, we compare these two analyses. In Tables \ref{Tab.1} and \ref{Tab.2}, we summarize our findings for the three regions.

\begin{table*}[tb]
\centering
\addtolength{\tabcolsep}{-4pt}
\begin{tabular}{lcccccccccc}
region & star & stellar type & $T_{\mathrm{eff}}\,[\mathrm{K}]$ & $L_{\star}\,[L_{\sun}]$ & $L_{\mathrm{w}}\,\mathrm{[erg\,s^{-1}]}$ & $M_{\mathrm{shell}}\,[M_{\sun}]$ & $v_{\mathrm{exp}}\,\mathrm{[km\,s^{-1}]}$ & $E_{\mathrm{kin}}\,\mathrm{[erg]}$ & $t_0\,\mathrm{[Myr]}$ & $E_{\mathrm{kin}}/(L_{\mathrm{w}}t_0)$ \\\hline
M42 & $\theta^1$ Ori C & O7V & $3.9\cdot 10^4$ & $2.0\cdot 10^5$ & $8\cdot 10^{35}$ & 1500 & 13 & $2.5\cdot 10^{48}$  & 0.2 & 0.5 \\
M43 & NU Ori & B0.5V & $3.1\cdot 10^4$& $2.6\cdot 10^4$ & $\sim 3\cdot 10^{31}$ & 7 & 6 & $3\cdot 10^{45}$ & 0.02 & 50 \\
NGC 1977 & 42 Ori & B1V & $2.5\cdot 10^4$ & $1.1\cdot 10^4$ & $\sim 3\cdot 10^{31}$ & 700 & 1.5 & $2\cdot 10^{46}$ & 0.4 & 40 \\
\end{tabular}
\caption{Comparison of stellar parameters with bubble energetics of the three regions. In the last column, we take for $t_0$ the value derived from the stellar wind models; in case of M43 the lifetime derived from pressure-driven expansion is a third of that value, increasing the ratio $E_{\mathrm{kin}}/(L_{\mathrm{w}}t_0)$ by a factor of three. Stellar parameters of $\theta^1$ Ori C are from \cite{Simon-Diaz2006}, of NU Ori from \cite{Simon-Diaz2011}, and of 42 Orionis from \cite{Hohle2010}.}
\label{Tab.1}
\end{table*}

\begin{table*}[tb]
\centering
\begin{tabular}{lcccc}
 & M42 (Veil shell) & M43 & NGC 1977 & reference \\\hline
$\mathbb{N}_{\mathrm{Lyc}}\;[10^{47}\,\mathrm{s^{-1}}]$ & 70 & 1.5 & 1 & 1, 2, 8 \\
$L_{\mathrm{w}}\;[L_{\sun}]$ & 400 & $\sim 1.5\cdot 10^{-2}$ & $\sim 1.5\cdot 10^{-2}$ & 3, 4 \\
mass of neutral gas $[M_{\sun}]$ & 1500 & 7 & 700 & 5, 8\\
mass of ionized gas $[M_{\sun}]$ & 24 & 0.3 & 16 & 6, 8 \\
$E_{\mathrm{kin}}$ of neutral gas $[10^{46}\,\mathrm{erg}]$ & 250 & 0.3 & 2 & 5, 8\\
$E_{\mathrm{kin}}$ of ionized gas $[10^{46}\,\mathrm{erg}]$ & 6 & -- & -- & 6, 7 \\
$E_{\mathrm{th}}$ of ionized gas $[10^{46}\,\mathrm{erg}]$ & 3 & 0.7 & 5 & 5, 8\\
$E_{\mathrm{th}}$ of hot gas $[10^{46}\,\mathrm{erg}]$ & 10 & -- & -- & 3 \\
$L_{\mathrm{FIR}}\;[L_{\sun}]$ & $3.2\cdot 10^4$ & $8.5\cdot 10^3$ & $1.5\cdot 10^4$ & 8 \\
$L_{\mathrm{[C\,\textsc{ii}]}}\;[L_{\sun}]$ & $170$ & $24$ & $140$ & 8\\
\end{tabular}
\caption{Masses, energetics, and luminosities in the expanding shells of M42 (Veil shell), M43, and NGC 1977. There are no X-ray observations of the hot gas in M43 and NGC 1977. References: (1) \cite{ODell2017}, (2) \cite{Simon-Diaz2011}, (3) \cite{Guedel2008}, (4) \cite{Oskinova2011}, (5) \cite{Pabst2019}, (6) \cite{Wilson1997}, (7) \cite{ODell2001}, (8) this work.}
\label{Tab.2}
\end{table*}

\begin{figure}[tb]
\includegraphics[width=0.5\textwidth, height=0.33\textwidth]{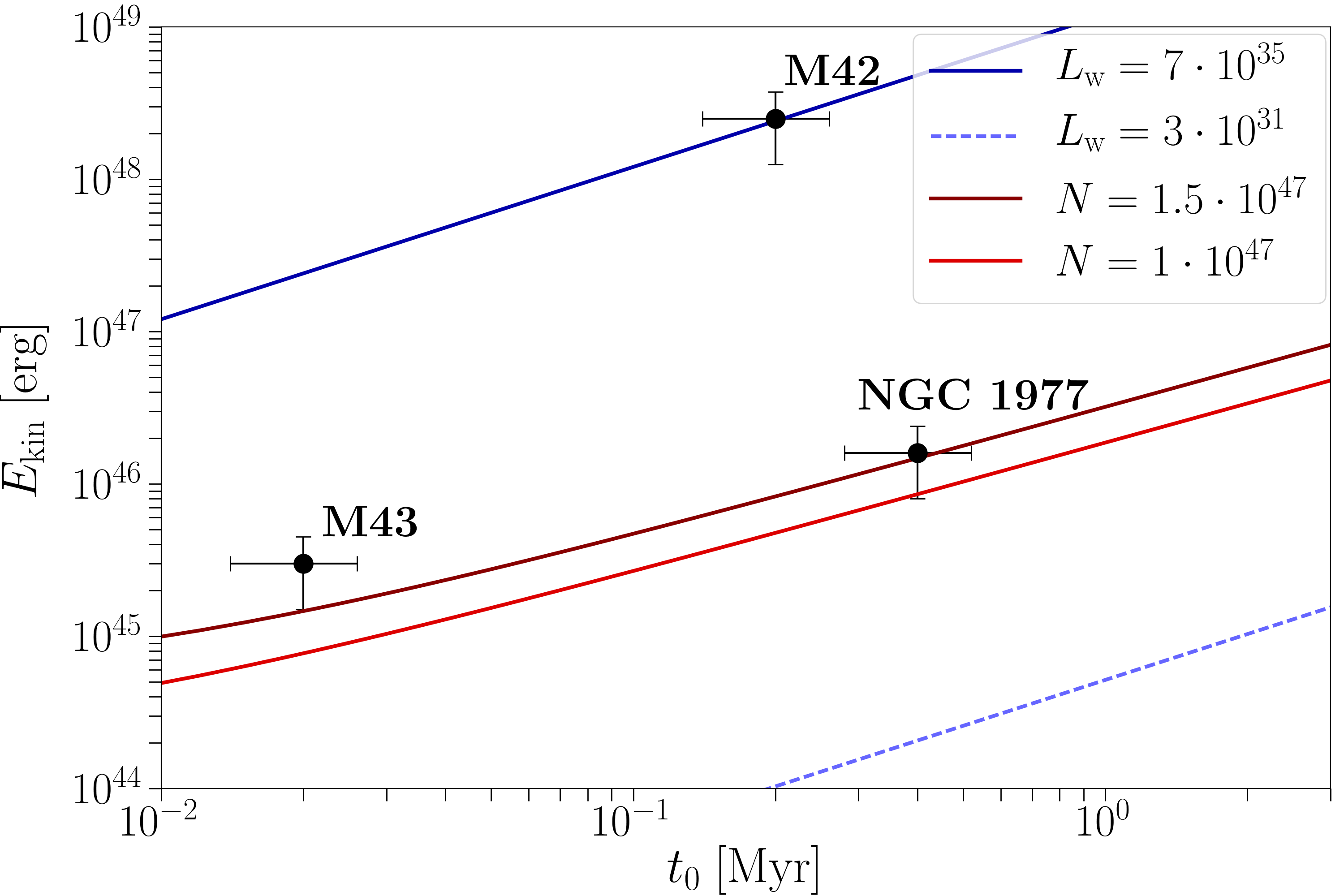}
\caption{Kinetic energy of expanding bubble shells versus expansion time. The lines are the predictions of wind models (blue) and models of pressure-driven expansion (red) with the parameters compiled in Tables \ref{Tab.1} and \ref{Tab.2}.}
\label{Fig.Ekin-t}
\end{figure}

\begin{figure}[tb]
\includegraphics[width=0.5\textwidth, height=0.33\textwidth]{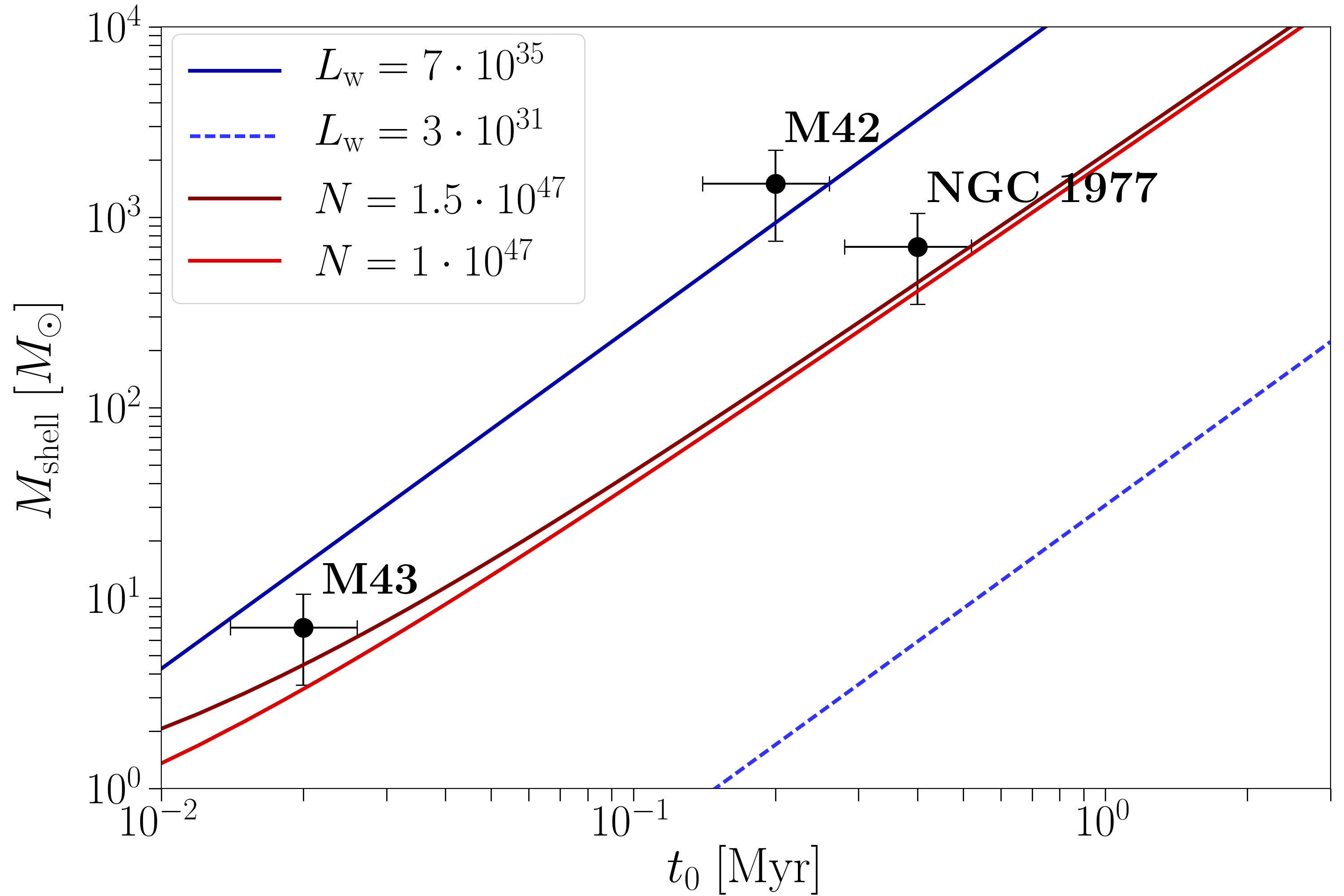}
\caption{Masses of expanding bubble shells versus expansion time. The lines are the predictions of wind models (blue) and models of pressure-driven expansion (red) with the parameters compiled in Tables \ref{Tab.1} and \ref{Tab.2}.}
\label{Fig.M-t}
\end{figure}

For M42, the momentum and energy in the ionized gas in the champagne-flow model is about two orders of magnitude less than observed for the Veil shell: $22\,M_{\sun}$ of ionized gas moving at $17\,\mathrm{km\,s^{-1}}$ \citep{ODell2001} versus a (half)shell of $1500\,M_{\sun}$ expanding at $13\,\mathrm{km\,s^{-1}}$ (Table \ref{Tab.2}), the neutral Veil shell carrying a total momentum of $2\cdot 10^4\,M_{\sun}\,\mathrm{km\,s^{-1}}$. For M43 and NGC 1977, on the other hand, the stellar wind is too feeble to drive the expansion of a large shell and the results are in good agreement with analytical models for the thermal expansion of H\,{\sc ii} regions. Figures \ref{Fig.Ekin-t} and \ref{Fig.M-t}  summarize this graphically. In this depiction, we have calculated the expected kinetic energy of a stellar wind bubble using the analytical model developed by \cite{Weaver1977},
\begin{align}
E_{\mathrm{kin}} \simeq \frac{6}{11}L_{\mathrm{w}}t,
\end{align}
and the pressure-driven expansion of ionized gas \citep{Spitzer1978},
\begin{align}
M_{\mathrm{shell}} &\simeq \frac{4\pi}{3}R_{S,0}^3 \mu n_0\left(1+\frac{7t}{4t_s}\right)^{12/7},\\%\left(1-\left(1+\frac{7t}{4t_s}\right)^{-6/7}\right),\\
E_{\mathrm{kin}} &\simeq \frac{1}{2}M_{\mathrm{shell}} c_s^2 \left(1+\frac{7t}{4t_s}\right)^{-6/7},
\end{align}
where $R_{S,0} = \left(\frac{3\mathbb{N}}{4\pi n_0^2 \alpha_{\mathrm{B}}}\right)^{1/3}$ is the initial Str\"{o}mgren radius and $t_s = R_{S,0}/c_s$. Comparing the predicted shell masses, where we assume that all the swept-up material is in the shell, to the measured masses as shown in Fig. \ref{Fig.M-t}, we observe that the measured masses of all three shells are slightly above the predictions, by a factor of about two. The mass discrepancy in M42 and NGC 1977 might be due to our choice of $\beta$ in the SEDs and a contamination from background material that is not expanding. The mass of M43, estimated from the gas density in the shell, is a rather rough estimate as well.

For the Veil shell, the kinetic energy is comparable to the integrated wind luminosity of a wind-blown bubble. In M43 and NGC 1977, the observed kinetic energy of the shells much exceeds the integrated wind luminosity. For M43 and NGC 1977, the kinetics of the shell are in good agreement with the thermal expansion of the H\,{\sc ii} region, while for M42 this falls short by two orders of magnitude.

From models of O stars and comparison with real O stars \citep{Pauldrach2001}, we obtain an upper limit of $L_w \simeq 1\cdot 10^{34} \,\mathrm{erg\,s^{-1}}$. In their models of earlier B stars, however, even a B0.5V star has a significant stellar wind with $L_w \simeq 1\cdot 10^{33}\,\mathrm{erg\,s^{-1}}$ \citep{Sternberg2003}. Since no X-ray observations are available for M43 and NGC 1977, that might reveal gas shock-heated from stellar winds in the bubble interior as in M42, we cannot judge conclusively on this matter.

\section{Conclusion}
We have observed the [C\,{\sc ii}] emission from the Orion Nebula M42, M43, and NGC 1973, 1975, and 1977. Velocity-resolved observations of the [C\,{\sc ii}] fine-structure line provide a unique tool to accurately quantify the kinematics of expanding bubbles in the ISM, in our case the large bubble associated with the Orion Nebula, the Orion Veil, and the bubbles of M43 and NGC 1973, 1975, and 1977. Our extended map of [C\,{\sc ii}] emission can be further used to disentangle dynamic structures within the same region on larger scales than previously possible. From the consistent arc structures in the pv diagrams and the line profiles, we are able to show that the gas is indeed expanding coherently on large scales as a bubble is blown into the ISM by stellar feedback.

We have compared three regions within the Orion Nebula complex that host different central stars causing the disruption of the molecular cloud cores in which these stars are born. The Orion Nebula M42 hosts the most massive star, an O-type star, which has a strong wind, that drives the expansion of the surrounding medium. M43 and NGC 1977 host less massive stars of early B type with much weaker winds. Still, both are effective in blowing bubbles of significant extent.

\cite{Pabst2019} have demonstrated that the Veil shell is a coherent, half-spherical structure, enclosing the layers of ionized gas observed towards that region. The stellar wind from $\theta^1$ Ori C seems to be very effective in sweeping up its surrounding gas into a large-scale dense shell, leaving the bubble interior filled with dilute, hot X-ray emitting gas. This precise structure, a thin, rather cold shell of swept-up gas enveloping a hot interior medium, is characteristic of wind-blown bubbles. The bubble of the Orion Nebula, though still closed, may be on the brink of entering the bursting phase, thus rejuvenating the medium of the Orion-Eridanus superbubble.

The bubbles associated with M43 and NGC 1977 also exhibit spherical symmetry and possess a rather thin shell. In these two cases, however, the masses and kinetics of these shells are incompatible with an origin in a wind-driven shell. Instead, these shells are due to the pressure-driven expansion of the ionized gas and the observations are in good agreement with simple analytical estimates. All structures we have investigated here are rather young compared to the total lifetime of the driving (massive) stars. As stars in the Orion Nebula cluster will go off as supernovae, these structures will be disrupted as they age.

\vspace{1cm}

\begin{acknowledgements}
This work is based on observations made with the NASA/DLR Stratospheric Observatory for Infrared Astronomy (SOFIA). SOFIA is jointly operated by the Universities Space Research Association, Inc. (USRA), under NASA contract NNA17BF53C, and the Deutsches SOFIA Institut (DSI) under DLR contract 50 OK 0901 to the University of Stuttgart.

JRG thanks the ERC and the Spanish MCIU for funding support under grants ERC-2013-Syg-610256-NANOCOSMOS and AYA2017-85111-P, respectively. Research on the interstellar medium at Leiden Observatory is supported through a Spinoza award of the Dutch Science Organisation (NWO). We are grateful for the dedicated work during the upGREAT square-degree survey of Orion of the USRA and NASA staff of the Armstrong Flight Research Center in Palmdale and of the Ames Research Center in Mountain View, and the DSI. We thank C.R. O'Dell for commenting on an earlier version of the manuscript and helpful discussions. We thank the anonymous referee for comments that helped improve the manuscript.
\end{acknowledgements}

\bibliographystyle{aa} % style aa.bst
\bibliography{article2} % your references Yourfile.bib

\begin{thebibliography}{112}
\expandafter\ifx\csname natexlab\endcsname\relax\def\natexlab#1{#1}\fi

\bibitem[{{Abel} {et~al.}(2019){Abel}, {Ferland}, \& {O'Dell}}]{Abel2019}
{Abel}, N.~P., {Ferland}, G.~J., \& {O'Dell}, C.~R. 2019, \apj, 881, 130

\bibitem[{{Abel} {et~al.}(2016){Abel}, {Ferland}, {O'Dell}, \&
  {Troland}}]{Abel2016}
{Abel}, N.~P., {Ferland}, G.~J., {O'Dell}, C.~R., \& {Troland}, T.~H. 2016,
  \apj, 819, 136

\bibitem[{{Allers} {et~al.}(2005){Allers}, {Jaffe}, {Lacy}, {Draine}, \&
  {Richter}}]{Allers2005}
{Allers}, K.~N., {Jaffe}, D.~T., {Lacy}, J.~H., {Draine}, B.~T., \& {Richter},
  M.~J. 2005, \apj, 630, 368

\bibitem[{{Arthur}(2012)}]{Arthur2012}
{Arthur}, S.~J. 2012, MNRAS, 421, 1283

\bibitem[{{Bedijn} \& {Tenorio-Tagle}(1981)}]{BedijnTenorio-Tagle1981}
{Bedijn}, P.~J. \& {Tenorio-Tagle}, G. 1981, \aap, 98, 85

\bibitem[{{Bernard-Salas} {et~al.}(2012){Bernard-Salas}, {Habart}, {Arab},
  {Abergel}, {Dartois}, {Martin}, {Bontemps}, {Joblin}, {White}, {Bernard}, \&
  {Naylor}}]{Bernard-Salas2012}
{Bernard-Salas}, J., {Habart}, E., {Arab}, H., {et~al.} 2012, \aap, 538, A37

\bibitem[{{Bern{\'e}} {et~al.}(2010){Bern{\'e}}, {Marcelino}, \&
  {Cernicharo}}]{Berne2010}
{Bern{\'e}}, O., {Marcelino}, N., \& {Cernicharo}, J. 2010, \nat, 466, 947

\bibitem[{{Bern\'{e}} {et~al.}(2014){Bern\'{e}}, {Marcelino}, \&
  {Cernicharo}}]{Berne2014}
{Bern\'{e}}, O., {Marcelino}, N., \& {Cernicharo}, N. 2014, ApJ, 795, 13

\bibitem[{{Bern{\'e}} \& {Matsumoto}(2012)}]{Berne2012}
{Bern{\'e}}, O. \& {Matsumoto}, Y. 2012, \apjl, 761, L4

\bibitem[{{Bohlin} {et~al.}(1978){Bohlin}, {Savage}, \& {Drake}}]{Bohlin1978}
{Bohlin}, R.~C., {Savage}, B.~D., \& {Drake}, J.~F. 1978, \apj, 224, 132

\bibitem[{{Boreiko} \& {Betz}(1996)}]{BoreikoBetz1996}
{Boreiko}, R.~T. \& {Betz}, A.~L. 1996, \apjl, 467, L113

\bibitem[{{Bron} {et~al.}(2018){Bron}, {Ag{\'u}ndez}, {Goicoechea}, \&
  {Cernicharo}}]{Bron2018}
{Bron}, E., {Ag{\'u}ndez}, M., {Goicoechea}, J.~R., \& {Cernicharo}, J. 2018,
  arXiv e-prints, arXiv:1801.01547

\bibitem[{{Carlyle} \& {Hillier}(2017)}]{Carlyle2017}
{Carlyle}, J. \& {Hillier}, A. 2017, \aap, 605, A101

\bibitem[{{Castor} {et~al.}(1975){Castor}, {McCray}, \& {Weaver}}]{Castor1975}
{Castor}, J., {McCray}, R., \& {Weaver}, R. 1975, ApJ, 200, L107

\bibitem[{{Chevalier} {et~al.}(1992){Chevalier}, {Blond\'{i}n}, \&
  {Emmering}}]{Chevalier1992}
{Chevalier}, R.~A., {Blond\'{i}n}, J.~M., \& {Emmering}, R.~T. 1992, ApJ, 392,
  118

\bibitem[{{Churchwell} {et~al.}(2006){Churchwell}, {Povich}, {Allen}, {Taylor},
  {Meade}, \& {et al.}}]{Churchwell2006}
{Churchwell}, E., {Povich}, M.~S., {Allen}, D., {et~al.} 2006, ApJ, 649, 759

\bibitem[{{Chuss} {et~al.}(2019){Chuss}, {Andersson}, {Bally}, {Dotson},
  {Dowell}, {Guerra}, {Harper}, {Houde}, {Jones}, {Lazarian}, {Lopez
  Rodriguez}, {Michail}, {Morris}, {Novak}, {Siah}, {Staguhn}, {Vaillancourt},
  {Volpert}, {Werner}, {Wollack}, {Benford}, {Berthoud}, {Cox}, {Crutcher},
  {Dale}, {Fissel}, {Goldsmith}, {Hamilton}, {Hanany}, {Henning}, {Looney},
  {Moseley}, {Santos}, {Stephens}, {Tassis}, {Trinh}, {Van Camp},
  {Ward-Thompson}, \& {HAWC + Science Team}}]{Chuss2019}
{Chuss}, D.~T., {Andersson}, B.-G., {Bally}, J., {et~al.} 2019, \apj, 872, 187

\bibitem[{{Cuadrado} {et~al.}(2019){Cuadrado}, {Salas}, {Goicoechea},
  {Cernicharo}, {Tielens}, \& {B{\'a}ez-Rubio}}]{Cuadrado2019}
{Cuadrado}, S., {Salas}, P., {Goicoechea}, J.~R., {et~al.} 2019, \aap, 625, L3

\bibitem[{{Da Rio} {et~al.}(2009){Da Rio}, {Robberto}, {Soderblom}, {Panagia},
  {Hillenbrand}, {Palla}, \& {Stassun}}]{DaRio2009}
{Da Rio}, N., {Robberto}, M., {Soderblom}, D.~R., {et~al.} 2009, ApJ SS, 183,
  261

\bibitem[{{Da Rio} {et~al.}(2014){Da Rio}, {Tan}, \& {Jaehing}}]{DaRio2014}
{Da Rio}, N., {Tan}, J.~C., \& {Jaehing}, K. 2014, ApJ, 795, 55

\bibitem[{{Dale} {et~al.}(2013){Dale}, {Ngoumou}, {Ercolano}, \&
  {Bonnell}}]{Dale2013}
{Dale}, J.~E., {Ngoumou}, J., {Ercolano}, B., \& {Bonnell}, I. 2013, MNRAS,
  436, 3430

\bibitem[{{Diaz-Miller} {et~al.}(1998){Diaz-Miller}, {Franco}, \&
  {Shore}}]{Diaz-Miller1998}
{Diaz-Miller}, I.~R., {Franco}, J., \& {Shore}, S.~N. 1998, ApJ, 501, 192

\bibitem[{{Duffell}(2016)}]{Duffell2016}
{Duffell}, P.~C. 2016, ApJ, 821, 76

\bibitem[{{Freyer} {et~al.}(2006){Freyer}, {Hensler}, \& {Yorke}}]{Freyer2006}
{Freyer}, T., {Hensler}, G., \& {Yorke}, H. 2006, ApJ, 638, 262

\bibitem[{{Garc{\'\i}a-D{\'\i}az} \& {Henney}(2007)}]{Garcia-Diaz2007}
{Garc{\'\i}a-D{\'\i}az}, M.~T. \& {Henney}, W.~J. 2007, \aj, 133, 952

\bibitem[{{Garc{\'\i}a-D{\'\i}az} {et~al.}(2008){Garc{\'\i}a-D{\'\i}az},
  {Henney}, {L{\'o}pez}, \& {Doi}}]{Garcia-Diaz2008}
{Garc{\'\i}a-D{\'\i}az}, M.~T., {Henney}, W.~J., {L{\'o}pez}, J.~A., \& {Doi},
  T. 2008, \rmxaa, 44, 181

\bibitem[{{Goicoechea} {et~al.}(2020){Goicoechea}, {Pabst}, {Kabanovic},
  {Santa-Maria}, {Marcelino}, {Tielens}, {Hacar}, {Berne}, {Buchbender},
  {Cuadrado}, {Higgins}, {Kramer}, {Stutzki}, {Suri}, {Teyssier}, \&
  {Wolfire}}]{Goicoechea2020}
{Goicoechea}, J.~R., {Pabst}, C.~H.~M., {Kabanovic}, S., {et~al.} 2020, arXiv
  e-prints, arXiv:2004.12938

\bibitem[{{Goicoechea} {et~al.}(2016){Goicoechea}, {Pety}, {Cuadrado},
  {Cernicharo}, {Chapillon}, {Fuente}, {Gerin}, {Joblin}, {Marcelino}, \&
  {Pilleri}}]{Goicoechea2016}
{Goicoechea}, J.~R., {Pety}, J., {Cuadrado}, S., {et~al.} 2016, \nat, 537, 207

\bibitem[{{Goicoechea} {et~al.}(2015){Goicoechea}, {Teyssier}, {Etxaluze},
  {Goldsmith}, \& {Ossenkopf}}]{Goicoechea2015}
{Goicoechea}, J.~R., {Teyssier}, D., {Etxaluze}, M., {Goldsmith}, P.~F., \&
  {Ossenkopf}, V. 2015, ApJ, 812, 75

\bibitem[{{Goldsmith} {et~al.}(2012){Goldsmith}, {Langer}, {Pineda}, \&
  {Velusamy}}]{Goldsmith2012}
{Goldsmith}, P.~F., {Langer}, W.~D., {Pineda}, J.~L., \& {Velusamy}, T. 2012,
  ApJ Supplement Series, 203, 13

\bibitem[{{Griffin} {et~al.}(2010){Griffin}, {Abergel}, {Abreu}, {Ade},
  {Andr{\'e}}, {Augueres}, {Babbedge}, {Bae}, {Baillie}, {Baluteau}, \& {et
  al}}]{Griffin2010}
{Griffin}, M.~J., {Abergel}, A., {Abreu}, A., {et~al.} 2010, A\&A, 518, L3

\bibitem[{{Gro{\ss}schedl} {et~al.}(2018){Gro{\ss}schedl}, {Alves}, {Meingast},
  {Ackerl}, {Ascenso}, {Bouy}, {Burkert}, {Forbrich}, {F{\"u}rnkranz},
  {Goodman}, {Hacar}, {Herbst-Kiss}, {Lada}, {Larreina}, {Leschinski},
  {Lombardi}, {Moitinho}, {Mortimer}, \& {Zari}}]{Grossschedl2018}
{Gro{\ss}schedl}, J.~E., {Alves}, J., {Meingast}, S., {et~al.} 2018, \aap, 619,
  A106

\bibitem[{{Gro{\ss}schedl} {et~al.}(2019){Gro{\ss}schedl}, {Alves}, {Teixeira},
  {Bouy}, {Forbrich}, {Lada}, {Meingast}, {Hacar}, {Ascenso}, {Ackerl},
  {Hasenberger}, {K{\"o}hler}, {Kubiak}, {Larreina}, {Linhardt}, {Lombardi}, \&
  {M{\"o}ller}}]{Grossschedl2019}
{Gro{\ss}schedl}, J.~E., {Alves}, J., {Teixeira}, P.~S., {et~al.} 2019, \aap,
  622, A149

\bibitem[{{G\"{u}del} {et~al.}(2008){G\"{u}del}, {Briggs}, {Montmerle},
  {Audard}, {Rebull}, \& {Skinner}}]{Guedel2008}
{G\"{u}del}, M., {Briggs}, K.~R., {Montmerle}, T., {et~al.} 2008, Science, 319,
  309

\bibitem[{{Haid} {et~al.}(2018){Haid}, {Walch}, {Seifried}, {W\"{u}nsch},
  {Dinnbier}, \& {Naab}}]{Haid2018}
{Haid}, S., {Walch}, S., {Seifried}, D., {et~al.} 2018, MNRAS, 478, 4799

\bibitem[{{Hillenbrand}(1997)}]{Hillenbrand1997}
{Hillenbrand}, L.~A. 1997, ApJ, 113, 1733

\bibitem[{{Hohle} {et~al.}(2010){Hohle}, {Neuhauser}, \& {Schutz}}]{Hohle2010}
{Hohle}, M.~M., {Neuhauser}, R., \& {Schutz}, B.~F. 2010, Astron. Nach., 331,
  349

\bibitem[{{Hollenbach} {et~al.}(1991){Hollenbach}, {Takahashi}, \&
  {Tielens}}]{HollenbachTT1991}
{Hollenbach}, D.~J., {Takahashi}, T., \& {Tielens}, A.~G.~G.~M. 1991, ApJ, 377,
  192

\bibitem[{{Hollenbach} \& {Tielens}(1999)}]{HollenbachTielens1999}
{Hollenbach}, D.~J. \& {Tielens}, A.~G.~G.~M. 1999, RvMP, 71, No. 1

\bibitem[{{Hopkins} {et~al.}(2012){Hopkins}, {Quataert}, \&
  {Murray}}]{Hopkins2012}
{Hopkins}, P.~F., {Quataert}, E., \& {Murray}, N. 2012, \mnras, 421, 3488

\bibitem[{{Howarth} \& {Prinja}(1989)}]{Howarth1989}
{Howarth}, I.~D. \& {Prinja}, R.~K. 1989, ApJS, 69, 527

\bibitem[{{Hummer} \& {Storey}(1987)}]{HummerStorey1987}
{Hummer}, D.~G. \& {Storey}, P.~J. 1987, MNRAS, 224, 801

\bibitem[{{Joblin} {et~al.}(2018){Joblin}, {Bron}, {Pinto}, {Pilleri}, {Le
  Petit}, {Gerin}, {Le Bourlot}, {Fuente}, {Berne}, {Goicoechea}, {Habart},
  {K{\"o}hler}, {Teyssier}, {Nagy}, {Montillaud}, {Vastel}, {Cernicharo},
  {R{\"o}llig}, {Ossenkopf-Okada}, \& {Bergin}}]{Joblin2018}
{Joblin}, C., {Bron}, E., {Pinto}, C., {et~al.} 2018, \aap, 615, A129

\bibitem[{{Kaufman} {et~al.}(2006){Kaufman}, {Wolfire}, \&
  {Hollenbach}}]{Kaufman2006}
{Kaufman}, M.~J., {Wolfire}, M.~G., \& {Hollenbach}, D.~J. 2006, \apj, 644, 283

\bibitem[{{Kim} {et~al.}(2016){Kim}, {Clarke}, {Fang}, \& {Facchini}}]{Kim2016}
{Kim}, J.~S., {Clarke}, C.~J., {Fang}, M., \& {Facchini}, S. 2016, ApJL, 826,
  L15

\bibitem[{{Knyazeva} \& {Kharitonov}(1998)}]{Knyazeva1998}
{Knyazeva}, L.~N. \& {Kharitonov}, A.~V. 1998, ARep, 42, 60

\bibitem[{{Kounkel} {et~al.}(2017){Kounkel}, {Hartmann}, {Loinard},
  {Ortiz-Le{\'o}n}, {Mioduszewski}, {Rodr{\'\i}guez}, {Dzib}, {Torres}, {Pech},
  {Galli}, {Rivera}, {Boden}, {Evans}, {Brice{\~n}o}, \& {Tobin}}]{Kounkel2017}
{Kounkel}, M., {Hartmann}, L., {Loinard}, L., {et~al.} 2017, \apj, 834, 142

\bibitem[{{Kraus} {et~al.}(2009){Kraus}, {Weigelt}, {Balega}, {Docobo},
  {Hofmann}, {Preibisch}, {Schertl}, {Tamazian}, {Driebe}, {Ohnaka}, {Petrov},
  {Sch{\"o}ller}, \& {Smith}}]{Kraus2009}
{Kraus}, S., {Weigelt}, G., {Balega}, Y.~Y., {et~al.} 2009, A\&A, 497, 195

\bibitem[{{Krumholz} \& {Matzner}(2009)}]{Krumholz2009}
{Krumholz}, M.~R. \& {Matzner}, C.~D. 2009, ApJ, 703, 1352

\bibitem[{{Langer} \& {Penzias}(1990)}]{Langer1990}
{Langer}, W.~D. \& {Penzias}, A.~A. 1990, ApJ, 357, 477

\bibitem[{{Lehmann} {et~al.}(2010){Lehmann}, {Vitrichenko}, {Bychkov},
  {Bychkova}, \& {Klochkova}}]{Lehmann2010}
{Lehmann}, H., {Vitrichenko}, E., {Bychkov}, V., {Bychkova}, L., \&
  {Klochkova}, V. 2010, A\&A, 514, A34

\bibitem[{{Li} \& {Draine}(2001)}]{Draine2001}
{Li}, A. \& {Draine}, B.~T. 2001, ApJ, 544, 778

\bibitem[{{Lombardi} {et~al.}(2014){Lombardi}, {Bouy}, {Alves}, \&
  {Lada}}]{Lombardi2014}
{Lombardi}, M., {Bouy}, H., {Alves}, J., \& {Lada}, C.~J. 2014, A\&A, 566, A45

\bibitem[{{Mac Low} \& {McCray}(1988)}]{MacLowMcCray1988}
{Mac Low}, M.-M. \& {McCray}, R. 1988, \apj, 324, 776

\bibitem[{{Marconi} {et~al.}(1998){Marconi}, {Testi}, {Natta}, \&
  {Walmsley}}]{Marconi1998}
{Marconi}, A., {Testi}, L., {Natta}, A., \& {Walmsley}, C.~M. 1998, A\&A, 330,
  696

\bibitem[{{McCray} \& {Kafatos}(1987)}]{McCray1987}
{McCray}, R. \& {Kafatos}, M. 1987, \apj, 317, 190

\bibitem[{{McKee} \& {Ostriker}(1977)}]{McKeeOstriker1977}
{McKee}, C.~F. \& {Ostriker}, J.~P. 1977, \apj, 218, 148

\bibitem[{{Megeath} {et~al.}(2012){Megeath}, {Gutermuth}, {Muzerolle},
  {Kryukova}, {Flaherty}, {Hora}, {Allen}, {Hartmann}, {Myers}, {Pipher},
  {Stauffer}, {Young}, \& {Fazio}}]{Megeath2012}
{Megeath}, S.~T., {Gutermuth}, R., {Muzerolle}, J., {et~al.} 2012, \aj, 144,
  192

\bibitem[{{Megeath} {et~al.}(2016){Megeath}, {Gutermuth}, {Muzerolle},
  {Kryukova}, {Hora}, {Allen}, {Flaherty}, {Hartmann}, {Myers}, {Pipher},
  {Stauffer}, {Young}, \& {Fazio}}]{Megeath2016}
{Megeath}, S.~T., {Gutermuth}, R., {Muzerolle}, J., {et~al.} 2016, \aj, 151, 5

\bibitem[{{Megier} {et~al.}(2005){Megier}, {Strobel}, {Bondar}, {Musaev},
  {Han}, {Krexowski}, \& {Galazutdinov}}]{Megier2005}
{Megier}, A., {Strobel}, A., {Bondar}, A., {et~al.} 2005, ApJ, 634, 451

\bibitem[{{Menten} {et~al.}(2007){Menten}, {Reid}, {Forbrich}, \&
  {Brunthaler}}]{Menten2007}
{Menten}, K.~M., {Reid}, M.~J., {Forbrich}, J., \& {Brunthaler}, A. 2007, A\&A,
  474, 515

\bibitem[{{Nakamura} {et~al.}(2006){Nakamura}, {McKee}, {Klein}, \&
  {Fisher}}]{Nakamura2006}
{Nakamura}, F., {McKee}, C.~F., {Klein}, R.~I., \& {Fisher}, R.~T. 2006, \apjs,
  164, 477

\bibitem[{{Norman} \& {Ikeuchi}(1989)}]{NormanIkeuchi1989}
{Norman}, C.~A. \& {Ikeuchi}, S. 1989, \apj, 345, 372

\bibitem[{{Ochsendorf} {et~al.}(2015){Ochsendorf}, {Brown}, {Bally}, \&
  {Tielens}}]{Ochsendorf2015}
{Ochsendorf}, B.~B., {Brown}, A.~G.~A., {Bally}, J., \& {Tielens}, A.~G.~G.~M.
  2015, ApJ, 808, 111

\bibitem[{{Ochsendorf} {et~al.}(2014){Ochsendorf}, {Verdolini}, {Cox},
  {Bern\'{e}}, {Kaper}, \& {Tielens}}]{Ochsendorf2014}
{Ochsendorf}, B.~B., {Verdolini}, S., {Cox}, N.~L.~J., {et~al.} 2014, A\&A,
  566, A75

\bibitem[{{O'Dell}(2001)}]{ODell2001}
{O'Dell}, C.~R. 2001, ARAA, 39, 99

\bibitem[{{O'Dell}(2018)}]{ODell2018}
{O'Dell}, C.~R. 2018, MNRAS, 478, 1017

\bibitem[{{O'Dell} \& {Goss}(2009)}]{ODellGoss2009}
{O'Dell}, C.~R. \& {Goss}, W.~M. 2009, \aj, 138, 1235

\bibitem[{{O'Dell} \& {Harris}(2010)}]{ODell2010}
{O'Dell}, C.~R. \& {Harris}, J.~A. 2010, ApJ, 140, 985

\bibitem[{{O'Dell} {et~al.}(2009){O'Dell}, {Henney}, {Abel}, {Ferland}, \&
  {Arthur}}]{ODell2009}
{O'Dell}, C.~R., {Henney}, W.~J., {Abel}, N.~P., {Ferland}, G.~J., \& {Arthur},
  S.~J. 2009, ApJ, 137, 367

\bibitem[{{O'Dell} {et~al.}(2017){O'Dell}, {Kollatschny}, \&
  {Ferland}}]{ODell2017}
{O'Dell}, C.~R., {Kollatschny}, W., \& {Ferland}, G.~J. 2017, ApJ, 837, 151

\bibitem[{{O'Dell} \& {Yusef-Zadeh}(2000)}]{ODell2000}
{O'Dell}, C.~R. \& {Yusef-Zadeh}, F. 2000, ApJ, 120, 382

\bibitem[{{Oskinova} {et~al.}(2011){Oskinova}, {Todt}, {Ignace}, {Brown},
  {Cassinelli}, \& {Hamann}}]{Oskinova2011}
{Oskinova}, L.~M., {Todt}, H., {Ignace}, R., {et~al.} 2011, Mon. Not. R.
  Astron. Soc., 416, 1456

\bibitem[{{Ossenkopf} {et~al.}(2013){Ossenkopf}, {R\"{o}llig}, {Neufeld},
  {Pilleri}, {Lis}, {Fuente}, {van der Tak}, \& {Bergin}}]{Ossenkopf2013}
{Ossenkopf}, V., {R\"{o}llig}, M., {Neufeld}, D.~A., {et~al.} 2013, A\&A, 550,
  A57

\bibitem[{{Ostriker} \& {McKee}(1988)}]{OstrikerMcKee1988}
{Ostriker}, J.~P. \& {McKee}, C.~F. 1988, Reviews of Modern Physics, 60, 1

\bibitem[{{Pabst} {et~al.}(2019){Pabst}, {Higgins}, {Goicoechea}, {Teyssier},
  {Berne}, {Chambers}, {Wolfire}, {Suri}, {Guesten}, {Stutzki}, {Graf},
  {Risacher}, \& {Tielens}}]{Pabst2019}
{Pabst}, C., {Higgins}, R., {Goicoechea}, J.~R., {et~al.} 2019, \nat, 565, 618

\bibitem[{{Pabst} {et~al.}(2017){Pabst}, {Goicoechea}, {Teyssier}, {Bern{\'e}},
  {Ochsendorf}, {Wolfire}, {Higgins}, {Riquelme}, {Risacher}, {Pety}, {Le
  Petit}, {Roueff}, {Bron}, \& {Tielens}}]{Pabst2017}
{Pabst}, C.~H.~M., {Goicoechea}, J.~R., {Teyssier}, D., {et~al.} 2017, \aap,
  606, A29

\bibitem[{{Pauldrach} {et~al.}(2001){Pauldrach}, {Hoffmann}, \&
  {Lennon}}]{Pauldrach2001}
{Pauldrach}, A.~W.~A., {Hoffmann}, T.~L., \& {Lennon}, M. 2001, A\&A, 375, 161

\bibitem[{{Pellegrini} {et~al.}(2009){Pellegrini}, {Baldwin}, {Ferland},
  {Shaw}, \& {Heathcote}}]{Pellegrini2009}
{Pellegrini}, E.~W., {Baldwin}, J.~A., {Ferland}, G.~J., {Shaw}, G., \&
  {Heathcote}, S. 2009, ApJ, 693, 285

\bibitem[{{Peterson} \& {Megeath}(2008)}]{Peterson2008}
{Peterson}, D.~E. \& {Megeath}, S.~T. 2008, Handbook of Star Forming Regions
  Vol. I

\bibitem[{{Petit} {et~al.}(2012){Petit}, {Owocki}, {Oksala}, \& {the MiMeS
  Collaboration}}]{Petit2012}
{Petit}, V., {Owocki}, S.~P., {Oksala}, M.~E., \& {the MiMeS Collaboration}.
  2012, ASP Conference Series, 465, 48

\bibitem[{{Poglitsch} {et~al.}(2010){Poglitsch}, {Waelkens}, {Geis},
  {Feuchtgruber}, {Vandenbussche}, {Rodriguez}, {Krause}, {Renotte}, {van
  Hoof}, {Saraceno}, {Cepa}, {Kerschbaum}, \& {et al}}]{Poglitsch2010}
{Poglitsch}, A., {Waelkens}, C., {Geis}, N., {et~al.} 2010, A\&A, 518, L2

\bibitem[{{Pound} \& {Wolfire}(2008)}]{PoundWolfire2008}
{Pound}, M.~W. \& {Wolfire}, M.~G. 2008, Astronomical Society of the Pacific
  Conference Series, Vol. 394, {The Photo Dissociation Region Toolbox}, ed.
  R.~W. {Argyle}, P.~S. {Bunclark}, \& J.~R. {Lewis}, 654

\bibitem[{{Prosser} {et~al.}(1994){Prosser}, {Stauffer}, {Hartmann},
  {Soderblom}, {Jones}, {Werner}, \& {McCaughrean}}]{Prosser1994}
{Prosser}, C.~F., {Stauffer}, J.~R., {Hartmann}, L., {et~al.} 1994, ApJ, 421,
  517

\bibitem[{{Risacher} {et~al.}(2016){Risacher}, {G\"{u}sten}, {Stutzki},
  {H\"{u}bers}, {Bell}, {Buchbender}, {B\"{u}chel}, \&
  {Csengeri}}]{Risacher2016}
{Risacher}, C., {G\"{u}sten}, R., {Stutzki}, J., {et~al.} 2016, A\&A, 595, A34

\bibitem[{{Salgado} {et~al.}(2016){Salgado}, {Bern\'e}, {Adams}, {Herter},
  {Keller}, \& {Tielens}}]{Salgado2016}
{Salgado}, F., {Bern\'e}, O., {Adams}, J.~D., {et~al.} 2016, ApJ, 830, 118

\bibitem[{{Schulz} {et~al.}(2006){Schulz}, {Testa}, {Huenemoerder},
  {Ishibashi}, \& {Canizares}}]{Schulz2006}
{Schulz}, N.~S., {Testa}, P., {Huenemoerder}, D.~P., {Ishibashi}, K., \&
  {Canizares}, C. 2006, ApJ, 653, 636

\bibitem[{{Silich} \& {Tenorio-Tagle}(2013)}]{Silich2013}
{Silich}, S. \& {Tenorio-Tagle}, G. 2013, \apj, 765, 43

\bibitem[{{Sim\'{o}n-D\'{i}az} {et~al.}(2011){Sim\'{o}n-D\'{i}az},
  {Garc\'{i}a-Rojas}, {Esteban}, {Stasi\'{n}ska}, {L\'{o}pez-S\'{a}nchez}, \&
  {Morisset}}]{Simon-Diaz2011}
{Sim\'{o}n-D\'{i}az}, S., {Garc\'{i}a-Rojas}, J., {Esteban}, C., {et~al.} 2011,
  A\&A, 530, A57

\bibitem[{{Sim\'{o}n-D\'{i}az} {et~al.}(2006){Sim\'{o}n-D\'{i}az}, {Herrero},
  {Esteban}, \& {Najarro}}]{Simon-Diaz2006}
{Sim\'{o}n-D\'{i}az}, S., {Herrero}, A., {Esteban}, C., \& {Najarro}, F. 2006,
  A\&A, 448, 351

\bibitem[{{Sofia} {et~al.}(2004){Sofia}, {Lauroesch}, {Meyer}, \&
  {Cartledge}}]{Sofia2004}
{Sofia}, U.~J., {Lauroesch}, J.~T., {Meyer}, D.~M., \& {Cartledge}, S.~I.~B.
  2004, ApJ, 605, 272

\bibitem[{{Spitzer}(1978)}]{Spitzer1978}
{Spitzer}, L. 1978, Physical Processes in the Interstellar Medium (New York:
  Wiley-Interscience)

\bibitem[{{Stahl} {et~al.}(1996){Stahl}, {Kaufer}, {Rivinius}, {Szeifert},
  {Wolf}, {Gaeng}, {Gummersbach}, {Jankovics}, {Kovacs}, {Mandel}, {Pakull}, \&
  {Peitz}}]{Stahl1996}
{Stahl}, O., {Kaufer}, A., {Rivinius}, T., {et~al.} 1996, A\&A, 312, 539

\bibitem[{{Stahl} {et~al.}(2008){Stahl}, {Wade}, {Petit}, {Stober}, \&
  {Schanne}}]{Stahl2008}
{Stahl}, O., {Wade}, G., {Petit}, V., {Stober}, B., \& {Schanne}, L. 2008,
  A\&A, 487, 323

\bibitem[{{Sternberg} {et~al.}(2003){Sternberg}, {Hoffmann}, \&
  {Pauldrach}}]{Sternberg2003}
{Sternberg}, A., {Hoffmann}, T.~L., \& {Pauldrach}, A.~W.~A. 2003, ApJ, 599,
  1333

\bibitem[{{Stone} \& {Gardiner}(2007)}]{Stone2007}
{Stone}, J.~M. \& {Gardiner}, T. 2007, \apj, 671, 1726

\bibitem[{{Storey} \& {Hummer}(1995)}]{StoreyHummer1995}
{Storey}, P.~J. \& {Hummer}, D.~G. 1995, MNRAS, 272, 41

\bibitem[{{St{\"o}rzer} \& {Hollenbach}(1999)}]{Stoerzer1999}
{St{\"o}rzer}, H. \& {Hollenbach}, D. 1999, ApJ, 515, 669

\bibitem[{{Subrahmanyan} {et~al.}(2001){Subrahmanyan}, {Goss}, \&
  {Malin}}]{Subrahmanyan2001}
{Subrahmanyan}, R., {Goss}, W.~M., \& {Malin}, D.~F. 2001, \aj, 121, 399

\bibitem[{{Tielens}(2010)}]{Tielens}
{Tielens}, A.~G.~G.~M. 2010, The Physics and Chemistry of the Interstellar
  Medium (New York: Cambridge University Press)

\bibitem[{{Tielens} {et~al.}(1993){Tielens}, {Meixner}, {van der Werf},
  {Bregman}, {Tauber}, {Stutzki}, \& {Rank}}]{Tielens1993}
{Tielens}, A.~G.~G.~M., {Meixner}, M.~M., {van der Werf}, P.~P., {et~al.} 1993,
  Science, 262, 86

\bibitem[{{Troland} {et~al.}(2016){Troland}, {Goss}, {Brogan}, {Crutcher}, \&
  {Roberts}}]{Troland2016}
{Troland}, T.~H., {Goss}, W.~M., {Brogan}, C.~L., {Crutcher}, R.~M., \&
  {Roberts}, D.~A. 2016, ApJ, 825, 2

\bibitem[{{Van der Werf} {et~al.}(2013){Van der Werf}, {Goss}, \&
  {O'Dell}}]{VanderWerf2013}
{Van der Werf}, P.~P., {Goss}, W.~M., \& {O'Dell}, C.~R. 2013, ApJ, 762, 101

\bibitem[{{Vitrichenko}(2002)}]{Vitrichenko2002}
{Vitrichenko}, E.~A. 2002, AstL, 28, 324

\bibitem[{{Walch} {et~al.}(2013){Walch}, {Whitworth}, {Bisbas}, {W\"{u}nsch},
  \& {Hubber}}]{Walch2013}
{Walch}, S., {Whitworth}, A.~P., {Bisbas}, T.~G., {W\"{u}nsch}, R., \&
  {Hubber}, D. 2013, MNRAS, 435, 917

\bibitem[{{Walch} {et~al.}(2012){Walch}, {Whitworth}, {Bisbas}, {W\"{u}nsch},
  \& {Hubber}}]{Walch2012}
{Walch}, S.~K., {Whitworth}, A.~P., {Bisbas}, T., {W\"{u}nsch}, R., \&
  {Hubber}, D. 2012, MNRAS, 427, 625

\bibitem[{{Wareing} {et~al.}(2018){Wareing}, {Pittard}, {Wright}, \&
  {Falle}}]{Wareing2018}
{Wareing}, C.~J., {Pittard}, J.~M., {Wright}, N.~J., \& {Falle}, S.~A.~E.~G.
  2018, \mnras, 475, 3598

\bibitem[{{Weaver} {et~al.}(1977){Weaver}, {McCray}, {Castor}, {Shapiro}, \&
  {Moore}}]{Weaver1977}
{Weaver}, R., {McCray}, R., {Castor}, J., {Shapiro}, P., \& {Moore}, R. 1977,
  ApJ, 218, 377

\bibitem[{{Weilbacher} {et~al.}(2015){Weilbacher}, {Monreal-Ibero},
  {Kollatschny}, {Ginsburg}, {McLeod}, {Kamann}, {Sandin}, \& {et
  al.}}]{Weilbacher2015}
{Weilbacher}, P.~M., {Monreal-Ibero}, A., {Kollatschny}, W., {et~al.} 2015,
  A\&A, 582, A114

\bibitem[{{Weingartner} \& {Draine}(2001)}]{Weingartner2001}
{Weingartner}, J.~C. \& {Draine}, B.~T. 2001, ApJ, 548, 296

\bibitem[{{Williams} \& {McKee}(1997)}]{WilliamsMcKee1997}
{Williams}, J.~P. \& {McKee}, C.~F. 1997, \apj, 476, 166

\bibitem[{{Wilson} {et~al.}(1997){Wilson}, {Filges}, {Codella}, {Reich}, \&
  {Reich}}]{Wilson1997}
{Wilson}, T.~L., {Filges}, L., {Codella}, C., {Reich}, W., \& {Reich}, P. 1997,
  A\&A, 327, 1177

\end{thebibliography}

\newpage

\appendix

\section{SED results}

\begin{figure}[htb]
\includegraphics[width=0.5\textwidth, height=0.33\textwidth]{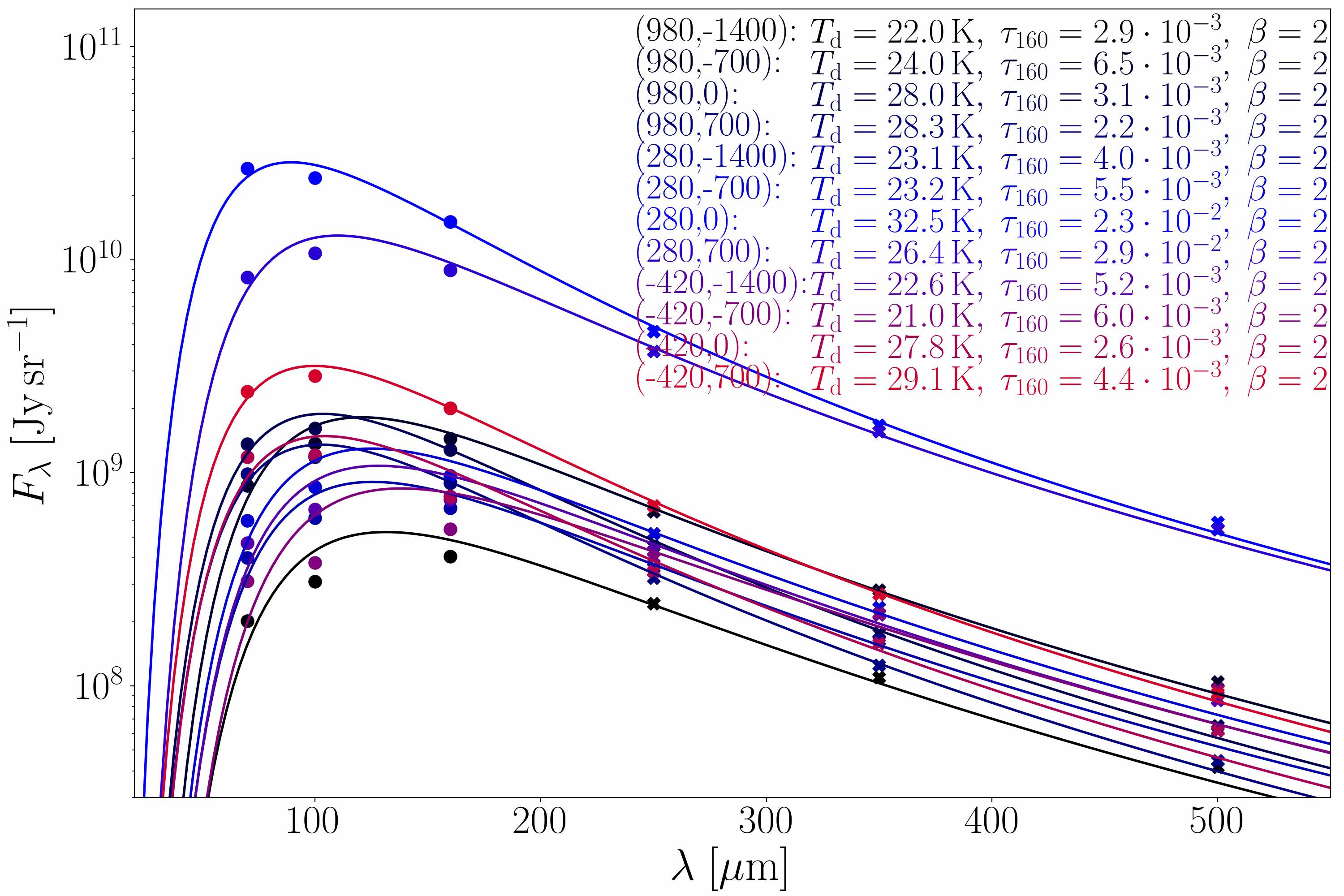}
\caption{SED fits to individual points throughout the Orion Nebula complex. The flux in the PACS bands is plotted as circles; fluxes of the SPIRE bands are plotted as crosses. Numbers in brackets are the R.A. and Dec. offset, respectively, in arc seconds.}
\label{Fig.SED-points}
\end{figure}

Figure \ref{Fig.SED-points} shows the results from an SED fit in individual points with the emissivity index $\beta=2$. The derived dust temperature decreases with increasing $\beta$, while the dust optical depth increases significantly. The integrated FIR intensity does not depend strongly on $\beta$.

\section{The expanding Rim}
\label{App.east-shell}

\begin{figure*}[tb]
\begin{minipage}[t]{0.49\textwidth}
\includegraphics[width=\textwidth, height=0.91\textwidth]{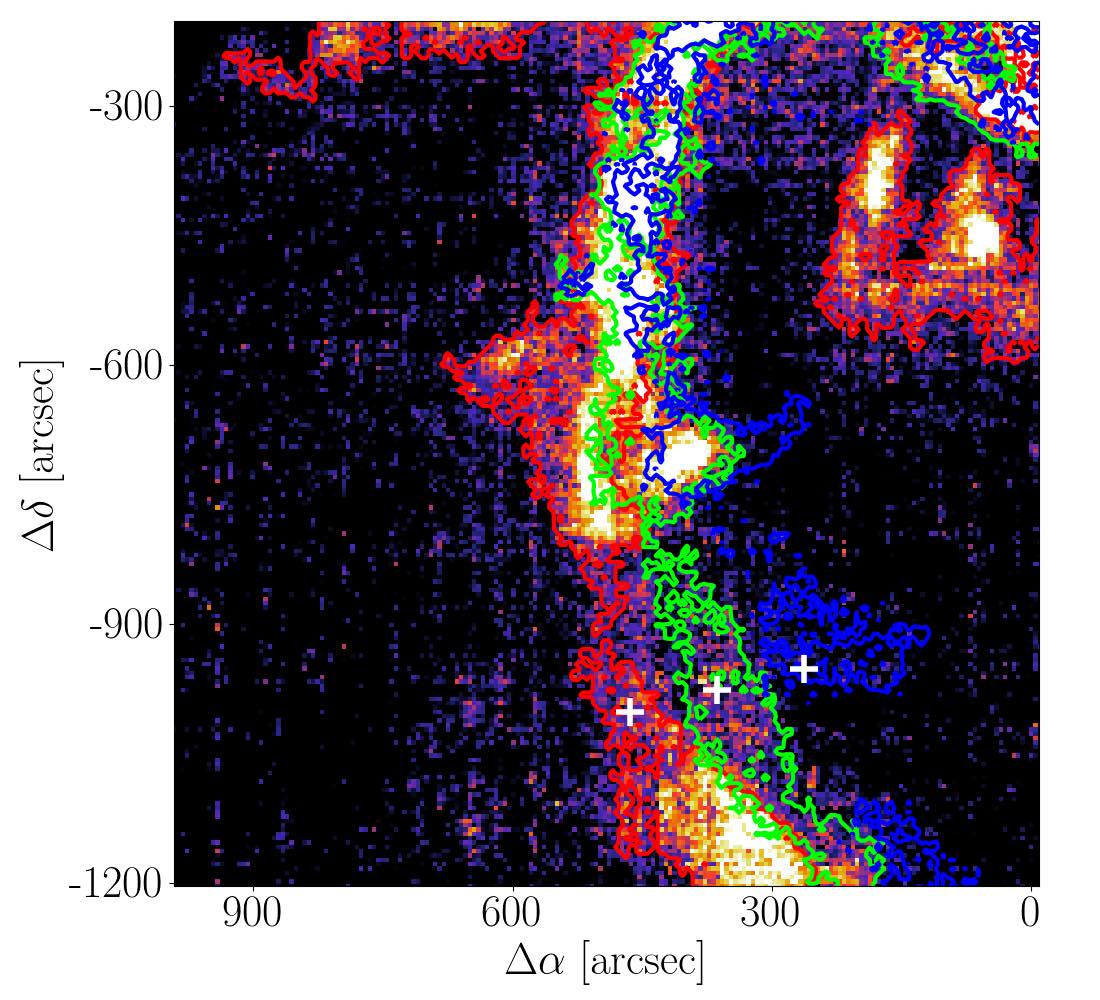}
\caption{Color map of [C\,{\sc ii}] emission in velocity interval $v_{\mathrm{LSR}}=0\mhyphen 5\,\mathrm{km\,s^{-1}}$, with contours of the channel emission in $v_{\mathrm{LSR}}=0\mhyphen 1\,\mathrm{km\,s^{-1}}$, $v_{\mathrm{LSR}}=2\mhyphen 3\,\mathrm{km\,s^{-1}}$, and $v_{\mathrm{LSR}}=4\mhyphen 5\,\mathrm{km\,s^{-1}}$ (blue, green, and red, respectively). From the displacement of the structure of brightest emission in these channels with respect to each other, points marked by the white crosses, we can estimate the expansion velocity to be $v_{\mathrm{exp}}\simeq 4.25\,\mathrm{km\,s^{-1}}$. The projected distance between the crosses is $0.2\,\mathrm{pc}$, and the projected distance from the Trapezium stars of the outermost filament, traced by blue contours, is $1.75\,\mathrm{pc}$.}
\label{Fig.east_shell_0-5}
\end{minipage}
\hspace{0.01\textwidth}
\begin{minipage}[t]{0.49\textwidth}
\includegraphics[width=\textwidth, height=0.91\textwidth]{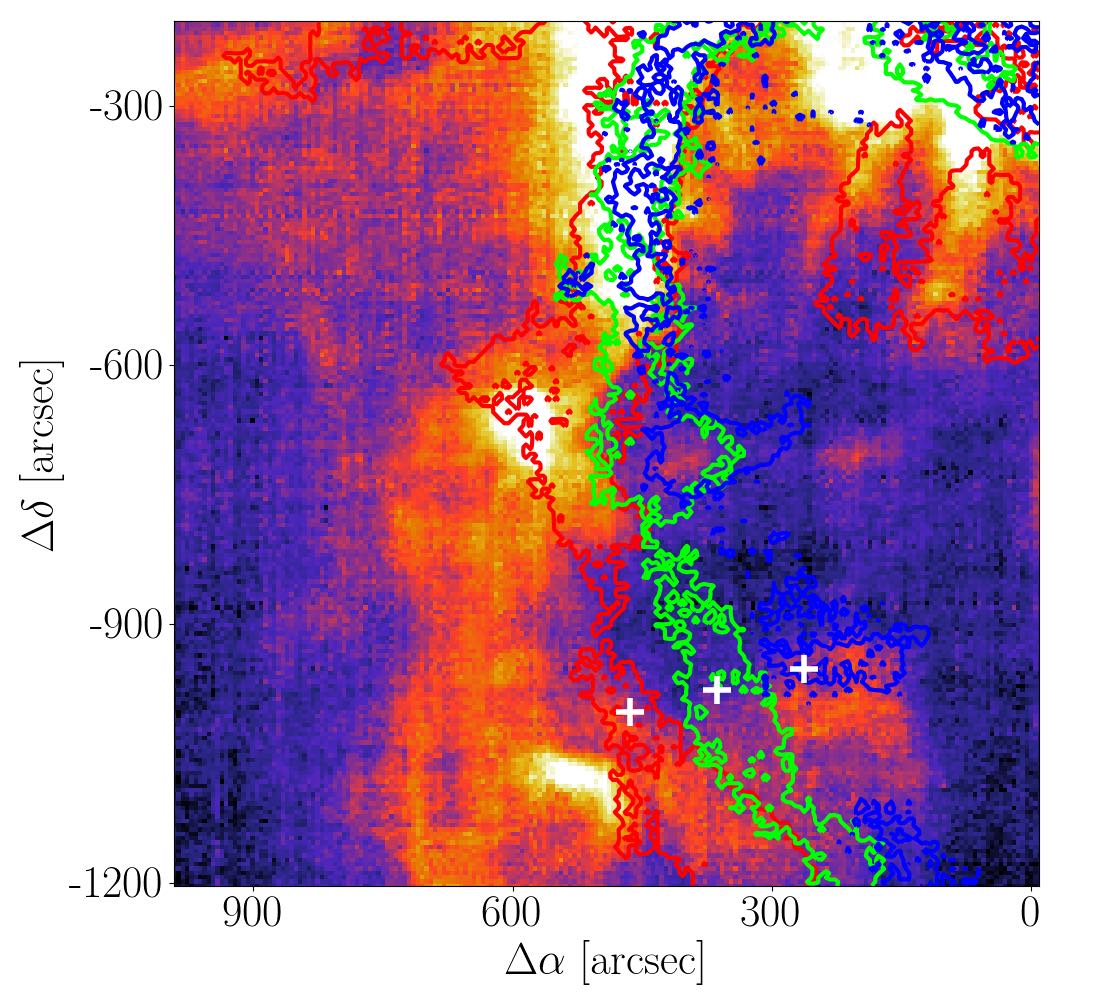}
\caption{Color map of [C\,{\sc ii}] emission in velocity interval $v_{\mathrm{LSR}}=0\mhyphen 15\,\mathrm{km\,s^{-1}}$, with contours of the channel emission in $v_{\mathrm{LSR}}=0\mhyphen 1\,\mathrm{km\,s^{-1}}$, $v_{\mathrm{LSR}}=2\mhyphen 3\,\mathrm{km\,s^{-1}}$, and $v_{\mathrm{LSR}}=4\mhyphen 5\,\mathrm{km\,s^{-1}}$ (blue, green, and red, respectively). The expanding filament grows brighter as it moves outward, probably mainly due to increasing limb-brightening, so the color map is dominated by the most outward filament.}
\label{Fig.east_shell_0-15}
\end{minipage}
\end{figure*}

We present here our analysis of the kinematics in the eastern bright arm, the so-called Eastern Rim, of the expanding shell. To estimate the three-dimensional expansion velocity, we make use of the following formula:
\begin{align}
v_{\mathrm{exp}} = \frac{v_2-v_1}{\sqrt{1-\frac{r_1^2}{r_2^2}}},
\end{align}
where $v_2-v_1$ is the velocity difference of the velocity channels considered; $r_1$ and $r_2$ are the respective distances of the emission filament in these two channels. $r_2$ is the further outward lying filament, for an accurate calculation it must be the outermost. We can write $r_1=r_2-\Delta r$, where $\Delta r$ is the separation between the two channel filaments. If $v_2>v_1$, the shell is moving toward us and we assign a positive expansion velocity; else it is expanding away from us. Here, we use the [C\,{\sc ii}] channels at $v_{\mathrm{LSR}}=0\mhyphen 1\,\mathrm{km\,s^{-1}}$, $v_{\mathrm{LSR}}=2\mhyphen 3\,\mathrm{km\,s^{-1}}$ and $v_{\mathrm{LSR}}=4\mhyphen 5\,\mathrm{km\,s^{-1}}$. With $v_2-v_1=2\,\mathrm{km\,s^{-1}}$, $r_2\simeq 1.75\,\mathrm{pc}$ and $\Delta r\simeq 0.2\,\mathrm{pc}$ (cf Fig. \ref{Fig.east_shell_0-5}), we obtain $v_{\mathrm{exp}}\simeq 4.25\,\mathrm{km\,s^{-1}}$. We note that only the distance of the furthest filament (cf. Fig. \ref{Fig.east_shell_0-15}) gives an accurate measure of the extent of the expanding shell, which renders the estimate from the three channels used here somewhat inaccurate; however, in higher-velocity channels the displacement between the filamentary structures is not as clear-cut as it is in these channels.

\begin{figure}[tb]
\includegraphics[width=0.5\textwidth, height=0.267\textwidth]{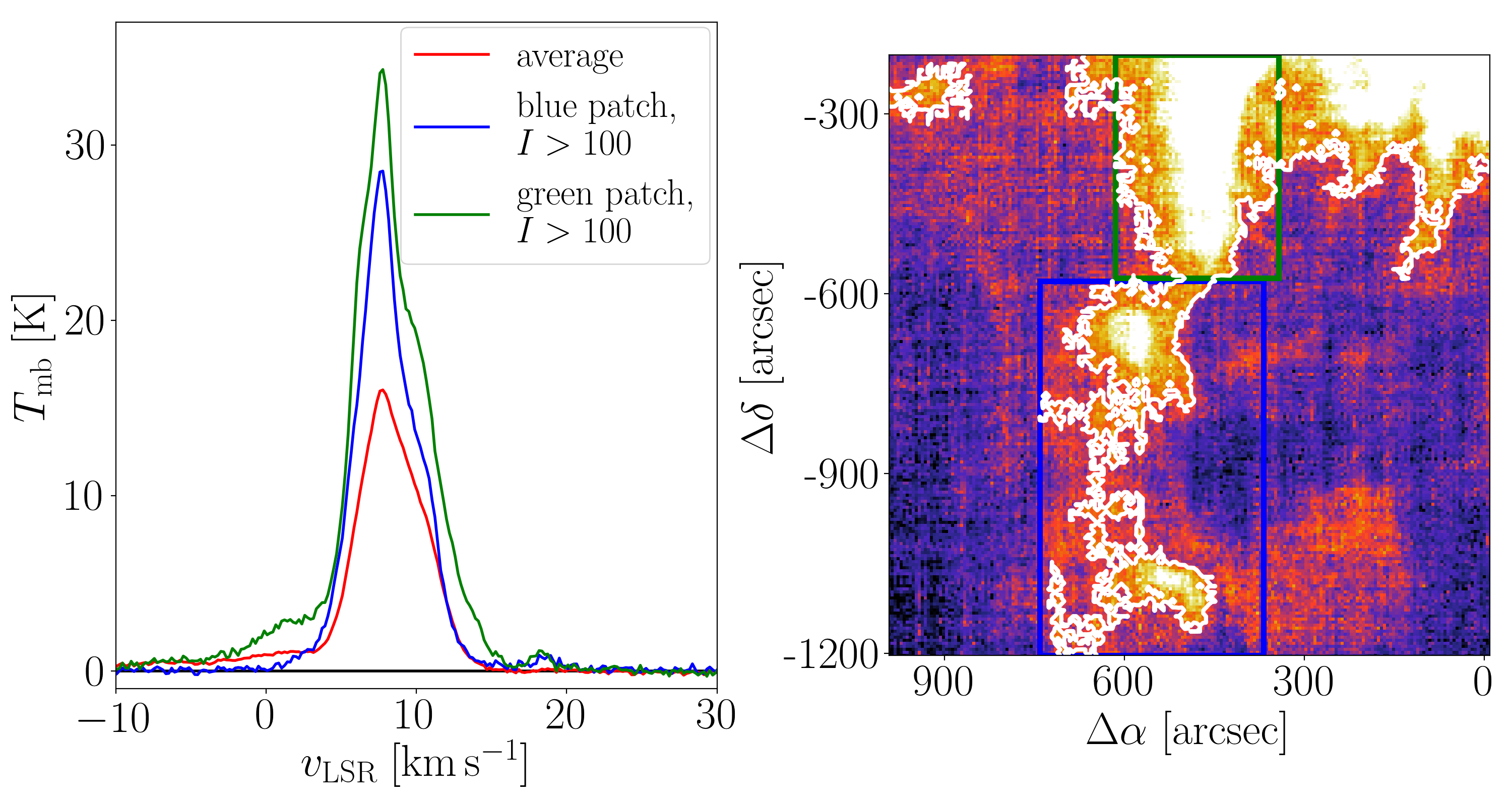}
\caption{Right: [C\,{\sc ii}] line-integrated intensity in the bright eastern Rim. White contours indicate $I_{\mathrm{[C\,\textsc{ii}]}} > 100\,\mathrm{K\;km\,s^{-1}}$. Left: Averaged spectra from this region as indicated. Results of the analysis of the peak temperatures are given in Table \ref{Tab.results_Rim}.}
\label{Fig.east-shell-13CII}
\end{figure}

We detect the [$^{13}$C\,{\sc ii}] $F=2\mhyphen 1$ line in averaged spectra in this region as shown in Fig. \ref{Fig.east-shell-13CII}. From the peak temperatures of the [$^{12}$C\,{\sc ii}] and [$^{13}$C\,{\sc ii}] line (Table \ref{Tab.results_Rim}), we compute the [C\,{\sc ii}] optical depth by Eqs. \ref{eq.tauCII} and \ref{eq.Tex}. The C$^+$ column density is given by Eq. \ref{eq.tauC+}. The analysis yields $\tau_{\mathrm{[C\,\textsc{ii}]}}\simeq 3$. From the column density, we estimate a gas density of the limb-brightened shell from $n=N_{\mathrm{C^+}}/(\mathrm{[C/H]} l)$, where $\mathrm{[C/H]}\simeq 1.6\cdot 10^{-4}$ \citep{Sofia2004} and the extent $l$ along the line of sight is $l\simeq r/2 \simeq 1\,\mathrm{pc}$. This yields $n\simeq 1\cdot 10^4\,\mathrm{cm^{-3}}$.

\begin{figure}[tb]
\includegraphics[width=0.5\textwidth, height=0.33\textwidth]{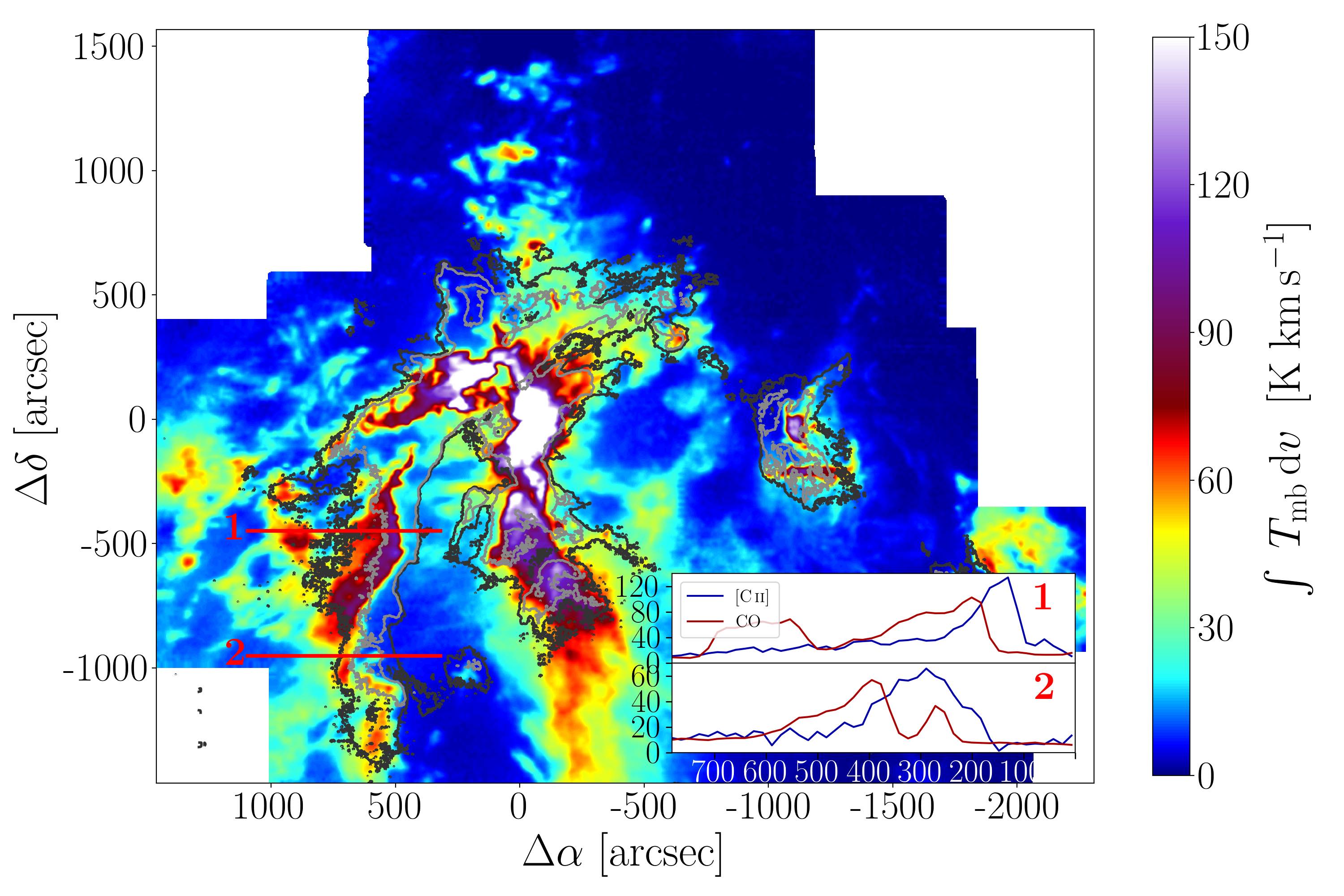}
\caption{$^{12}$CO(2-1) emission in velocity interval $v_{\mathrm{LSR}}=5\mhyphen 8\,\mathrm{km\,s^{-1}}$ with [C\,{\sc ii}] emission in same velocity interval in contours (dark grey: $30\,\mathrm{K\,km\,s^{-1}}$, light grey: $50\,\mathrm{K\,km\,s^{-1}}$). The inlaid panels show the line cuts along the two red lines indicated in the figure.}
\label{Fig.M42_CO21_CII}
\end{figure}

Figure \ref{Fig.M42_CO21_CII} compares the CO(2-1) emission with the [C\,{\sc ii}] emission in the expanding Rim, that is observed in the velocity range $v_{\mathrm{LSR}}=5\mhyphen 8\,\mathrm{km\,s^{-1}}$. From the peak separation in line cut 1, $d\simeq 70\arcsec\simeq 0.14\,\mathrm{pc}$, we derive a density of $n\simeq 9\cdot 10^3\,\mathrm{cm^{-3}}$ in the Rim. From line cut 2 ($d\simeq 110\arcsec\simeq 0.22\,\mathrm{pc}$), we derive a density of $n\simeq 6\cdot 10^3\,\mathrm{cm^{-3}}$ in the south of the bright Rim.

\begin{table}[htb]
\addtolength{\tabcolsep}{-1pt}
\begin{tabular}{l|ccccc}
 & $T_{\mathrm{P}}(\mathrm{[^{12}C\,\textsc{ii}])}$ &$T_{\mathrm{P}}(\mathrm{[^{13}C\,\textsc{ii}])}$ & $\tau_{\mathrm{[C\,\textsc{ii}]}}$ & $T_{\mathrm{ex}}$ & $N_{\mathrm{C^+}}$ \\
 & $\mathrm{[K]}$ & $\mathrm{[K]}$ &  & $\mathrm{[K]}$ & $\mathrm{[cm^{-3}]}$ \\ \hline
green & 35 & 1 & 2.9 & 48 & $5.7\cdot 10^{18}$ \\
blue & 30 & 1 & 3.4 & 44 & $6.4\cdot 10^{18}$
\end{tabular}
\caption{Gaussian fit parameters of spectra in Fig. \ref{Fig.east-shell-13CII}.}
\label{Tab.results_Rim}
\end{table}

From the dust optical depth, we estimate a mean $\tau_{160}\simeq 4\cdot 10^{-3}$ in the [C\,{\sc ii}]-bright regions. This translates to a gas column density of $N_{\mathrm{H}}\simeq 2\cdot 10^{22}\,\mathrm{cm^{-2}}$. Assuming that all carbon is ionized, this results in a C$^+$ column density of $N_{\mathrm{C^+}}\simeq 3\cdot 10^{18}\,\mathrm{cm^{-2}}$, which is a factor of 2 lower than what we estimate from the [C\,{\sc ii}] spectra. Since $\tau_{160}$ strongly depends on the choice of the grain emissivity index $\beta$, the agreement might be satisfactory. However, it might be a hint that we overestimate the [C\,{\sc ii}] optical depth and underestimate the excitation temperature. With these givens the physical temperature of the gas would be $T\simeq 50\,\mathrm{K}$, which is lower than expected from the PDR models for the [C\,{\sc ii}]-emitting (surface) layers. At these high opacities opacity broadening of the line might also become significant, which is not what we observe. The fitting procedure is complicated by the presence of multiple components. The main line can be fitted by two emission components or by one emission and one absorption component. However, the latter would increase the result for $\tau_{\mathrm{[C\,\textsc{ii}]}}$ even further. In conclusion, the values obtained here can only be a crude estimate due to the complex emission structure from this region, which might be inadmissibly averaged together in the spectra.

\section{PV diagrams}
\label{App.pv-diagrams}

\begin{figure*}[tb]
\includegraphics[width=\textwidth, height=0.67\textwidth]{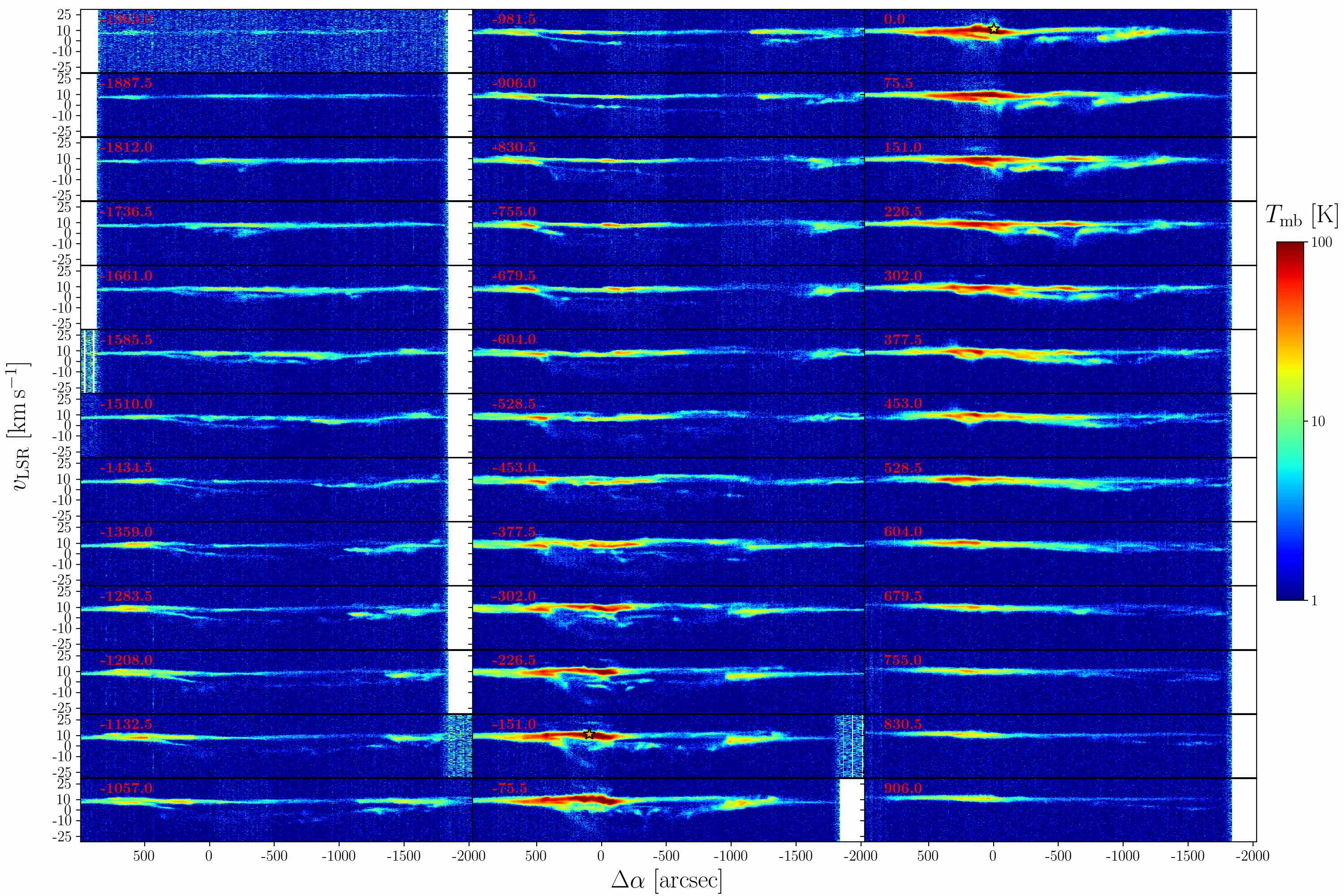}
\caption{[C\,{\sc ii}] pv diagrams from the Veil shell, sliced along R.A. axis. Red numbers are the respective Dec. offsets in arc seconds.}
\label{Fig.pv-diagrams-x_wolines}
\end{figure*}

\begin{figure*}[tb]
\includegraphics[width=\textwidth, height=0.67\textwidth]{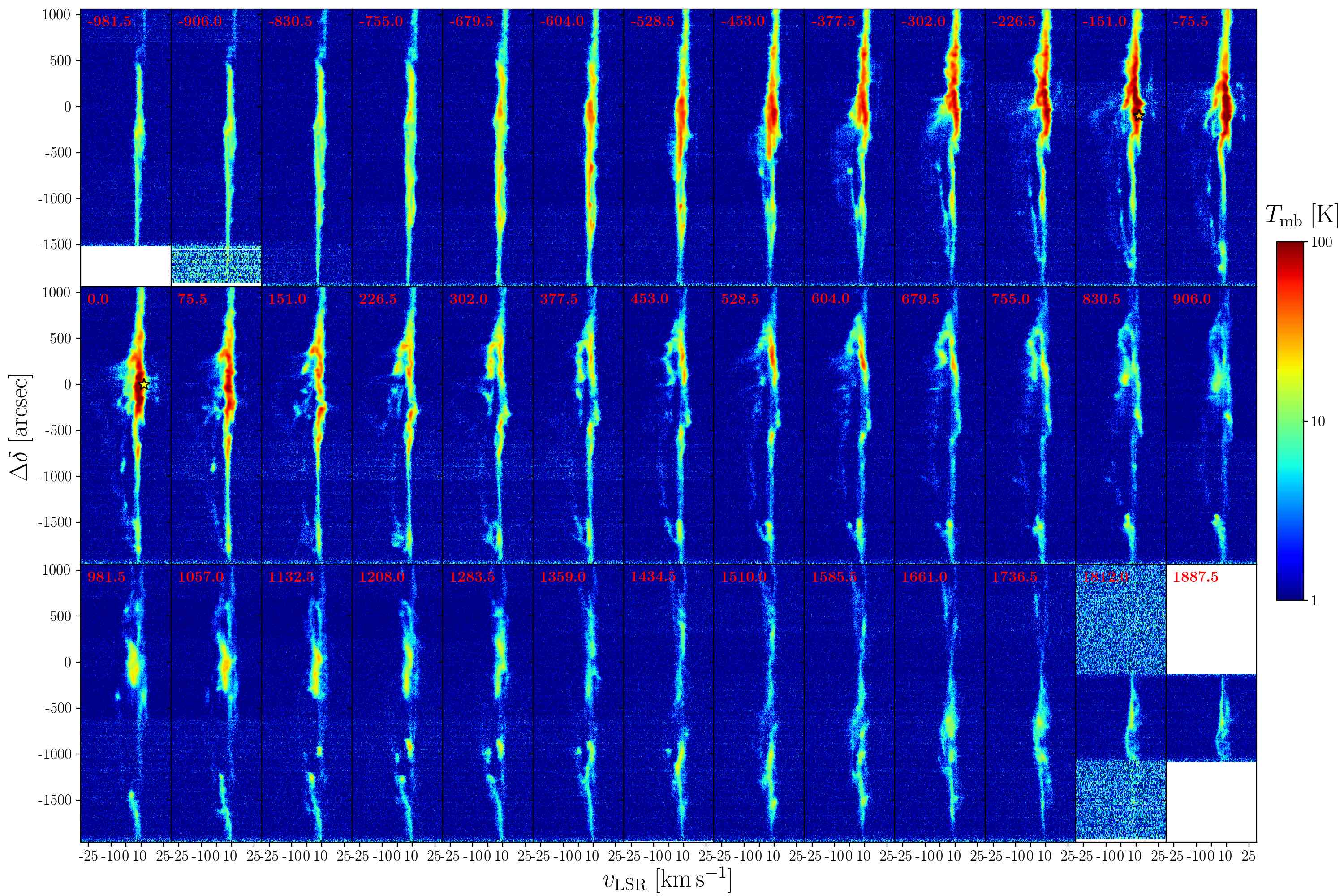}
\caption{[C\,{\sc ii}] pv diagrams from the Veil shell, sliced along Dec. axis. Red numbers are the respective R.A. offsets in arc seconds.}
\label{Fig.pv-diagrams-y_wolines}
\end{figure*}

\begin{figure*}[tb]
\includegraphics[width=\textwidth, height=0.67\textwidth]{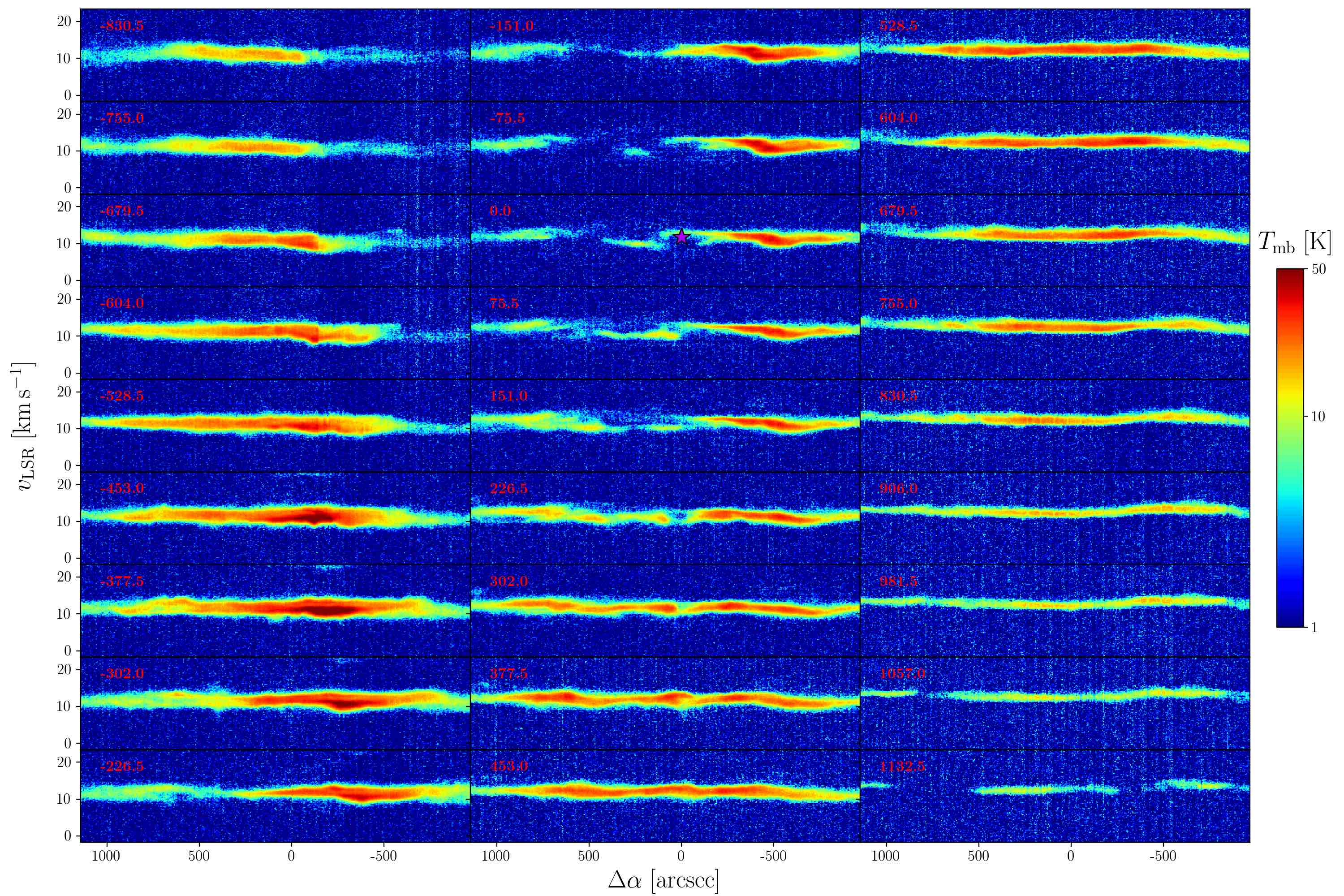}
\caption{[C\,{\sc ii}] pv diagrams from NGC 1977, sliced along R.A. axis. Red numbers are the respective Dec. offsets in arc seconds. Coordinate offsets are given with respect to the position of 42 Orionis.}
\label{Fig.pv-diagrams-x}
\end{figure*}

\begin{figure*}[tb]
\includegraphics[width=\textwidth, height=0.67\textwidth]{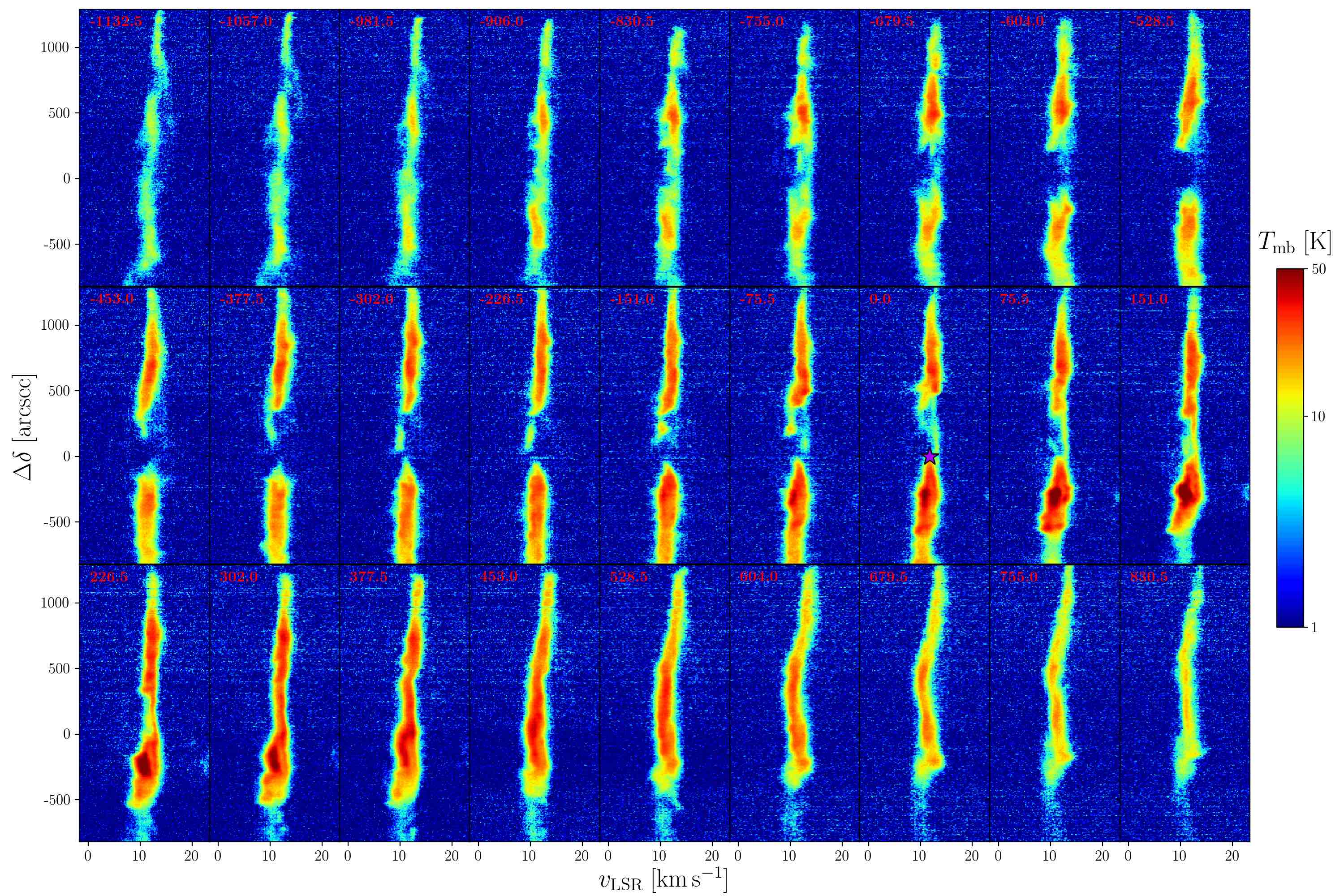}
\caption{[C\,{\sc ii}] pv diagrams from NGC 1977, sliced along Dec. axis. Red numbers are the respective R.A. offsets in arc seconds. Coordinate offsets are given with respect to the position of 42 Orionis.}
\label{Fig.pv-diagrams-y}
\end{figure*}

Figures \ref{Fig.pv-diagrams-x_wolines} to \ref{Fig.pv-diagrams-y} show the pv diagrams resulting from slicing the velocity-resolved [C\,{\sc ii}] map from the Extended Orion Nebula with M43 and NGC 1977 along the R.A. and Dec. axis, respectively. The coherent arc structure indicative of the expansion of the Veil shell can be seen in almost all pv diagrams. The expanding structure of NGC 1977 is more difficult to recognize.

\end{document}